\documentclass[8pt]{article}
\pdfoutput=1
     
\usepackage[usenames,dvipsnames,svgnames,table]{xcolor} 
\usepackage[obeyspaces,hyphens,spaces]{url}
\usepackage{jcapmod}
\usepackage{verbatim}
\usepackage{graphics}
\usepackage{epsfig}
\usepackage{booktabs}
\usepackage{booktabs}
\usepackage[english]{babel}
\usepackage{amsmath, amssymb, amsbsy, amstext, amsthm, simplewick}
\usepackage{hyperref}
\usepackage{graphicx}
\usepackage{amsfonts}
\usepackage{upgreek}
\usepackage{framed}
\usepackage{tensor}
\usepackage{pifont}
\usepackage{latexsym, mathrsfs}
\usepackage{array}
\usepackage{hyperref}
\usepackage{xspace}
\usepackage{longtable}
\usepackage{multirow}
\usepackage{cases}
\usepackage{empheq}

\usepackage{bm}
\usepackage{bbm}
\usepackage{float}
\usepackage{afterpage}
\usepackage{setspace}
\usepackage{caption}
\usepackage[load-configurations=astronomy, range-units=brackets, range-phrase=-, per-mode=reciprocal, mode=math]{siunitx}
\usepackage{threeparttable}
\usepackage{subfigure}
\usepackage{tcolorbox}
\allowdisplaybreaks 

\usepackage{braket}

\usepackage{ragged2e}

\usepackage{hhline}
\newcolumntype{P}[1]{>{\centering\arraybackslash}p{#1}}
\usepackage{booktabs}
\usepackage{tikz}
\usetikzlibrary{calc,shadings,patterns,tikzmark,fadings}
\usepackage{cleveref} 
\Crefname{equation}{Eq.}{Eqs.}
\Crefname{section}{Sec.}{Secs.}
\Crefname{figure}{Fig.}{Figs.}
\Crefname{table}{Table}{Tables}

\definecolor{Blue}{rgb}{0.25, 0.41, 0.88}
\definecolor{Red}{rgb}{0.92,0.,0.}
\definecolor{darkorange}{rgb}{1.0,0.549,0.}
\definecolor{cobalt}{RGB}{44, 98, 120}
\definecolor{Mathematica1}{rgb}{0.368417, 0.506779, 0.709798}
\definecolor{Mathematica2}{rgb}{0.880722, 0.611041, 0.142051}
\definecolor{Mathematica3}{rgb}{0.560181, 0.691569, 0.194885}
\definecolor{Mathematica4}{rgb}{0.922526, 0.385626, 0.209179}
\definecolor{Mathematica5}{rgb}{0.528488, 0.470624, 0.701351}
\definecolor{Mathematica6}{rgb}{0.772079, 0.431554, 0.102387}
\definecolor{Mathematica7}{rgb}{0.363898, 0.618501, 0.782349}
\definecolor{Mathematica8}{rgb}{1, 0.75, 0}
\definecolor{Mathematica9}{rgb}{0.647624, 0.37816, 0.614037}
\definecolor{plotBlue}{RGB}{94, 130, 181}
\definecolor{plotRed}{RGB}{233, 85, 54}
\definecolor{plotGreen}{RGB}{142, 176, 50}
\definecolor{plotPurple}{RGB}{135, 120, 178}

\newcolumntype{C}[1]{>{\centering\let\newline\\\arraybackslash\hspace{0pt}}m{#1}}

\def\d{{\rm d}}

\def\x{{\bf x}}

\makeatletter
\newlength{\apb@width}
\newcommand{\autoparbox}[2][c]{\settowidth{\apb@width}{#2}\parbox[#1]{\apb@width}{#2}}

\makeatother

\makeatletter
\newsavebox\myboxA
\newsavebox\myboxB
\newlength\mylenA

\newcommand*\xoverline[2][0.75]{
	\sbox{\myboxA}{$\m@th#2$}%
	\setbox\myboxB\null
	\ht\myboxB=\ht\myboxA%
	\dp\myboxB=\dp\myboxA%
	\wd\myboxB=#1\wd\myboxA
	\sbox\myboxB{$\m@th\overline{\copy\myboxB}$}
	\setlength\mylenA{\the\wd\myboxA}
	\addtolength\mylenA{-\the\wd\myboxB}%
	ifdim\wd\myboxB<\wd\myboxA%
	\rlap{\hskip 0.5\mylenA\usebox\myboxB}{\usebox\myboxA}%
	\else
	\hskip -0.5\mylenA\rlap{\usebox\myboxA}{\hskip 0.5\mylenA\usebox\myboxB}%
	\fi}
\makeatother

\numberwithin{equation}{section}
\numberwithin{figure}{section}
\numberwithin{table}{section}

\def\beq{\begin{equation}}
\def\eeq{\end{equation}}

\def\bea{\begin{eqnarray}}
\def\eea{\end{eqnarray}}

\def\d{{\rm d}}
\def\dd{{\rm d}}

\def\beq{\begin{equation}}
\def\eeq{\end{equation}}
\def\bea{\begin{eqnarray}}
\def\eea{\end{eqnarray}}

\def\d{{\rm d}}
\def\dd{{\rm d}}

\numberwithin{equation}{section}
\def\beq{\begin{equation}}
\def\eeq{\end{equation}}

\def\bea{\begin{eqnarray}}
\def\eea{\end{eqnarray}}

\def\d{{\rm d}}

\DeclareRobustCommand{\SkipTocEntry}[4]{}

\usepackage{colortbl}
\definecolor{blue2}{cmyk}{1, 0.1, 0.1, 0.1}

\definecolor{pyBlue}{RGB}{31, 119, 180}
\definecolor{pyRed}{RGB}{214, 39, 40}
\definecolor{pyGreen}{RGB}{44, 160, 44}
\definecolor{pyBlue2}{RGB}{0, 111, 237}
\definecolor{pyRed2}{RGB}{224, 52, 36}

\newcolumntype{P}[1]{>{\centering\arraybackslash}p{#1}}
\newcolumntype{M}[1]{>{\centering\arraybackslash}m{#1}}

\begin{document}

\pagenumbering{roman}
\begin{titlepage}
\baselineskip=14.5pt \thispagestyle{empty}

\bigskip\

\vspace{0cm}
\begin{center}
{\fontsize{40}{40}\selectfont  \bfseries \textcolor{Sepia}{${\cal C}$osmological  ${\cal G}$eometric ${\cal P}$hase}}\\
{\fontsize{40}{40}\selectfont  \bfseries \textcolor{Sepia}{~${\cal F}$rom~}}\\
{\fontsize{40}{40}\selectfont  \bfseries \textcolor{Sepia}{ ${\cal P}$ure~${\cal Q}$uantum~${\cal S}$tates}}\\  \vspace{0.25cm}
	{\fontsize{13}{16}\selectfont  \bfseries \textcolor{Sepia}{A study without/ with having Bell's inequality violation}}
\end{center}
\vspace{0.1cm}
\begin{center}
{\fontsize{15}{18}\selectfont Sayantan Choudhury$^{a,b}$
			\footnote{{\it NOTE: This project is the part of the non-profit virtual international research consortium ``Quantum Structures of the Space-Time \& Matter" }. ${}^{}$}} 
\end{center}

\begin{center}
\vskip8pt
\textit{${}^{a}$Centre For Cosmology and Science Popularization (CCSP),
SGT University, Gurugram, Delhi- NCR, Haryana- 122505, India}\\
	\textit{${}^{b}$School of Physical Sciences,  \\ National Institute of Science Education and Research, Bhubaneswar, Odisha - 752050, India}\\
		\textit{${}^{c}$Homi Bhabha National Institute, Training School Complex, Anushakti Nagar, Mumbai - 400085, India}\\ 
	
	\text{Email:~\textcolor{blue}{sayantan\_ccsp@sgtuniversity.org, sayantan.choudhury@niser.ac.in}, \textcolor{blue}{sayanphysicsisi@gmail.com} }


\end{center}

\vspace{0.09cm}
\hrule \vspace{0.09cm}
\begin{center}
\noindent {\bf Abstract}\\

\end{center} 
In this paper,  using the concept of  Lewis Riesenfeld invariant quantum operator method for finding continuous eigenvalues of quantum mechanical wave functions we derive the analytical expressions for the cosmological geometric phase,  which is commonly identified to be the {\it Pancharatnam Berry phase} from primordial cosmological perturbation scenario.  We compute this cosmological geometric phase from two possible physical situations,  (1) In the absence of Bell's inequality violation and (2) In the presence of Bell's inequality violation having the contributions in the sub Hubble region ($-k\tau\gg 1$),  super Hubble region ($-k\tau\ll 1$) and at the horizon crossing point ($-k\tau= 1$) for  massless field ($m/{\cal H}\ll 1$),  partially massless field ($m/{\cal H}\sim 1$) and massive/heavy field ($m/{\cal H}\gg 1$),  in the background of quantum field theory of spatially flat quasi De Sitter geometry.  The prime motivation for this work is to investigate the various unknown quantum mechanical features of primordial universe.  To give the realistic interpretation of the derived theoretical results we express everything initially in terms of slowly varying conformal time dependent parameters,  and then to connect with cosmological observation we further express the results in terms of cosmological observables,  which are spectral index/tilt of scalar mode power spectrum ($n_{\zeta}$) and tensor-to-scalar ratio ($r$).  Finally,  this identification helps us to provide the stringent numerical constraints on the {\it Pancharatnam Berry phase},  which confronts well with recent cosmological observation.

\vskip10pt
\hrule
\vskip10pt

\text{Keywords:~~Geometric~Phase, ~Bell's~inequality~violation,~QFT of De Sitter space,}\\
\text{~~~~~~~~~~~~~~~~~~ Cosmology,~Quantum Information Theory.}

\end{titlepage}

\thispagestyle{empty}
\setcounter{page}{2}
\begin{spacing}{1.03}
\tableofcontents
\end{spacing}

\clearpage
\pagenumbering{arabic}
\setcounter{page}{1}

\clearpage

\section{Introduction and Summary}
The {\it Pancharatnam Berry phase} is appearing within the framework of quantum mechanics particularly for pure quantum states where the physical system under consideration undergo slowly varying cyclic evolution \cite{Pancharatnam:1956jv,Berry:1984jv}.  For mixed quantum states the more generalised version of the geometric phase is commonly known as {\it Uhlmann phase} \cite{1986RpMP...24..229U}.  These geometric phases are treated as the remarkable generalization of {\it quantum adiabatic theorem} \cite{1928ZPhy...51..165B,Simon:1983mh,Aharonov:1987gg,Samuel:1988zz} and this concept is very nearly related to the well known {\it  Born–Oppenheimer approximation} \cite{PhysRevLett.23.880,Rohrlich2009}. It has a very diverse application in the different areas of theoretical physics including atomic physics,  condensed matter physics,  quantum optics, nuclear physics,  elementary particle physics and now in the context of cosmology which is our prime area of interest in this paper.  Initially the concept of geometric phase was appeared in the framework of the well known {\it Aharonov-Bohm effect} \cite{PhysRev.115.485,PhysRevLett.89.210401,PhysRevA.86.040101,Peshkin:1981hz,Li:2018ujc}.  But later the more generalized extended version of this geometric phase,  which is the {\it Pancharatnam Berry phase} was appeared to describe a quantum state which is influenced by a slowly varying adiabatic change in a parameter dependent quantum mechanical Hamiltonian of the system under consideration.  Most importantly,  this adiabatic change in the parameter of the corresponding Hamiltonian happened in a closed cyclic path for which after traversing the complete trajectory it returns to it initial value at which it starts the adiabatic change in the parameter.  As an outcome,  the corresponding wave function of the quantum mechanical Hamiltonian of the system under consideration pick up a geometric phase,  which is identified to be the {\it Pancharatnam Berry phase} and it basically depends on the closed path on which the adiabatic slow change was performed.  To avoid any further confusion it is important to note that,  this geometric phase is completely different from the usual {\it dynamical phase} which is appearing as an out come of unitary time evolution of the Hamiltonian of the quantum system under consideration.  Apart from having a very strong quantum mechanical origin such possibilities also appear in the classical phenomena where it is commonly identified as {\it Hannay angle} \cite{Hannay_1985}.  There exists a semi-classical map which connects the concept of the geometric phases in quantum mechanical and classical processes where the adiabatic evolution play significant role.  It has a very deeper connection with experiments as well where the appearance,  its validity and justifiability can be tested in the context of quantum optics-photonics experiments,  quantum computational experiments and many areas related to the fundamental and foundational aspects of quantum mechanics.  Apart from having the huge progress in the many related areas of quantum mechanics there are various progress has been made within the framework of primordial aspects of cosmology where the the quantum mechanical effects are dominant in the time line of our universe.  See refs.~\cite{Mazur:1986gb,Corichi:1994ga,Cai:1989iq,Page:1987ca,Pal:2011xx,Pal:2012me,RevModPhys.82.1959,Cohen:2019hip} for details on this issue.

Next we come to the prime subject area of this paper, which is primordial cosmology within which we compute and estimate the {\it Pancharatnam Berry phase}.  The underlying physics of primordial cosmological perturbation theory \cite{Guth:1980zm,Guth:1982ec,10.1143/PTPS.78.1,Mukhanov:1990me,liddle_lyth_2000,Riotto:2002yw,Durrer:2004fx,Nakamura:2004rm,lyth_liddle_2009,Baumann:2009ds,Sriramkumar:2009kg,Senatore:2013roa,Peter:2013woa,Baumann:2014nda,Baumann:2018muz} and its connection with the quantum field theory of De Sitter space \cite{Spradlin:2001pw,Akhmedov:2013vka,Baumann:2009ds,Senatore:2013roa,Baumann:2018muz} is one of the most useful framework using which one can able to address various underlying unknown physical phenomena appearing at the quantum regime,  classical regime and at the quantum to classical transition region.  The great success of this set up is that,  it is successfully capable to generate the seeds which are appearing from the metric perturbations in spatially flat de Sitter background geometry and finally generate scalar and tensor modes of fluctuations out of that.  These fluctuations can be tested via the quantum correlations in various observational probes related to CMB experiments.  Though from the measured data and from the tested theories of primordial universe it is very clear that it is impossible at present to break the degeneracy among the predictions of various models and rule out them with high statistical accuracy.  To to do that either one need to increase the statistical accuracy of various upcoming or running observational probes or to look for some new physics by incorporating various unknown physical aspects appearing in the context of primordial cosmology.  Upgrading the observational probes is a very costly way to address this issue and we are not very that very sure when exactly that can be easily can be implemented in a cost effective fashion.  On the other hand,  we believe the second possibility is very effective in the realistic physical ground.  The series of possibilities can appear which one can incorporate to study the effect of new physics and its impact in the cosmological observables that can be tested in ongoing or near future CMB related experiments,  which are:
\begin{enumerate}
\item Measurement of tensor-to-scalar ratio is the one of the important issues in this list which can solely confirm the signature of primordial gravitational waves \cite{Baumann:2015xxa,Baumann:2014nda,Baumann:2009mq,Choudhury:2013iaa,Choudhury:2013jya,Choudhury:2014kma,Choudhury:2014wsa,Choudhury:2014sua,Choudhury:2015pqa,Choudhury:2015hvr,Choudhury:2017glj,Choudhury:2017cos,Creminelli:2014nqa,Creminelli:2014wna,Cheung:2007st,Choudhury:2012yh,Choudhury:2003vr,Choudhury:2011jt,Choudhury:2011sq,Choudhury:2012ib,Choudhury:2013zna,Choudhury:2014sxa,Choudhury:2015yna,Mazumdar:2001mm,Panda:2007ie,Ali:2010jx,Ali:2008ij,Panda:2006mw,Panda:2005sg,Chingangbam:2004ng,Panda:2010pj,Moniz:2009ax} and also the detection of which confirms the scale of inflationary paradigm,
\item Measurement of primordial non-Gaussian features \cite{Maldacena:2002vr,Maldacena:2011nz,Senatore:2009gt,Senatore:2008wk,Baumann:2019oyu,Baumann:2020dch,Baumann:2020ksv,Baumann:2009ds,Baumann:2018muz,Meerburg:2019qqi,Senatore:2016aui,Creminelli:2005hu,Creminelli:2006rz,Smith:2009jr,Senatore:2013roa,Green:2013rd,Smith:2015uia,Flauger:2016idt,Creminelli:2003iq,Creminelli:2004yq,Creminelli:2010qf,Assassi:2012zq,Behbahani:2012be,Green:2020whw,Creminelli:2006gc,Choudhury:2015yna,Bordin:2016ruc,Mirbabayi:2014zpa} in the cosmological correlations from equal time correlation functions at late time scale is another important aspects detection of which can serve as a strong probe to rule out various models of primordial universe,
  \item Measuring the various features of primordial power spectrum having spectral running and sufficient non negligible deviation from the scale invariance \cite{Choudhury:2013jya,Choudhury:2013zna,Lidsey:2003cq,Zarei:2014bta,Li:2018iwg} is another important feature which can serve the purpose,
    \item Finding out new cosmological consistency relations \cite{Gruzinov:2004jx,Gong:2017wgx,Hui:2018cag,Choudhury:2015pqa,Choudhury:2014kma,Choudhury:2013iaa} also going to help to address this important issue.
\item Impact of Bell's inequality violation \cite{Maldacena:2015bha,Choudhury:2016cso,Choudhury:2016pfr,Martin:2017zxs,Martin:2016nrr,Kanno:2017teu,Kanno:2017dci} in cosmological observables and the direct comparison with the possibility where Bell's inequality violation is not appearing is another important aspects which can suffice the above mentioned purpose.
\item Detecting the direct signatures or putting stringent constraints from the present data from various types of geometric phases at the different cosmological scales are the another important issue which also can address the above mentioned motivation of incorporating new physics in primordial cosmological paradigm.  There are two possible phases appearing in the present literature,  A.  {\it Pancharatnam Berry phase} for the pure quantum states \cite{Pancharatnam:1956jv,Berry:1984jv,Simon:1983mh,Aharonov:1987gg,Samuel:1988zz} and B.  {\it Uhlmann phase} for the mixed quantum states \cite{1986RpMP...24..229U,Kirklin:2020zic,Kirklin:2020qtv,Kirklin:2019ror,Kirklin:2021ipb,PhysRevLett.112.130401,PhysRevA.67.032110}.

\item Quantification of correlations when out-of-equilibrium aspects are dominant and this can be done using the concept of out-of-time-ordered-correlation (OTOC) functions \cite{Larkin:1969,Maldacena:2015waa, Hashimoto:2017oit, Chakrabarty:2018dov,Chaudhuri:2018ymp,Chaudhuri:2018ihk, Haehl:2017eob, Akutagawa:2020qbj, Bhagat:2020pcd, Romero-Bermudez:2019vej,Choudhury:2020yaa,Choudhury:2021tuu}.
\end{enumerate}

In this paper,  we have actually studied the possibility of having {\it Pancharatnam Berry phase} for the pure quantum states in absence and in presence of Bell's inequality violation within the framework of primordial cosmology.  The detailed outcomes and the related interpretations are appended below point-wise:
\begin{enumerate}

 \item \underline{Massless field case:}~~
 In this possibility we have explored the possibility of having either $m(\tau)/{\cal H}\ll 1$ or $m(\tau)/{\cal H}\rightarrow 0$,  so that one can able to completely neglect the contribution of the mass in the primordial cosmological perturbation of scalar modes completely.  This is identified as the {\it Massless field case}.  
 
 \item \underline{Partially Massless field case:}~~
 In this possibility we have explored the possibility of having either $m(\tau)/{\cal H}=1$ exactly or $m(\tau)/{\cal H}\approx 1$,  so that one can able to take care of the contribution of the mass in the primordial cosmological perturbation of scalar modes.  This is identified to be the {\it Partially Massless field case}.
 
 \item \underline{Massive field case:}~~
 Last but not the least,  in this possibility we have explored the possibility of having $m(\tau)/{\cal H}\gg 1$,  so that one can able to significantly take care of the contribution of the mass in the primordial cosmological perturbation of scalar modes.  This is a very interesting situation in the principal mass series and is identified to be the partially {\it Heavy/Massive field case}.  
 
 \end{enumerate}
The prime highlighting points of the present work are appended below point-wise:
\begin{itemize}
\item In this paper we have provided the analytical computation of the {\it Pancharatnam Berry phase} within the framework of primordial cosmology 
by incorporating the possibility of having no Bell violation and the Bell violation for the massless,  partially massless and massive/heavy field case.  

\item We also have provided the numerical constraints on the amplitude of the {\it Pancharatnam Berry phase} derived in this paper which confronts well the present observational cosmological data.

\end{itemize}

 The organization of this paper is as follows point-wise:
 \begin{itemize}
 \item In section (\ref{QM}),  we provide a brief review on the subject of geometric phase within the framework of quantum mechanics.  In this section we have made a clear statement regarding the differences between the usual dynamical phase appearing as an outcome of unitary time of evolution of the Hamiltonian and the {\it Pancharatnam Berry phase}.  
 
 \item Before going to the further technical details of the computation of {\it Pancharatnam Berry phase} within the framework of primordial cosmology in section (\ref{IHO}),  we provide a very simple example of computing the corresponding phase in the context of inverted harmonic oscillator model having time dependent frequency.    
 
 \item In section (\ref{COSMOS}),  we provide the explicit computation of the relevant eigenfunction needed to derive the {\it Pancharatnam Berry phase} within the framework of primordial cosmology by following {\it Lewis Riesenfeld invariant trick} as discussed in the previous section.

 \item Next in section (\ref{GP}),  we provide the further details of the analytical computation of the {\it Pancharatnam Berry phase} within the framework of primordial cosmology where we have expressed the results in terms of the slowly varying conformal time dependent parameters.

 \item Further in section (\ref{OC}),  we first express the derived theoretical results of the {\it Pancharatnam Berry phase} in terms of cosmological observables,   such as,  scalar spectral index/tilt and tensor-to-scalar ratio respectively both for non Bell violating and for the Bell violating cases.  
 
 \item Finally in section (\ref{CON}),  we conclude with relevant future prospects/directions from our analysis performed in this paper.
 \end{itemize}  

\section{Brief review on Geometric Phase in Quantum Mechanics}
\label{QM}
In this section we briefly review on the geometric phase appearing within the framework of quantum mechanics.  We believe this discussion will help us the rest of the content of the paper.  In general prescription the geometric phase treated to be phase difference which is appearing in a cycle,  where the quantum system under consideration is participating in a cyclic adiabatic process.  This is actually originated from the pure geometric properties of the quantum mechanical Hamiltonian under consideration.  In the corresponding literature this geometric phase is commonly known as the {\it Pacharatnam - Berry phase} \cite{Pancharatnam:1956jv,Berry:1984jv}.   Later in this construction a more generalized version,  a topological extended version which is the {\it  Aharonov–Anandan phase} \cite{PhysRev.115.485,PhysRevLett.89.210401,PhysRevA.86.040101,Peshkin:1981hz,Li:2018ujc} was introduced to deal with the  gauge-invariant generalization of the previously mentioned {\it Pacharatnam - Berry phase}.   {\it  Aharonov–Anandan phase} was actually computed in terms of closed loops of the quantum mechanical system,  which is not necessarily follow the adiabatic process.  Most importantly,  in this computation,  no parameters from the Hamiltonian from the quantum mechanical system have been used \cite{Cohen:2019hip}.  Later a further generalized version of this computation was proposed,  where both the unitary and cyclic evolution features of the quantum system was not considered \cite{PhysRevA.86.022105,Marzlin:2004zz,Ma:2006zzb,Pati2004InconsistenciesOT,Cohen:2019hip}.  

It was pointed in the computation of {\it Pacharatnam - Berry phase} \cite{Pancharatnam:1956jv,Berry:1984jv} that when the parameters of a quantum mechanical system are adiabatically changing in a cyclic closed path,  the corresponding quantum phase of its eigenstate not necessarily return to its initial value from which it started changing.  In this construction suppose a quantum mechanical system starts adiabatically changing from its $n$-th eigenstate $|n ({\bf R}), 0\rangle$ of the Hamiltonian $H({\bf R})$,  where ${\bf R}$ represent the parameters in the Hamiltonian.  Now,  following the principle of adiabatic theorem one can explicitly write down the expression for the eigenstate $|n ({\bf R}), t\rangle$ of the instantaneous Hamiltonian at any time $t$.  In this computation when the representative parameters ${\bf R}$ of the Hamiltonian $H({\bf R})$ complete a cycle ${\cal C}$,  the final state of the Hamiltonian $H({\bf R})$ will return to its initial value from which it was started changing and along with this an additional time dependent phase factor will appear which is not at all appearing from the 
dynamical time evolution of the Hamiltonian $H({\bf R})$.  This additional time dependent phase factor depends only on the geometry
of the path ${\cal C}$ considered in this computation.  In this context,  one can explicitly show that under approximation the coefficient of the $n$-th quantum mechanical eigen state can be expressed as \cite{Pancharatnam:1956jv,Berry:1984jv}$g_n(t)=g_n(0)~\exp(i\gamma_n\left[{\cal C}\right](t))~~\forall ~~n=0,1,\cdots,\infty,$
where $\gamma_n\left[{\cal C}\right](t)$ is the well known {\it Pacharatnam - Berry phase} \cite{Pancharatnam:1956jv,Berry:1984jv} which is defined in terms of the eigenstate $|n,t\rangle$ of the instantaneous Hamiltonian for the adiabatic change in the  parameters ${\bf R}$ of the Hamiltonian $H({\bf R})$ over a complete cyclic path ${\cal C}$ as,   
$\gamma_n\left[{\cal C}\right](t):=i \oint_{{\cal C}}~{\cal A}({\bf R})~.~d{\bf R},$
where the newly introduced symbol ${\cal A}({\bf R})$ signifies the {\it Pacharatnam - Berry connection},  which is defined as,  
${\cal A}({\bf R}):=\langle n ({\bf R}), t|\nabla_{\bf R}|n ({\bf R}), t\rangle.$
In this context,  the {\it Pacharatnam - Berry connection} can be interpreted as the generalized vector potential for the quantum mechanical system under consideration.  Here,  $\nabla_{\bf R}$ represents the gradient vector differential operator defined in the parameter space ${\bf R}$ of the Hamiltonian $H({\bf R})$ and $i=\sqrt{-1}$.  This expression actually represents the holonomy in a line bundle,   which further implies the geometrical measure of the collapse of preserving the quantum mechanical data when it is 
parallel transported around a closed path ${\cal C}$ under the influence in the adiabatic change in the  parameters ${\bf R}$ of the Hamiltonian $H({\bf R})$.    In the table \ref{tab1},  we have pointed different sources of Geometric Phases appearing in the context of Physics including the recent development within the framework of Cosmology including our contribution in the list.  We believe this information will be helpful to know about the previous remarkable contributions in this direction.
\begin{table}[!t]
\footnotesize
 \begin{tabular}{||p{4cm}||p{3cm}||p{5cm}||p{3cm}||  }
 \hline\hline
 \multicolumn{4}{|c|}{Different sources of Geometric Phases in Physics} \\
 \hline\hline
 Phases invented& Year & Physical context & Type of phase\\
 \hline
 Pancharatnam   & 1956 \cite{Pancharatnam:1956jv}    & Optics&   Adiabatic\\
      &2021 & Cosmology with Bell's inequality violation (present work)&  \\
 \hline
 Aharonov–Bohm &  1959 \cite{PhysRev.115.485} & QED & Topological \\
 \hline
 Abelian anyons &1977 \cite{Cohen:2019hip} & Condensed Matter&  Topological \\
 &1982 \cite{Cohen:2019hip}&  &  \\
 &1984 \cite{Cohen:2019hip}&  &  \\
 \hline
 Berry   &1984 \cite{Berry:1984jv}  & Quantum Mechanics &  Adiabatic\\
      &2013  \cite{Pal:2011xx,Pal:2012me}& Cosmology &  \\
      &2021 & Cosmology with Bell's inequality violation (present work)&  \\
    \hline
 Aharonov–Casher &   1984  \cite{Aharonov:1984xb} & QED & Topological\\
 \hline
 Hannay angle & 1985 \cite{Hannay_1985} & Classical Mechanics   & Adiabatic\\
 \hline
 Uhlmann & 1986  \cite{1986RpMP...24..229U} & Quantum Mechanics   & Adiabatic\\
 \hline
 Aharonov–Anandan & 1987 \cite{Aharonov:1987gg} & Quantum Mechanics &Topological\\
 \hline
 Zak & 1989 \cite{PhysRevLett.62.2747} & Condensed Matter& Non-topological, non-adiabatic\\
 \hline\hline
\end{tabular}
\caption{In this table we have pointed different sources of Geometric Phases appearing in the context of Physics.  We have taken the help from ref.~\cite{Cohen:2019hip} to give the correct information regarding the year of invention.  Also we have updated the list by including the contributions from the cosmology literature in this connection.}\label{tab1}
 \end{table}

The concept of geometric phases are actually deep rooted to the topological aspects of physics which is appearing in the context of a quantum system having Hilbert space  ${\cal H}$,  and the corresponding parameter space of this Hilbert space adiabatically changes the previously mentioned parameters ${\bf R}$ in the differential manifold ${\cal M}$. This is technically identified to be a
{\it vector bundle},  which basically consists of the differential manifold ${\cal M}$ which is embedded within the product Hilbert space  ${\cal M}\otimes{\cal H}$,  where the definition of the differential vector operator $\nabla_{\bf R}$ is not very simple.  In this context,  the fundamental computable quantity is the {\it Pacharatnam - Berry connection},  ${\cal A}({\bf R})$,  which physicists use to compute frequently in different contexts.  
This {\it Pacharatnam - Berry connection} allows to perform the differentiation of ket vector $|\psi({\bf R}(t))\rangle$ over the parameters ${\bf R}$ by providing a trick to promote them from the Hilbert space ${\cal H}$ to the product Hilbert space  ${\cal M}\otimes{\cal H}$.  This process is technically identified to be the {\it parallel transport phenomena},  where the corresponding {\it Pacharatnam - Berry phase} $\gamma$ is smoothly defined.  This is particularly a very important condition which helps to parallel transport of $|\psi({\bf R}(t))\rangle$.  Parallel transport operation in the product Hilbert space  ${\cal M}\otimes{\cal H}$, 
along a closed path ${\cal C}$ maps the quantum mechanical state $\psi({\bf R})$ to the following \cite{Cohen:2019hip}:
\bea |\psi({\bf R}(t))\rangle \longrightarrow \widetilde{|\psi({\bf R}(t))\rangle}:={\cal O}_{\bf Lin}(\gamma(t),{\cal A}({\bf R}))|\psi({\bf R}(t))\rangle,\eea
where ${\cal O}_{\bf Lin}(\gamma(t),{\cal A}({\bf R}))$ is identified to be a linear map and commonly in this literature is known as the holonomy of
the closed path $\widetilde{{\cal C}}$.  It can be explicitly established that
the {\it Pacharatnam - Berry curvature} is the outcome of the holonomy of an infinitesimally small loop $\widetilde{{\cal C}}$ in this construction. 
Also it is important to note that,  the {\it Pacharatnam - Berry curvature} integral over a surface ${\cal S}$ bounded by the previously mentioned holonomy’s closed small loop $\widetilde{{\cal C}}$ give rise to finally the expression for the {\it Pacharatnam - Berry phase} in this construction, as given by \cite{Cohen:2019hip}:
\bea \gamma(t):&=&i\int^{t}_{0}~dt^{\prime}~~\widetilde{\langle\psi({\bf R}(t^{\prime}))|}\left(\frac{d}{dt^{\prime}}\right)\widetilde{|\psi({\bf R}(t^{\prime}))\rangle}~~~~{\rm where}~~~~.\widetilde{|\psi({\bf R}(t^{\prime}=t))\rangle}=\widetilde{|\psi({\bf R}(t^{\prime}=0))\rangle}. ~~~~~\nonumber\\
&=&i \oint_{\widetilde{{\cal C}}}\widetilde{\langle\psi({\bf R}(t^{\prime}))|}\left({\cal D}_{\widetilde{{\cal C}}}\right)\widetilde{|\psi({\bf R}(t^{\prime}))\rangle}.\eea
Here ${\cal D}_{\widetilde{{\cal C}}}$ is the differential operator defined on the  small loop $\widetilde{{\cal C}}$.  The above mentioned integral have deeper understandings within the framework of topological physics which finally corresponds to an integer,  which determines {\it Chern Class} \cite{Cohen:2019hip}
 of the topological bundle's {\it Pacharatnam - Berry connection}.  This is completely different with respect to the dynamical phase appearing in the quantum mechanical system.  In this construction the dynamical phase is given by \cite{Cohen:2019hip}:
\bea  \delta(t):&=&-\int^{t}_{0}~dt^{\prime}~~\langle\psi({\bf R}(t^{\prime}))|H({\bf R}(t^{\prime}))|\psi({\bf R}(t^{\prime}))\rangle~~~~~~{\rm where}~~~H({\bf R}(t^{\prime}))|\psi({\bf R}(t^{\prime}))=E(t^{\prime})|\psi({\bf R}(t^{\prime}))
.~~~\eea 
 
In this paper by following the same notion we will first show how the geometric phase can be computed for a given simple toy model of inverted harmonic oscillator (IHO) in quantum mechanics.  Next,  we will extend this idea to estimate the geometric phase by doing a detailed computation within the context of Cosmological Perturbation Theory in terms of scalar and tensor modes of quantum mechanical fluctuations in the early universe.  As we proceed it will be more clear how the mentioned geometric phases can be explicitly computed and physically interpreted.
\section{Time-dependent inverted harmonic oscillator in Quantum Mechanics}
\label{IHO}
In this section,  we explicitly discuss the important role of time-dependent inverted harmonic oscillator within the framework of Quantum Mechanics in finding the expression for the geometric phase.  Particularly,  in this paper we are interested in time-dependent inverted harmonic oscillator because in Cosmological Perturbation Theory the scalar and tensor mode of quantum fluctuations can also be expressed in terms of time-dependent inverted harmonic oscillator in conformal time scale.  This analogy will help us to promote the computational technique of finding the geometric phase within the framework of Quantum Mechanics to Cosmology.
\subsection{The toy model in Quantum Mechanics}
Let us start with a toy model of a general time-dependent inverted harmonic oscillator,  which is described by the following action:
\bea S=\int dt~L\left(q(t),\left(\frac{dq(t)}{dt}\right)\right),~~~~~~\eea
where the Lagrangian of the system is given by the following expression:
\bea L\left(q(t),\left(\frac{dq(t)}{dt}\right)\right)=m(t)~\bigg[
\frac{1}{2}\left(\frac{dq(t)}{dt}\right)^2+\frac{1}{2}\left(\omega^2(t)+f^2(t)\right)q^2(t)-f(t)q(t)\left(\frac{dq(t)}{dt}\right)\bigg].~~~~~~~\eea
where $q(t)$ is the generalized coordinate,  $m(t)$ is the time-dependent mass,  $\omega(t)$ is the time-dependent frequency and $f(t)$ is a general time-dependent coupling parameter of the inverted harmonic oscillator in this construction.  Here it is important to note that the second term which is representing the potential energy of the inverted harmonic oscillator appearing with a negative sign compared to the usual harmonic oscillator model.  The prime reason is in this construction the frequency of the harmonic oscillator $\omega(t)$ is actually replaced by $i\omega(t)$ within the context of inverted harmonic oscillator and this explain the origin of an additional negative signature in the potential energy. At this point,  we have not chosen any particular time dependence in the coupling parameter $y(t)$, time-dependent mass $m(t)$ and in the time-dependent frequency $\omega(t)$.  But later if it is required we will choose some specific models for the time-dependence to solve the problem. 

After varying the stated action for the general time-dependent inverted harmonic oscillator,  we get the following Euler-Lagrange equation:
\bea \bigg[\frac{d^2}{dt^2}+{\cal G}(t)\frac{d}{dt}-\omega^2_{\rm eff}(t)\bigg]q(t)=0.\eea
The effective time-dependent frequency $\omega^2_{\rm eff}(t)$ is defined as:
\bea \omega^2_{\rm eff}(t):=\bigg(\omega^2(t)+f^2(t)+{\cal G}(t)f(t)+\frac{df(t)}{dt}\bigg)~~~~~~ {\rm where}~~~~~~ {\cal G}(t):=\frac{d\ln m(t)}{dt}.\eea
For a given model of the time-dependence of $m(t)$,  $\omega(t)$ and $f(t)$ it may possible to find analytical or numerical solutions of the equation of motion of the mentioned model of general time-dependent inverted harmonic oscillator.  In the next section,  we will also clear the one to one correspondence between the above mentioned general time-dependent inverted harmonic oscillator model with the quantum fluctuating scalar and tensor modes which are appearing in the Fourier space analysis of Cosmological Perturbation Theory set up. 
\subsection{Construction of Hamiltonian for general time-dependent inverted harmonic oscillator in Quantum Mechanics}
In this subsection we construct the classical as well as the quantum mechanical Hamiltonian of the general time-dependent inverted harmonic oscillator model constructed in the previous subsection.  

To serve this purpose,  first of all we compute the canonically conjugate momentum $\Pi_q (t)$ for the generalized coordinate $q(t)$ for the general time-dependent inverted harmonic oscillator,  which is given by the following expression:
\bea \Pi_q (t)=\frac{\displaystyle \partial L\left(q(t),\left(\frac{dq(t)}{dt}\right)\right)}{\displaystyle \partial \left(\frac{dq(t)}{dt}\right)}=m(t)\bigg[\left(\frac{dq(t)}{dt}\right)-f(t)q(t)\bigg]. \eea
Using this equation one can able to express the generalized velocity with respect to the canonically conjugate momentum $\Pi_q (t)$ and substituting it in the corresponding Legendre transformation the classical Hamiltonian for the general time-dependent inverted harmonic oscillator model is derived as:
\bea H\left(q(t),\Pi_q (t)\right)=\bigg[\frac{1}{2m(t)}\Pi^2_q(t)-\frac{1}{2}m(t)\omega^2(t)q^2(t)+f(t)\Pi_q(t) q(t)\bigg].\eea
Now to construct the quantum mechanical Hamiltonian operator for the general time-dependent inverted harmonic oscillator model we promote the classical solutions $(q(t),\Pi_q (t))$ as the quantum operator $(\hat{q}(t),\hat{\Pi}_q (t))$ and also use the following equal time commutation relation, 
$\left[\hat{q}(t),\hat{\Pi}_q(t)\right]=i.$
Consequently the quantum Hamiltonian operator can be expressed as:
\bea \hat{H}\left(\hat{q}(t),\hat{\Pi}_q (t)\right)&=&\bigg[\frac{1}{2m(t)}\hat{\Pi}^2_q(t)-\frac{1}{2}m(t)\omega^2(t)\hat{q}^2(t)+\frac{1}{2}f(t)\left(\hat{\Pi}_q(t) \hat{q}(t)+\hat{q}(t) \hat{\Pi}_q(t)\right)\bigg]\nonumber\\
&=&\bigg[-\frac{1}{2m(t)}\hat{\partial}^2_{q}-\frac{1}{2}m(t)\omega^2(t)\hat{q}^2(t)-\frac{i}{2}f(t)-\frac{i}{2}f(t)\hat{q}(t) \hat{\partial}_q\bigg].~~~~~~~~~~~~\eea
where we have used,  $\hat{\Pi}_q(t)=-i\hat{\partial}_q$.  
\subsection{Lewis Riesenfeld invariant trick in Quantum Mechanics}
In this subsection we discuss about the Lewis Riesenfeld invariant trick for computing continuous eigenvalues in the context of general time-dependent inverted harmonic oscillator model mentioned in the previous subsection.  See refs.~\cite{doi:10.1063/1.1664991,doi:10.1063/1.1664532,doi:10.1142/S0217979204024732} for more details to know about this formalism.
The well known statement of this good old trick is as follows.  Let us consider a Hermitian operator $\widehat{\cal I}^{\bf LR}(t)$ and this is identified to be an {\it invariant operator} for a given quantum system when the following condition is satisfied explicitly \cite{doi:10.1063/1.1664991,doi:10.1063/1.1664532,doi:10.1142/S0217979204024732}:
\bea \frac{d \widehat{\cal I}^{\bf LR}(t)}{dt}=\frac{\partial \widehat{\cal I}^{\bf LR}(t)}{\partial t}+i\left[\hat{H}(t),\widehat{\cal I}^{\bf LR}(t)\right]=0,\eea 
where $\hat{H}(t)$ is the time-dependent general Hamiltonian of a quantum mechanical system.  The most general solution of the corresponding time-dependent Schrödinger equation of the quantum system can be written as \cite{doi:10.1063/1.1664991,doi:10.1063/1.1664532,doi:10.1142/S0217979204024732}:
\bea \Psi(q,t)=\int d\lambda~{\cal J}(\lambda)~\exp\left(i\gamma^{\bf LR}_{\lambda}(t)\right)~\Phi_{\lambda}(q,t),\eea
where we have taken integration over all possible continuous eigenvalues in this construction to find out the expression for the most general solution.  Here the eigenvalue dependent function,  ${\cal J}(\lambda)$ is defined in terms of the following phase dependent overlap as \cite{doi:10.1063/1.1664991,doi:10.1063/1.1664532,doi:10.1142/S0217979204024732}:
\bea {\cal J}(\lambda)=\exp\left(-i\gamma^{\bf LR}_{\lambda}(0)\right)\langle \Phi_{\lambda}(q,0)|\Psi(q,0)\rangle. \eea
Also it is important to note that,  the function $\Phi_{\lambda}(q,t)$ is identified to be eigenfunction of the {\it invariant operator} $\widehat{\cal I}^{\bf LR}(t)$ which satisfies the following eigenvalue equation in the present context \cite{doi:10.1063/1.1664991,doi:10.1063/1.1664532}:
\bea \widehat{\cal I}^{\bf LR}(t)\Phi_{\lambda}(q,t)=\lambda \Phi_{\lambda}(q,t).\eea
Most importantly,  the {\it Lewis Riesenfeld phase} factor $\gamma^{\bf LR}_{\lambda}(t)$ can be computed from the following expression:
\bea \label{LRP} \gamma^{\bf LR}_{\lambda}(t):&=&
\gamma^{\bf PB}_{\lambda}(t)+\delta^{\bf Dynamical}_{\lambda}(t),\eea 
where,  the {\it Pancharatnam-Berry phase} and the {\it dynamical phase} for the continuous eigenvalue $\lambda$ in this construction can be expressed as:
\bea \gamma^{\bf PB}_{\lambda}(t):&=&i\int^{t}_{0} dt^{\prime}~\langle\Phi_{\lambda}(q,t^{\prime})|\bigg(\frac{\partial}{\partial t^{\prime}}\bigg)|\Phi_{\lambda}(q,t^{\prime})\rangle,\\
\delta^{\bf Dynamical}_{\lambda}(t)&=&-\int^{t}_{0} dt^{\prime}~\langle\Phi_{\lambda}(q,t^{\prime})|\hat{H}(t^{\prime})|\Phi_{\lambda}(q,t^{\prime})\rangle.\eea
By following this trick one can construct the following {\it invariant operator} for the general time-dependent inverted harmonic oscillator model Hamiltonian constructed in the previous section \cite{doi:10.1063/1.1664991,doi:10.1063/1.1664532,doi:10.1142/S0217979204024732}:
\bea \widehat{\cal I}^{\bf LR}(t):=-\frac{1}{2}\bigg[\left(\frac{\hat{q}(t)}{{\cal K}(t)}\right)^2-\left[i{\cal K}(t)\hat{\partial}_q+m(t)\left(\frac{d {\cal K}(t)}{dt}-f(t){\cal K}(t)\right)\hat{q}(t)\right]^2\bigg],\eea
where the function ${\cal K}(t)$ representing a {\it c-number} which satisfy the following quantum auxiliary equation:
\bea \bigg[{\cal D}_t-\omega^2_{\rm eff}(t)\bigg]{\cal K}(t)={\cal W}(t)~~~~~~{\rm where}~~~~~{\cal W}(t)=-\frac{1}{m^2(t){\cal K}^3(t)}.\eea
Here the differential operator ${\cal D}_t$ and the effective-time dependent frequency factor $\omega^2_{\rm eff}(t)$ are defined in the earlier subsection of this paper.  Now it is important to note that,  the mathematical structure of this derived equation is like an equation of forced general time-dependent inverted harmonic oscillator and can be explicitly derived from the structure of the equation of motioned derived for this system in the earlier subsection. 

Now if we closely look into the mathematical structure of the above mentioned {\it invariant operator} for the general time-dependent inverted harmonic oscillator model Hamiltonian then we found that it is not in a diagonal form and for this reason it is very difficult to finally derive the expression for the eigenfunction $\Phi_{\lambda}(q,t)$ and the most general solution $\Psi(q,t)$ of the corresponding time-dependent Schrödinger equation for this system.  Now,  using the following unitary transformation on the {\it invariant operator}  one can express it in terms of a desirable diagonal form:
\bea \widetilde{\widehat{\cal I}^{\bf LR}(t)}:=\hat{\cal U}_{\bf LR}(t)~\widehat{\cal I}^{\bf LR}(t)~\hat{\cal U}^{\dagger}_{\bf LR}(t),\eea
where the corresponding unitary operator for this general time-dependent inverted harmonic oscillator model can be expressed as:
\bea \hat{\cal U}_{\bf LR}(t):=\exp\bigg(-\frac{i}{2}\frac{m(t)}{{\cal K}(t)}\left(\frac{d {\cal K}(t)}{dt}-f(t){\cal K}(t)\right)\hat{q}^2(t)\bigg).\eea
Consequently,  the diagonal simplified form of the {\it invariant operator} for the general time-dependent inverted harmonic oscillator model Hamiltonian is given by:
\bea \widetilde{\widehat{\cal I}^{\bf LR}(t)}=-\frac{1}{2}\bigg(\left(\frac{\hat{q}(t)}{{\cal K}(t)}\right)^2+{\cal K}^2(t)\hat{\partial}^2_q\bigg). \eea
After doing the above mentioned unitary transformation the new eigenfunction of the diagonal {\it invariant operator} can be written as \cite{doi:10.1063/1.1664991,doi:10.1063/1.1664532,doi:10.1142/S0217979204024732}: 
\bea \Phi_{\lambda}(q,t)~~\longrightarrow ~~\widetilde{\Phi_{\lambda}(q,t)}:&=&\hat{\cal U}_{\bf LR}(t)~\Phi_{\lambda}(q,t)=\exp\bigg(-\frac{i}{2}\frac{m(t)}{{\cal K}(t)}\left(\frac{d {\cal K}(t)}{dt}-f(t){\cal K}(t)\right)\hat{q}^2(t)\bigg)~\Phi_{\lambda}(q,t).~~~~~~~~~~\eea 
But the good part is that the continuous eigenvalues of the {\it invariant operator} will remain completely unchanged before and after performing the unitary transformation in this construction.  This implies:
\bea  {\widehat{\cal I}^{\bf LR}(t)}{\Phi_{\lambda}(q,t)}=\lambda~{\Phi_{\lambda}(q,t)} ~~\longrightarrow~~\widetilde{\widehat{\cal I}^{\bf LR}(t)}\widetilde{\Phi_{\lambda}(q,t)}=\lambda~\widetilde{\Phi_{\lambda}(q,t)}.\eea
Consequently,  we get the following solutions of the eigenfunctions of the {\it invariant operator} before and after performing the unitary transformation \cite{doi:10.1063/1.1664991,doi:10.1063/1.1664532,doi:10.1142/S0217979204024732}:
\bea \widetilde{\Phi_{\lambda}(q,t)}:&=&\frac{1}{\sqrt{{\cal K}(t)}}~{\cal P}\bigg(\lambda-\frac{1}{2},\sqrt{2}~\frac{q(t)}{{\cal K}(t)}\bigg),\\
{\Phi_{\lambda}(q,t)}:&=&\hat{\cal U}^{\dagger}_{\bf LR}(t)\widetilde{\Phi_{\lambda}(q,t)}=\frac{1}{\sqrt{{\cal K}(t)}}~\exp\bigg(\frac{i}{2}\frac{m(t)}{{\cal K}(t)}\left(\frac{d {\cal K}(t)}{dt}-f(t){\cal K}(t)\right){q}^2(t)\bigg){\cal P}\bigg(\lambda-\frac{1}{2},\sqrt{2}~\frac{q(t)}{{\cal K}(t)}\bigg).~~~~~~~~~~\eea
Here ${\cal P}(\nu,x)$ represents the {\it Parabolic Cylinder function} in this context.   In the next subsection,  we derive the corresponding {\it Lewis Riesenfeld phase} factor $\gamma^{\bf LR}_{\lambda}(t)$ and also the related most general solution of the time-dependent Schrödinger equation for the present model. 
\subsection{Computation of Geometric Phase in Quantum Mechanics} 
In this subsection we explicitly compute the expression for the {\it Lewis Riesenfeld phase} factor $\gamma^{\bf LR}_{\lambda}(t)$ for the general time-dependent inverted harmonic oscillator model.  To serve this purpose we first of all use the expression for the Hamiltonian operator stated in equation \ref{ham} in equation \ref{LRP},  and consequently we get \cite{doi:10.1063/1.1664991,doi:10.1063/1.1664532,doi:10.1142/S0217979204024732}: 
\bea \gamma^{\bf LR}_{\lambda}(t):&=&\int^{t}_{0} dt^{\prime}~\langle\Phi_{\lambda}(q,t^{\prime})|\bigg(i\frac{\partial}{\partial t^{\prime}}+\frac{1}{2m(t^{\prime})}\hat{\partial}^2_{q}+\frac{1}{2}m(t^{\prime})\omega^2(t^{\prime})\hat{q}^2(t^{\prime})\nonumber\\
&&~~~~~~~~~~~~~~~~~~+\frac{i}{2}f(t^{\prime})+\frac{i}{2}f(t^{\prime})\hat{q}(t)^{\prime} \hat{\partial}_q\bigg)|\Phi_{\lambda}(q,t^{\prime})\rangle =-\lambda\int^{t}_{0} dt^{\prime}~\frac{1}{m(t^{\prime}){\cal K}^2(t^{\prime})}.~~~~~~~~~\eea
Then the general solution for the time-dependent Schrödinger equation for a given continuous eigenvalue $\lambda$ is given by the following expression \cite{doi:10.1063/1.1664991,doi:10.1063/1.1664532,doi:10.1142/S0217979204024732}:
\bea \Psi_{\lambda}(q,t)&=&\exp(i\gamma^{\bf LR}_{\lambda}(t))~{\Phi_{\lambda}(q,t)}\nonumber\\
&=&\frac{1}{\sqrt{{\cal K}(t)}}~\exp\left(-i\lambda\int^{t}_{0} dt^{\prime}~\frac{1}{m(t^{\prime}){\cal K}^2(t^{\prime})}\right)~\nonumber\\
&&~~~~~~~~~~~~~~~~~~~~~~~\exp\bigg(\frac{i}{2}\frac{m(t)}{{\cal K}(t)}\left(\frac{d {\cal K}(t)}{dt}-f(t){\cal K}(t)\right){q}^2(t)\bigg){\cal P}\bigg(\lambda-\frac{1}{2},\sqrt{2}~\frac{q(t)}{{\cal K}(t)}\bigg).\eea 
Further integrating over all possible eigenvalues we get the following general solution for wave function of the time-dependent Schrödinger equation for the general time-dependent inverted harmonic oscillator model \cite{doi:10.1063/1.1664991,doi:10.1063/1.1664532,doi:10.1142/S0217979204024732}:
\bea \Psi(q,t)&=&\frac{1}{\sqrt{{\cal K}(t)}}~\int d\lambda~{\cal J}(\lambda)~\exp\left(-i\lambda\int^{t}_{0} dt^{\prime}~\frac{1}{m(t^{\prime}){\cal K}^2(t^{\prime})}\right)~~\nonumber\\
&&~~~~~~~~~~~~~~~~~~~~~~~\exp\bigg(\frac{i}{2}\frac{m(t)}{{\cal K}(t)}\left(\frac{d {\cal K}(t)}{dt}-f(t){\cal K}(t)\right){q}^2(t)\bigg){\cal P}\bigg(\lambda-\frac{1}{2},\sqrt{2}~\frac{q(t)}{{\cal K}(t)}\bigg).\eea 
In the next section,  by following the same procedure we derive the expression for the geometric phases from scalar and tensor quantum fluctuations of the metric perturbation within the framework of Cosmological Perturbation Theory.  
\section{Time-dependent parametric oscillator in Cosmology}
\label{COSMOS} 
In this section our prime objective is to give an analogous interpretation of previously discussed general time-dependent inverted harmonic oscillator model to derive the expression for the geometric phases from scalar and tensor counterpart of quantum fluctuations originated from the metric perturbation in spatially flat FLRW quasi De-Sitter space-time.
\subsection{The toy model in Cosmology}
Let us start with a  very well known simplest toy model of our universe described in terms of the scalar fields and minimally interacting with the the background classical gravitational space-time.  The corresponding effective action is given by:
\bea S=\frac{1}{2}\int d^4x~\sqrt{-g}\bigg[R-\left(\partial \phi\right)^2-2V(\phi)\bigg].\eea
Here we have fixed the reduced Planck mass $M_p=1$ for the further simplifications.  First term represent the usual Einstein Hilbert term,  second term signifies the the kinetic term of the scalar field $\phi$ and last term represents the effective potential for the scalar field.  The present analysis is true for any model of $V(\phi)$ for this reason we have not mention here any specific model in this construction. 

The background classical metric for this analysis is chosen to be spatially flat FLRW with quasi De Sitter solution and this is described by the following line element:
\bea ds^2=-dt^2+a^2(t)\delta_{ij}dx^i dx^j~~~~~{\rm where}~~~a(t)=\exp((Ht+\cdots),H=\frac{d\ln a(t)}{dt}=\frac{1}{a(t)}\frac{da(t)}{dt}\eea
where the $\cdots$ represents the small deviation from the exact De Sitter solution in the scale factor.  For the quasi De Sitter expansion an additional parameter have been introduced, 
$\epsilon=-\frac{1}{H^2}\frac{dH}{dt},$
which is taking care of small deviation from the exact De Sitter solution of the scale factor.  Further using the transformation equation, 
$ d\tau=\frac{1}{a(t)}~dt,$
one can further transform the above mentioned background metric in the following conformally flat form:
\bea ds^2=a^2(\tau)\left(-d\tau^2+\delta_{ij}dx^i dx^j\right)~~~~~{\rm where}~~~a(\tau)=-\frac{1}{H\tau}(1+\epsilon),\eea
where in terms of the conformal time coordinate this small deviation parameter from the exact De Sitter solution can be expressed as,  
$\epsilon(\tau)=1-\frac{1}{{\cal H}^2}\frac{d{\cal H}}{d\tau}$  where ${\cal H}=\frac{d\ln a(\tau)}{d\tau}=a(t)H.$
Now,  further varying the representative action with respect the metric and  the background homogeneous part of the field ${\phi}(\tau)$ before performing the metric perturbation we get the following classical field equations for the system:
\bea &&{\cal H}^2=\frac{1}{3}\bigg(a^2(\tau)V(\phi(\tau))+\frac{1}{2}\left(\frac{d\phi(\tau)}{d\tau}\right)^2\bigg),\\
&&\frac{d {\cal H}}{d\tau}=\frac{1}{3}\bigg(a^2(\tau)V(\phi(\tau))-\left(\frac{d\phi(\tau)}{d\tau}\right)^2\bigg),\\
&& \frac{d^2\phi(\tau)}{d\tau^2}+2{\cal H}\frac{d\phi(\tau)}{d\tau}+a^2(\tau)V(\phi(\tau))=0.\eea
\subsection{Scalar modes in Cosmological Perturbation Theory} 
In this subsection,  we discuss about the generation of scalar and tensor modes in Cosmological Perturbation Theory as an outcome of quantum fluctuations in the primordial universe.  To serve this purpose,  first of all we perturb the scalar field in the mentioned quasi De-Sitter spatially flat background,  which gives$\phi({\bf x},\tau)=\phi(\tau)+\delta\phi({\bf x},\tau),$
where,  $\phi(\tau)$ is the dynamical scalar field described in the homogeneous background before perturbation and $\delta\phi({\bf x},\tau)$ is the contribution in the perturbation which captures all sorts of inhomogeneous effects.  In this description the complete dynamics after perturbation is described in terms of the gauge invariant perturbation variable $\zeta({\bf x},\tau)$ which is given by,  
$\zeta({\bf x},\tau)=-
\frac{{\cal H}}{(d\phi(\tau)/d\tau}~\delta\phi({\bf x},\tau).$
Further,  considering the linear contributions in the Cosmological Perturbation Theory for the mentioned spatially flat quasi De Sitter FLRW background we finally have the following perturbed line element:
\bea  ds^2_{\rm perturbed}=a^2(\tau)\bigg(-d\tau^2+\left\{
\left[1+2\zeta({\bf x},\tau)\right]\delta_{ij}+2h_{ij}({\bf x},\tau)\right\}dx^i dx^j\bigg),\eea
where the solution for the scale factor in the conformal time coordinate for the quasi De Sitter space is mentioned in the previous subsection.  In this derived expression for the perturbed metric,  $\zeta({\bf x},\tau)$ representing the scalar perturbation and the corresponding tensor perturbation is described by $h_{ij}({\bf x},\tau)$ which satisfy the transverse and traceless properties,  
$\partial_i h_{ij}({\bf x},\tau)=0,$ $h^i_i({\bf x},\tau)=0.$
Here it is important to note that the perturbed line element is actually derived in the co-moving gauge, 
$\delta\phi({\bf x},\tau)=0.$  
 \subsubsection{Without Bell's inequality violation}~~~~~~~~~~~~~~~~~~~~~~~~~~~~~~~~~~~~~~~~~~~~~~~~~~~~~~~~\\
After gauge fixing in the absence of any kind of Bell's inequality violating contribution the second order perturbed action for the scalar fluctuations can be expressed in the conformal coordinate as:
\bea \delta^{(2)}S^{\bf NB}_{\bf Scalar}&=&\frac{1}{2}\int d\tau~d^3{\bf x}~\frac{a^2(\tau)}{{\cal H}^2}~\left(\frac{d\phi(\tau)}{d\tau}\right)^2~\bigg(\left(\partial_{\tau}\zeta({\bf x},\tau)\right)^2-\left(\partial_{j}\zeta({\bf x},\tau)\right)^2\bigg).~~~~\eea 
Further,  we re-parametrize the above mentioned action in terms of the newly introduce space-time dependent perturbation variable $v({\bf x},\tau):=z(\tau)~\zeta({\bf x},\tau),$
where conformal time dependent variable $z(\tau)$ is known as the {\it Mukhanov-Sasaki varibale} which is defined as, 
$z(\tau):=a(\tau)~\sqrt{2\epsilon(\tau)}$.
In terms of this newly introduced re-parametrized perturbation field variable $v({\bf x},\tau)$ the above mentioned gauge fixed second order action for the scalar part of the metric perturbation can be expressed as:
\bea \delta^{(2)}S^{\bf NB}_{\bf Scalar}&=&\int d\tau~d^3{\bf x}~\bigg(\left(\partial_{\tau}v({\bf x},\tau)\right)^2-\left(\partial_{j}v({\bf x},\tau)\right)^2-2\left(\frac{1}{z(\tau)}\frac{dz(\tau)}{d\tau}\right)v({\bf x},\tau)\left(\partial_{\tau}v({\bf x},\tau)\right)\nonumber\\
&&~~~~~~~~~~~~~~~~~~~~~~~~~~~~~~~~~~~~~~~~~~~~~~~~~~~~~~~+\left(\frac{1}{z(\tau)}\frac{dz(\tau)}{d\tau}\right)^2v^2({\bf x},\tau)^2\bigg).~~~~\eea
In the above mentioned action the contribution from the {\it Mukhanov-Sasaki varibale} can be further simplified to the following form:
\bea \left(\frac{1}{z(\tau)}\frac{dz(\tau)}{d\tau}\right):&=&\bigg[\left(\frac{1}{a(\tau)}\frac{da(\tau)}{d\tau}\right)+\frac{1}{2}\left(\frac{1}{\epsilon(\tau)}\frac{d\epsilon(\tau)}{d\tau}\right)\bigg]=\bigg[{\cal H}+\frac{1}{\displaystyle \bigg(1-\frac{1}{{\cal H}^2}\frac{d{\cal H}}{d\tau}\bigg)}\bigg\{\frac{1}{{\cal H}^3}\bigg(\frac{d{\cal H}}{d\tau}\bigg)^2-\frac{1}{2{\cal H}^2}\bigg(\frac{d^2{\cal H}}{d\tau^2}\bigg)\bigg\}\bigg].~~~~~~~\eea
To simplify further our next job is to introduce the following Fourier transformation in the newly introduced re-parametrized perturbation field variable $v({\bf x},\tau)$,  which is given by:
\bea v({\bf x},\tau):=\int \frac{d^3{\bf k}}{(2\pi)^3}~\exp(i {\bf k}.{\bf x})~v_{\bf k}(\tau).\eea 
Next substituting these expressions in the gauge fixed second order action and performing the integration over the space coordinates we get the following simplified result for the action in the Fourier space,  which is given by:
\bea \delta^{(2)}S^{\bf NB}_{\bf Scalar}&=&\int d\tau~d^3{\bf k}~\bigg(|\partial_{\tau}v_{\bf k}(\tau)|^2+\left(k^2+\left(\frac{1}{z(\tau)}\frac{dz(\tau)}{d\tau}\right)^2\right)|v_{\bf k}(\tau)|^2\nonumber\\
&&~~~~~-\left(\frac{1}{z(\tau)}\frac{dz(\tau)}{d\tau}\right)v_{-\bf k}(\tau)\left(\partial_{\tau}v_{\bf k}(\tau)\right)-\left(\frac{1}{z(\tau)}\frac{dz(\tau)}{d\tau}\right)v_{\bf k}(\tau)\left(\partial_{\tau}v_{-\bf k}(\tau)\right)\bigg),~~~~~~~~\eea
where we define the following quantities:
\bea &&|\partial_{\tau}v_{\bf k}(\tau)|^2:=\left(\partial_{\tau}v_{-\bf k}(\tau)\right)\left(\partial_{\tau}v_{\bf k}(\tau)\right),|v_{\bf k}(\tau)|^2:=v_{-\bf k}(\tau)v_{\bf k}(\tau).\eea
This in fact implies that in Fourier space the gauge fixed second order action actually expressed in terms of two modes,  one carries momentum ${\bf k}$ and the other one carries momentum $-{\bf k}$. 

Further,  we perform the integration by parts over the conformal time coordinate,  which will give rise to the following simplified form of the gauge fixed second order action in the Fourier space:
\bea \delta^{(2)}S^{\bf NB}_{\bf Scalar}&=&\int d\tau~d^3{\bf k}~{\cal L}^{\bf NB}_{\bf Scalar}(v_{\bf k}(\tau),\partial_{\tau}v_{\bf k}(\tau)),~~~~~~~~\eea
where the corresponding Lagrangian density in Fourier space can be written as:
\bea {\cal L}^{\bf NB}_{\bf Scalar}(v_{\bf k}(\tau),\partial_{\tau}v_{\bf k}(\tau)):=\bigg(|\partial_{\tau}v_{\bf k}(\tau)|^2-\omega^2_{\rm eff}(k,\tau)|v_{\bf k}(\tau)|^2\bigg).\eea
Here the effective conformal time dependent frequency in the Fourier space can be written as:
\bea \omega^2_{\rm eff}(k,\tau):=\bigg(k^2-\frac{1}{z(\tau)}\frac{d^2z(\tau)}{d\tau^2}\bigg)., ~~~~{\rm where}~~~~\frac{1}{z(\tau)}\frac{d^2z(\tau)}{d\tau^2}:=\frac{1}{\tau^2}\bigg(\nu^2_{\bf S}(\tau)-\frac{1}{4}\bigg),\eea
where the expression for mass parameter for the scalar modes $\nu_{\bf S}(\tau)$ is given by the following expression:
\bea \nu_{\bf S}(\tau):\approx \frac{3}{2}\bigg(1+2\epsilon(\tau)-\frac{2}{3}\eta(\tau)+\cdots\bigg),\eea 
where the parameters $\epsilon(\tau)$ and $\eta(\tau)$ are slowly time varying parameters.   Here the $\cdots$ terms are the sub-leading correction terms in this expression.  We have defined $\epsilon(\tau)$ earlier and $\eta(\tau)$ is defined in conformal time coordinate as$\eta(\tau)=2\epsilon(\tau)-\frac{1}{2{\cal H}}\frac{d\ln \epsilon(\tau)}{d\tau}.$
In the slowly time varying region it is expected to have,  $\epsilon(\tau)\ll 1$ and $|\eta(\tau)|\ll 1$.  For this reason,  in the mass parameter for the scalar modes $\nu_{\bf S}$ these parameters are appearing as a very slowly time varying correction parameter which is basically taking care of small deviation from the exact De Sitter solution.  This justifies the quasi De Sitter approximation justifiable in the present computation.  In this case after substituting the expression for the mass parameter in the slowly time varying limit we get the following simplified result for the above mentioned quantity:
\bea \frac{1}{z(\tau)}\frac{d^2z(\tau)}{d\tau^2}:
&\approx&\frac{2}{\tau^2}\bigg(1+\frac{9}{2}\bigg(\epsilon(\tau)-\frac{1}{3}\eta(\tau)\bigg)\bigg)+{\cal O}(\epsilon^2(\tau),\eta^2(\tau)).\eea

In terms of the slowly varying conformal time dependent parameters we get then the following simplified result in quasi de Sitter space which will going to be very useful in the rest of the computation performed in this paper:
\bea \left(\frac{1}{z(\tau)}\frac{dz(\tau)}{d\tau}\right)^2&\approx&\frac{1}{\tau^2}\bigg[1+2\bigg(2\epsilon(\tau)-\eta(\tau)\bigg)\bigg].~~~~~~~~~\eea

After varying the gauge fixed second order action in the Fourier space we get the following equation of motion for the perturbation field variable for the scalar modes:
\bea \bigg[\frac{d^2}{\d\tau^2}+\omega^2_{\rm eff}(k,\tau)\bigg]v_{\bf k}(\tau)=0.\eea
The most general solution of of the above mentioned equation of motion can be expressed as:
\bea v_{\bf k}(\tau)=\sqrt{-\tau}\left[{\cal C}_1~{\cal H}^{(1)}_{\nu_{\bf S}}(-k\tau)+{\cal C}_2~{\cal H}^{(2)}_{\nu_{\bf S}}(-k\tau)\right].~~~~~~~\eea
Here ${\cal H}^{(1)}_{\nu_{\bf S}}(-k\tau)$ and ${\cal H}^{(2)}_{\nu_{\bf S}}(-k\tau)$ are the Hankel functions of first and second kind with order  $\nu_{\bf S}$.  Also,  ${\cal C}_1$ and ${\cal C}_2$ are two arbitrary integration constants,  which can be fixed by choosing proper quantum initial conditions.  There are three possible choices,  which are commonly used to fix these coefficients:
\bea &&{\rm Bunch-Davies/Euclidean~ vacuum:}~~~~~~{\cal C}_1=1,~~~~{\cal C}_2=0,\\
&&{\rm \alpha~ vacua:}~~~~~~~~~~~~~~~~~~~~~~~~~~~~~~~~~~~{\cal C}_1=\cosh\alpha,~~~~{\cal C}_2=\sinh\alpha,\\
&&{\rm Mota-Allen~ vacua:}~~~~~~~~~~~~~~~~{\cal C}_1=\cosh\alpha,~~~~{\cal C}_2=\exp(i\gamma)\sinh\alpha.
\eea
These all are ${\rm SO}(1,4)$ isommetric De Sitter vacua.

Further combining the above the asymptotic solutions at $-k\tau\rightarrow 0,\infty$,  we get the following simplified solution of the mode function for the scalar modes:
\begin{align}
\nonumber
v_{\bf k}(\tau) &= -\frac{2^{\nu_{\bf S}-\frac{3}{2}}~(-k\tau)^{\frac{3}{2}-\nu_{\bf S}}}{\sqrt{2k}} \biggl|\frac{\Gamma(\nu_{\bf S})}{\Gamma\left(\frac{3}{2}\right)}\biggr|\times \biggl[{\cal C}_1\left(1-\frac{i}{k\tau}\right)\exp\biggl(-i\biggl\{k\tau+\frac{\pi}{2}\left(\nu_{\bf S}+\frac{1}{2}\right)\biggr\}\biggr) \\ & ~~~~~~~~~~~~~~~~~~~~~~~~~~~~~~~~+ {\cal C}_2 \left(1+\frac{i}{k\tau}\right)\exp\biggl(i\biggl\{k\tau+\frac{\pi}{2}\left(\nu_{\bf S}+\frac{1}{2}\right)\biggr\}\biggr)\biggr].	
\end{align} 
 \subsubsection{With Bell's inequality violation}~~~~~~~~~~~~~~~~~~~~~~~~~~~~~~~~~~~~~~~~~~~~~~~~~~~~~~~~\\
After gauge fixing in presence of Bell's inequality violating contribution the second order perturbed action for the scalar fluctuations can be expressed in the conformal coordinate as~\footnote{Here in the first order a non-zero term exist due to Bell's inequality violation:
\bea \delta^{(1)}S^{\bf B,int}_{\bf Scalar}&=&-\int d\tau~d^3{\bf x}~\frac{m(\tau)}{{\cal H}}~\partial_{\tau}\zeta({\bf x},\tau)~\delta^3({\bf x}).\eea 
which will not directly contribute to the final equation of motion,  but appearance of such term will lead to non-zero one point function at the end which is not present in the usual cosmological set up where Bell's inequality violation not implemented.  Due to the presence of the three dimensional Dirac Delta function $\delta^3({\bf x})$ one can treat the corresponding Bell's violating contribution is localized in space at ${\bf x}=0$.  }:
\bea \delta^{(2)}S^{\bf B}_{\bf Scalar}&=&\frac{1}{2}\int d\tau~d^3{\bf x}~\frac{a^2(\tau)}{{\cal H}^2}~\left(\frac{d\phi(\tau)}{d\tau}\right)^2~\bigg(\left(\partial_{\tau}\zeta({\bf x},\tau)\right)^2-\left(\partial_{j}\zeta({\bf x},\tau)\right)^2\bigg)\nonumber\\
&&~~~~~~~~~~~-\frac{1}{2}\int d\tau~d^3{\bf x}~\frac{a^2(\tau)}{{\cal H}^2}~\left(\frac{d\phi(\tau)}{d\tau}\right)^2~a^2(\tau)m^2(\tau)~\zeta^2({\bf x},\tau).~~~~\eea  
Here the first term is exactly same as appearing from the second order perturbed action without having any Bell's inequality violation in the previous case.  On the other hand,  the second term is appearing due to explicit Bell's inequality violating contribution in the cosmological perturbation theory.  Practically the the second term captures the effect of massive particle ($m\gg {\cal H}$) which in in general represented by the conformal time dependent mass $m(\tau)$.  Also it is important to note that,  in the interaction picture this can be identified as a coupling parameter.  In quasi De Sitter evolutionary picture implementing the concept of Bell's inequality violation itself a very complicated task.  One can treat the above mentioned framework to be a special one where such specific contribution is appearing,  which helps further to violate Bell's inequality explicitly.  

Further,  we re-parametrize the above mentioned action in terms of the newly introduce space-time dependent perturbation variable,  $v({\bf x},\tau)$,  which we have explicitly defined in the previous case,  and consequently we get the following simplified result for the gauge fixed second order action for the scalar part of the metric perturbation in presence of Bell's inequality violating contribution~\footnote{After re-parametrizing this interacting part of the action in terms of the space-time dependent perturbation variable,  $v({\bf x},\tau)$,  we get the following simplified version of the Bell's inequality violating action:
\bea \delta^{(1)}S^{\bf B,int}_{\bf Scalar}&=&-\int d\tau~d^3{\bf x}~\frac{m(\tau)}{{\cal H}}~\frac{1}{z(\tau)}~\bigg(\partial_{\tau}v({\bf x},\tau)-\left(\frac{1}{z(\tau)}\frac{dz(\tau)}{d\tau}\right)v({\bf x},\tau)\bigg)~\delta^3({\bf x}).\eea 
}:
\bea \delta^{(2)}S^{\bf B}_{\bf Scalar}&=&\frac{1}{2}\int d\tau~d^3{\bf x}~\bigg(\left(\partial_{\tau}v({\bf x},\tau)\right)^2-\left(\partial_{j}v({\bf x},\tau)\right)^2-2\left(\frac{1}{z(\tau)}\frac{dz(\tau)}{d\tau}\right)v({\bf x},\tau)\left(\partial_{\tau}v({\bf x},\tau)\right)\nonumber\\
&&~~~~~~~~~~+\left(\frac{1}{z(\tau)}\frac{dz(\tau)}{d\tau}\right)^2v^2({\bf x},\tau)^2\bigg)-\frac{1}{2}\int d\tau~d^3{\bf x}~a^2(\tau)m^2(\tau)~v^2({\bf x},\tau).~~~~\eea 
To simplify further we use the previously mentioned ansatz for Fourier transformation of the re-parametrized perturbation field variable $v({\bf x},\tau)$ using which the above mentioned action can be expressed as~\footnote{Further performing Fourier transformation we get the following simplified action:
\bea \delta^{(1)}S^{\bf B,int}_{\bf Scalar}&=&-\int d\tau~d^3{\bf k}~\frac{m(\tau)}{{\cal H}}~\frac{1}{z(\tau)}~\bigg(\partial_{\tau}v_{\bf k}(\tau)-\left(\frac{1}{z(\tau)}\frac{dz(\tau)}{d\tau}\right)v_{\bf k}(\tau)\bigg).\eea
This result will be further used during the construction of Hamiltonian. }:
\bea \delta^{(2)}S^{\bf B}_{\bf Scalar}&=&\int d\tau~d^3{\bf k}~\bigg(|\partial_{\tau}v_{\bf k}(\tau)|^2+\left(k^2+\left(\frac{1}{z(\tau)}\frac{dz(\tau)}{d\tau}\right)^2\right)|v_{\bf k}(\tau)|^2\nonumber\\
&&~~~~~~~~~~~~~~-\left(\frac{1}{z(\tau)}\frac{dz(\tau)}{d\tau}\right)v_{-\bf k}(\tau)\left(\partial_{\tau}v_{\bf k}(\tau)\right)-\left(\frac{1}{z(\tau)}\frac{dz(\tau)}{d\tau}\right)v_{\bf k}(\tau)\left(\partial_{\tau}v_{-\bf k}(\tau)\right)\bigg)\nonumber 
\\
&&~~~~~~~~~~~~~~~~~~~~~~~~~~~~~~~~~~~~~~~~~~-\int d\tau~d^3{\bf k}~a^2(\tau)m^2(\tau)~|v_{\bf k}(\tau)|^2.~~~~~~~~\eea 
Further,  we perform the integration by parts over the conformal time coordinate,  which will give rise to the following simplified form of the gauge fixed second order action in the Fourier space including Bell's inequality violating contribution:
\bea \delta^{(2)}S^{\bf B}_{\bf Scalar}&=&\int d\tau~d^3{\bf k}~{\cal L}^{\bf B}_{\bf Scalar}(v_{\bf k}(\tau),\partial_{\tau}v_{\bf k}(\tau)),~~~~~~~~\eea
where the corresponding Bell's inequality violating Lagrangian density in Fourier space can be written as:
\bea {\cal L}^{\bf B}_{\bf Scalar}(v_{\bf k}(\tau),\partial_{\tau}v_{\bf k}(\tau)):=\bigg(|\partial_{\tau}v_{\bf k}(\tau)|^2-\Omega^2_{\rm eff}(k,\tau)|v_{\bf k}(\tau)|^2\bigg).\eea
Here the effective conformal time dependent frequency in the Fourier space can be written as:
\bea \Omega^2_{\rm eff}(k,\tau):&=&\bigg(k^2+a^2(\tau)m^2(\tau)-\frac{1}{z(\tau)}\frac{d^2z(\tau)}{d\tau^2}\bigg)=\bigg(k^2+\frac{1}{\tau^2}\left[\frac{m^2(\tau)}{{\cal H}^2}-\bigg(\nu^2_{\bf S}(\tau)-\frac{1}{4}\bigg)\right]\bigg). \eea
Next we express the second term in the effective conformal time dependent frequency in terms of a mass parameter for the scalar modes $\nu_{\bf S}$ as defined in the earlier section. 

After varying the gauge fixed second order action in the Fourier space we get the following equation of motion for the perturbation field variable for the scalar modes:
\bea \bigg[\frac{d^2}{\d\tau^2}+\Omega^2_{\rm eff}(k,\tau)\bigg]v_{\bf k}(\tau)=0.\eea
The most general solution of of the above mentioned equation of motion may or may not be found dependent on a specific conformal time dependent mass profile $m(\tau)$.   One can try with different types of mass profiles to solve the above equation of motion.  Since this portion of our work is motivated from Bell's inequality violation in the primordial cosmology set up,  we are going the use the following profile which was used in previous refs.~\cite{Maldacena:2015bha,Choudhury:2016cso,Choudhury:2016pfr} $\frac{m^2(\tau)}{{\cal H}^2}=\gamma\left(\frac{\tau}{\tau_0}-1\right)^2+\delta.$
Here $\gamma$ and $\delta$ are the two constants appearing in mass profile which can be fixed from the following two boundary conditions:
\bea &&\underline{\rm At~past~\tau=\tau_0:}~~~~~\longrightarrow ~~~\delta=\frac{m^2(\tau_0)}{{\cal H}(\tau_0)},\\
&&\underline{\rm At~present~\tau=0:}~~~\longrightarrow ~~~\gamma=\frac{m^2(0)}{{\cal H}(0)}-\frac{m^2(\tau_0)}{{\cal H}(\tau_0)}.\eea
It is important to note that in this context,  $\gamma\gg1$ and $\delta\gg 1$ for which we have the massive/heavy mass profile i.e.  $|m/{\cal H}|\gg 1$.

Using this profile the equation of motion can be recast in the following form:
\bea \bigg[\frac{d^2}{\d\tau^2}+\bigg(k^2+{\cal E}_1-\frac{1}{\tau}{\cal E}_2+\frac{1}{\tau^2}\left[\gamma+\delta-\bigg(\nu^2_{\bf S}(\tau)-\frac{1}{4}\bigg)\right]\bigg)\bigg]v_{\bf k}(\tau)=0,\eea
where ${\cal E}_1$ and ${\cal E}_2$,  are defined as, 
${\cal E}_1=\frac{\gamma}{\tau^2_0},{\cal E}_2=\frac{2\gamma}{\tau_0}=2{\cal E}_1\tau_0.$ 

The most general solution of the above mentioned equation of motion for the given mass profile is given by the following expression:
\bea v_{\bf k}(\tau)=(-\tau)^{\frac{3}{2}}~\exp\bigg(-\frac{1}{2}{\cal P}_1\tau\bigg)~\left({\cal P}_1\tau\right)^{{\cal P}_2+\frac{{\cal E}_2}{{\cal P}_1}}~\bigg[{\cal C}_1~{}_1F_1\left({\cal P}_2;{\cal P}_3;{\cal P}_1\tau\right)+{\cal C}_2~{\cal U}\left({\cal P}_2;{\cal P}_3;{\cal P}_1\tau\right)\bigg],~~~~~~~~\eea
where we have introduced three quantities ${\cal P}_1$,  ${\cal P}_2$ and ${\cal P}_3$,  which are defined as:
\bea {\cal P}_1&=&2i~\sqrt{k^2+{\cal E}_1},
{\cal P}_2=-\frac{{\cal E}_2}{{\cal P}_1}+i\sqrt{(\gamma+\delta)-\nu^2_{\bf S}}+\frac{1}{2},
{\cal P}_3=2\bigg({\cal P}_2+\frac{{\cal E}_2}{{\cal P}_1}\bigg).\eea 
Also,  ${\cal C}_1$ and ${\cal C}_2$ are two arbitrary integration constants,  which can be fixed by choosing proper quantum initial conditions.   

Further combining the asymptotic solutions at $-k\tau\rightarrow 0,\infty$,  we get the following simplified solution of the mode function for the scalar modes: 
\bea
\nonumber
&&v_{\bf k}(\tau) = (-\tau)^{\frac{3}{2}}~\exp\bigg(-\frac{1}{2}{\cal P}_1\tau\bigg)~\left({\cal P}_1\tau\right)^{{\cal P}_2+\frac{{\cal E}_2}{{\cal P}_1}}~\nonumber\\
 &&~~~~~~~~~\times\bigg[{\cal C}_1~\bigg(1+\frac{{\cal P}_1{\cal P}_2}{{\cal P}_3}\tau+\frac{\Gamma\left({\cal P}_3\right)}{\Gamma\left({\cal P}_3-{\cal P}_2\right)}\left(-{\cal P}_1\tau\right)^{-{\cal P}_2}+\frac{\Gamma\left({\cal P}_3\right)}{\Gamma\left({\cal P}_2\right)}\left({\cal P}_1\tau\right)^{{\cal P}_2-{\cal P}_3}\exp\left({\cal P}_1\tau\right)\bigg)\nonumber\\
 &&~~~~~~~~~~~~+{\cal C}_2~\bigg(\frac{\Gamma\left(1-{\cal P}_3\right)}{\Gamma\left(1+{\cal P}_2-{\cal P}_3\right)}\bigg(1+\frac{{\cal P}_1{\cal P}_2}{{\cal P}_3}\tau\bigg)+\frac{\Gamma\left({\cal P}_3-1\right)}{\Gamma\left({\cal P}_2\right)}\left({\cal P}_1\tau\right)^{1-{\cal P}_3}+\left({\cal P}_1\tau\right)^{-{\cal P}_2}\bigg)\bigg],~~~~~~~~
\eea

One can further consider a case having $m/{\cal H}\sim 1$ or exactly $m/{\cal H}=1$ and this is identified to be the partially massless case.  In this situation the field equation can be recast as:
\bea \bigg[\frac{d^2}{\d\tau^2}+\bigg(k^2-\frac{1}{\tau^2}\bigg(\nu^2_{\bf S}(\tau)-\frac{5}{4}\bigg)\bigg)\bigg]v_{\bf k}(\tau)=0.\eea
The most general solution of of the above mentioned equation of motion can be expressed as:
\bea v_{\bf k}(\tau)=\sqrt{-\tau}\left[{\cal C}_1~{\cal H}^{(1)}_{\sqrt{\nu^2_{\bf S}-1}}(-k\tau)+{\cal C}_2~{\cal H}^{(2)}_{\sqrt{\nu^2_{\bf S}-1}}(-k\tau)\right].~~~~~~~\eea
Here ${\cal H}^{(1)}_{\sqrt{\nu^2_{\bf S}-1}}(-k\tau)$ and ${\cal H}^{(2)}_{\sqrt{\nu^2_{\bf S}-1}}(-k\tau)$ are the Hankel functions of first and second kind with order  $\sqrt{\nu^2_{\bf S}-1}$.  Also,  ${\cal C}_1$ and ${\cal C}_2$ are two arbitrary integration constants,  which can be fixed by choosing proper quantum initial conditions.

Further combining the asymptotic solutions at $-k\tau\rightarrow 0,\infty$,  we get the following simplified solution of the mode function for the partially massless modes:
\begin{align} 
v_{\bf k}(\tau) &= -\frac{2^{\sqrt{\nu^2_{\bf S}-1}-\frac{3}{2}}~(-k\tau)^{\frac{3}{2}-\sqrt{\nu^2_{\bf S}-1}}}{\sqrt{2k}} \biggl|\frac{\Gamma\left(\sqrt{\nu^2_{\bf S}-1}\right)}{\Gamma\left(\frac{3}{2}\right)}\biggr|\\&\nonumber
\times \biggl[{\cal C}_1\left(1-\frac{i}{k\tau}\right)\exp\biggl(-i\biggl\{k\tau+\frac{\pi}{2}\left(\sqrt{\nu^2_{\bf S}-1}+\frac{1}{2}\right)\biggr\}\biggr) + {\cal C}_2 \left(1+\frac{i}{k\tau}\right)\exp\biggl(i\biggl\{k\tau+\frac{\pi}{2}\left(\sqrt{\nu^2_{\bf S}-1}+\frac{1}{2}\right)\biggr\}\biggr)\biggr].	   
\end{align}   
\subsection{Construction of Hamiltonian in Cosmology}
In this subsection our objective is to construct the Hamiltonian within the framework of primordial cosmology with and without having Bell's inequality violation.
\subsubsection{Without Bell's inequality violation}
Here we start with the previously mentioned Lagrangian density of re-parametrized gauge invariant perturbation field variable in Fourier space without having effect of any Bell's inequality violation,  which can be written before performing integration by parts over conformal time as:
\bea &&{\cal L}^{\bf NB}_{\bf Scalar}(v_{\bf k}(\tau),\partial_{\tau}v_{\bf k}(\tau)):=\bigg(|\partial_{\tau}v_{\bf k}(\tau)|^2+\left(k^2+\left(\frac{1}{z(\tau)}\frac{dz(\tau)}{d\tau}\right)^2\right)|v_{\bf k}(\tau)|^2\nonumber\\
&&~~~~~~~~~~~~~~~~~~~~~~~~~~~~~~~~~~~-\left(\frac{1}{z(\tau)}\frac{dz(\tau)}{d\tau}\right)v_{-\bf k}(\tau)\left(\partial_{\tau}v_{\bf k}(\tau)\right)-\left(\frac{1}{z(\tau)}\frac{dz(\tau)}{d\tau}\right)v_{\bf k}(\tau)\left(\partial_{\tau}v_{-\bf k}(\tau)\right)\bigg).~~~~~~~~~~\eea
Next we compute the canonically conjugate momenta corresponding to the re-parametrized gauge invariant perturbation field variable in Fourier space,  which is given by:
\bea \Pi_{v_{\bf k}}(\tau)=\frac{\partial {\cal L}^{\bf NB}_{\bf Scalar}(v_{\bf k}(\tau),\partial_{\tau}v_{\bf k}(\tau))}{\partial\left(\partial_{\tau}v_{\bf k}(\tau)\right)}=\bigg(\partial_{
\tau}v^{*}_{{\bf k}}(\tau)-\left(\frac{1}{z(\tau)}\frac{dz(\tau)}{d\tau}\right)v_{\bf k}(\tau)\bigg).\eea
Using this result we perform the Legendre transformation.  Consequently,  we get the following simplified expression for the Hamiltonian in primordial cosmology without having any Bell's inequality violation:
\bea {H}^{\bf NB}_{\bf k}(\tau)
&=&\bigg[\Pi_{v_{-\bf k}}(\tau)\Pi_{v_{\bf k}}(\tau)+k^2v_{-\bf k}(\tau)v_{\bf k}(\tau)+\left(\frac{1}{z(\tau)}\frac{dz(\tau)}{d\tau}\right)\bigg(\Pi_{v_{-\bf k}}(\tau)v_{\bf k}(\tau)+v_{-\bf k}(\tau)\Pi_{v_{\bf k}}(\tau)\bigg)\bigg]~~~~~~~~\eea
where we have actually expressed the Hamiltonian in the Fourier as a two mode Hamiltonian carrying both the momenta ${\bf k}$ as well as $-{\bf k}$. 
Here it is important to note that the first two terms represent the contribution from the free part of the Hamiltonian in Fourier space and the last two term is the outcome of the interaction.  To write it in this particular form we have used the following crucial conditions which are satisfying in the present context, 
$\Pi^{*}_{v_{\bf k}}(\tau)=\Pi_{v_{-\bf k}}(\tau),
v^{*}_{\bf k}(\tau)=v_{-\bf k}(\tau).$

On the other hand,  if we start with the Lagrangian density of re-parametrized gauge invariant perturbation field variable in Fourier space without having effect of any Bell's inequality violation,  which can be written after performing integration by parts over conformal time as:
\bea {\cal L}^{\bf NB}_{\bf Scalar}(v_{\bf k}(\tau),\partial_{\tau}v_{\bf k}(\tau)):&=&\bigg(|\partial_{\tau}v_{\bf k}(\tau)|^2-\bigg(k^2-\frac{1}{\tau^2}\bigg(\nu^2_{\bf S}(\tau)-\frac{1}{4}\bigg)\bigg)|v_{\bf k}(\tau)|^2\bigg),\eea
where the expression for mass parameter for the scalar modes $\nu_{\bf S}(\tau)$ is defined in terms of the slowly varying parameters $\epsilon(\tau)$ and $\eta(\tau)$ in the previous section.    

Next we compute the canonically conjugate momenta corresponding to the re-parametrized gauge invariant perturbation field variable in Fourier space,  which is given by:
\bea \Pi_{v_{\bf k}}(\tau)=\frac{\partial {\cal L}^{\bf NB}_{\bf Scalar}(v_{\bf k}(\tau),\partial_{\tau}v_{\bf k}(\tau))}{\partial\left(\partial_{\tau}v_{\bf k}(\tau)\right)}=\partial_{
\tau}v^{*}_{{\bf k}}(\tau).\eea
Using this result we perform the Legendre transformation.  Consequently,  we get the following simplified expression for the Hamiltonian in primordial cosmology without having any Bell's inequality violation:
\bea {H}^{\bf NB}_{\bf k}(\tau)
&=&\bigg[\Pi_{v_{-\bf k}}(\tau)\Pi_{v_{\bf k}}(\tau)+\bigg(k^2-\frac{1}{\tau^2}\bigg(\nu^2_{\bf S}(\tau)-\frac{1}{4}\bigg)\bigg)v_{-\bf k}(\tau)v_{\bf k}(\tau)\bigg].\eea
where we have actually expressed the Hamiltonian in the Fourier as a two mode Hamiltonian carrying both the momenta ${\bf k}$ as well as $-{\bf k}$. 
Here it is important to note that the first two terms represent the contribution from the free part of the Hamiltonian in Fourier space and after integration by parts over the conformal time no contribution left from the the interaction part.  Though the contribution from the interaction part is absent in this case explicitly,  but there is an additional conformal time dependent term appears in this case for the second term of the free part of the Hamiltonian,  which was absent in the previous case.  The explicit effects of these two Hamiltonians with and without having the interaction term will be more clearly visible when we quantize them explicitly.  Though for the present discussion of the paper the explicit quantization of the Hamiltonian in terms of the creation and annihilation operators of the oscillators are not directly required.   For this reason we will not discuss about this issue in the later section of the paper.  But we will surely investigate the effects of the Hamiltonian before considering integration by parts and after integration by parts in detail to check that whether it will propagate in the final result of the geometric phase computed in this paper or not.  In the last two steps of the above mentioned derived Hamiltonian we have used the reality condition of the re-parametrized gauge invariant perturbation field variable and its associated canonically conjugate momenta in Fourier space. Additionally it is important to note that,  during quantization we will study the effect in absence and also in the presence of squeezed quantum states in our computation.

\subsubsection{With Bell's inequality violation}
Here we start with the previously mentioned Lagrangian density of re-parametrized gauge invariant perturbation field variable in Fourier space without having effect of any Bell's inequality violation,  which can be written before performing integration by parts over conformal time as:
\bea &&{\cal L}^{\bf B}_{\bf Scalar}(v_{\bf k}(\tau),\partial_{\tau}v_{\bf k}(\tau)):=\bigg(|\partial_{\tau}v_{\bf k}(\tau)|^2+\left(k^2-\frac{1}{\tau^2}\frac{m^2(\tau)}{{\cal H}^2}+\left(\frac{1}{z(\tau)}\frac{dz(\tau)}{d\tau}\right)^2\right)|v_{\bf k}(\tau)|^2\nonumber\\
&&-\left(\frac{1}{z(\tau)}\frac{dz(\tau)}{d\tau}\right)v_{-\bf k}(\tau)\left(\partial_{\tau}v_{\bf k}(\tau)\right)-\left(\frac{1}{z(\tau)}\frac{dz(\tau)}{d\tau}\right)v_{\bf k}(\tau)\left(\partial_{\tau}v_{-\bf k}(\tau)\right)\bigg)+{\cal L}^{\bf B,int}_{\bf Scalar}(\partial_{\tau}v_{\bf k}(\tau)).\eea
Here ${\cal L}^{\bf B,int}_{\bf Scalar}(\partial_{\tau}v_{\bf k}(\tau))$ is the interaction part of the Lagrangian density which is generated due to Bell's inequality violation and contributed to the computation of the one point function in present context,  which is usually absent without having Bell's inequality violation.  The explicit form of this term is given by~\footnote{For details see the previous section where this contribution appearing first time due to Bell's inequality violation.}:
\bea {\cal L}^{\bf B,int}_{\bf Scalar}(\partial_{\tau}v_{\bf k}(\tau)):=-\frac{m(\tau)}{{\cal H}}~\frac{1}{z(\tau)}~\bigg(\partial_{\tau}v_{\bf k}(\tau)-\left(\frac{1}{z(\tau)}\frac{dz(\tau)}{d\tau}\right)v_{\bf k}(\tau)\bigg).\eea
Next we compute the canonically conjugate momenta corresponding to the re-parametrized gauge invariant perturbation field variable in Fourier space,  which is given by:
\bea &&\Pi_{v_{\bf k}}(\tau)=\frac{\partial {\cal L}^{\bf B}_{\bf Scalar}(v_{\bf k}(\tau),\partial_{\tau}v_{\bf k}(\tau))}{\partial\left(\partial_{\tau}v_{\bf k}(\tau)\right)}=\bigg(\partial_{
\tau}v^{*}_{{\bf k}}(\tau)-\left(\frac{1}{z(\tau)}\frac{dz(\tau)}{d\tau}\right)v_{\bf k}(\tau)-\frac{m(\tau)}{{\cal H}}~\frac{1}{z(\tau)}\bigg).\eea
Using this result we perform the Legendre transformation.  Consequently,  we get the following simplified expression for the Hamiltonian in primordial cosmology without having any Bell's inequality violation:
\bea {H}^{\bf B}_{\bf k}(\tau)
&=&\bigg[\Pi_{v_{-\bf k}}(\tau)\Pi_{v_{\bf k}}(\tau)+\bigg(k^2+\frac{1}{\tau^2}\frac{m^2(\tau)}{{\cal H}^2}\bigg)v_{-\bf k}(\tau)v_{\bf k}(\tau)\nonumber\\
&&~~~~~~~~~+\left(\frac{1}{z(\tau)}\frac{dz(\tau)}{d\tau}\right)\bigg(\Pi_{v_{-\bf k}}(\tau)v_{\bf k}(\tau)+v_{-\bf k}(\tau)\Pi_{v_{\bf k}}(\tau)\bigg)+\frac{m(\tau)}{{\cal H}}~\frac{1}{z(\tau)}~\Pi_{v_{-\bf k}}(\tau)\bigg]\eea 
where we have actually expressed the Hamiltonian in the Fourier as a two mode Hamiltonian carrying both the momenta ${\bf k}$ as well as $-{\bf k}$ for the first four terms.  In the one last additional term linear contribution of the canonically conjugate momenta is appearing with momenta $-{\bf k}$.
Here it is important to note that the first two terms represent the contribution from the free part of the Hamiltonian in Fourier space and the last three term is the outcome of the interaction.  Particularly the last term and conformal time dependent mass squared term carrying the signature of Bell's inequality violation in the present context,  which was absent in the previous analysis.

On the other hand,  if we start with the Lagrangian density of re-parametrized gauge invariant perturbation field variable in Fourier space without having effect of any Bell's inequality violation,  which can be written after performing integration by parts over conformal time as:
\bea {\cal L}^{\bf B}_{\bf Scalar}(v_{\bf k}(\tau),\partial_{\tau}v_{\bf k}(\tau)):
&=&\bigg(|\partial_{\tau}v_{\bf k}(\tau)|^2-\bigg(k^2+\frac{1}{\tau^2}\bigg(\frac{m^2(\tau)}{{\cal H}^2}-\bigg(\nu^2_{\bf S}(\tau)-\frac{1}{4}\bigg)\bigg)\bigg)|v_{\bf k}(\tau)|^2\bigg)\nonumber\\
&&~~~~~~~~~~~~~~~~~~~~~~~~~~~~~~~~~~~~~~~~+{\cal L}^{\bf B,int}_{\bf Scalar}(\partial_{\tau}v_{\bf k}(\tau)),\eea
where the expression for mass parameter for the scalar modes $\nu_{\bf S}(\tau)$ is defined in terms of the slowly varying parameters $\epsilon(\tau)$ and $\eta(\tau)$ in the previous section.   

Next we compute the canonically conjugate momenta corresponding to the re-parametrized gauge invariant perturbation field variable in Fourier space,  which is given by:
\bea \Pi_{v_{\bf k}}(\tau)=\frac{\partial {\cal L}^{\bf B}_{\bf Scalar}(v_{\bf k}(\tau),\partial_{\tau}v_{\bf k}(\tau))}{\partial\left(\partial_{\tau}v_{\bf k}(\tau)\right)}=\bigg(\partial_{
\tau}v^{*}_{{\bf k}}(\tau)-\frac{m(\tau)}{{\cal H}}~\frac{1}{z(\tau)}\bigg).\eea 
Using this result we perform the Legendre transformation.  Consequently,  we get the following simplified expression for the Hamiltonian in primordial cosmology without having any Bell's inequality violation:
\bea {H}^{\bf B}_{\bf k}(\tau)
&=&\bigg[\Pi_{v_{-\bf k}}(\tau)\Pi_{v_{\bf k}}(\tau)+\bigg(k^2+\frac{1}{\tau^2}\bigg(\frac{m^2(\tau)}{{\cal H}^2}-\bigg(\nu^2_{\bf S}(\tau)-\frac{1}{4}\bigg)\bigg)\bigg)v_{-\bf k}(\tau)v_{\bf k}(\tau)\nonumber\\
&&~~~~~~~~~~~~~~~~~~~~~~~~~~~~~~~~~~~~~~~~~~+\frac{m(\tau)}{{\cal H}}~\frac{1}{z(\tau)}~\bigg(\Pi_{v_{-\bf k}}(\tau)-\bigg(\frac{1}{z(\tau)}\frac{dz(\tau)}{d\tau}\bigg)v_{\bf k}(\tau)\bigg)\bigg].\eea
where we have actually expressed the Hamiltonian in the Fourier as a two mode Hamiltonian carrying both the momenta ${\bf k}$ as well as $-{\bf k}$ for the first two terms.  In the one last additional term linear contribution of the canonically conjugate momenta is appearing with momenta $-{\bf k}$.  Particularly the last two terms in this case and conformal time dependent mass squared term carrying the signature of Bell's inequality violation in the present context,  which was absent in the previous analysis.

\subsection{Lewis Riesenfeld invariant trick in Cosmology} 
In this section our prime objective is to find out the solution of the conformal time dependent Schrödinger equation using the previously mentioned Hamiltonian of the scalar modes with and without having the effect of Bell's inequality violation using the Lewis Riesenfeld invariant trick within the framework of cosmology.  

In the present context of discussion the associated conformal time dependent Schrödinger equation can be written with and without having the effect of Bell's inequality violation can be written as:
\bea && \underline{\rm Without~Bell's~inequality~violation:}~~~~~~~~~~\hat{H}^{\bf NB}_{\bf k}(\tau)\Psi_{\bf NB}(v_{-\bf k},v_{\bf k},\tau)=i\frac{\partial \Psi_{\bf NB}(v_{-\bf k},v_{\bf k},\tau)}{\partial\tau},~~~~~~\\
&& \underline{\rm With~Bell's~inequality~violation:}~~~~~~~~~\hat{H}^{\bf B}_{\bf k}(\tau)\Psi_{\bf B}(v_{-\bf k},v_{\bf k},\tau)=i\frac{\partial \Psi_{\bf B}(v_{-\bf k},v_{\bf k},\tau)}{\partial\tau}.\eea
Here,  $\hat{H}^{\bf NB}_{\bf k}(\tau)$ and $\hat{H}^{\bf B}_{\bf k}(\tau)$ are the quantum mechanical Hamiltonians without and with having the effect of Bell's inequality violation,  which we have obtained from the previously derived classical Hamiltonians ${H}^{\bf NB}_{\bf k}(\tau)$ and ${H}^{\bf B}_{\bf k}(\tau)$ by promoting the momenta and field variable as quantum mechanical operator i.e.  
\bea && \Pi_{v_{-\bf k}}(\tau)\rightarrow \hat{\Pi}_{v_{-\bf k}}(\tau), \Pi_{v_{\bf k}}(\tau)\rightarrow \hat{\Pi}_{v_{\bf k}}(\tau), v_{-\bf k}(\tau)\rightarrow \hat{v}_{-\bf k}(\tau), v_{\bf k}(\tau)\rightarrow \hat{v}_{\bf k}(\tau).\eea
This allows us to treat the above mentioned Hamiltonians as:
\bea &&\hat{H}^{\bf NB}_{\bf k}(\tau)\equiv \hat{H}^{\bf NB}_{\bf k}( \hat{\Pi}_{v_{-\bf k}}(\tau),\hat{\Pi}_{v_{\bf k}}(\tau),\hat{v}_{-\bf k}(\tau),\hat{v}_{\bf k}(\tau)),
\hat{H}^{\bf B}_{\bf k}(\tau)\equiv \hat{H}^{\bf NB}_{\bf k}( \hat{\Pi}_{v_{-\bf k}}(\tau),\hat{\Pi}_{v_{\bf k}}(\tau),\hat{v}_{-\bf k}(\tau),\hat{v}_{\bf k}(\tau)).~~~~~~\eea
Here explicit quantization in terms of creation and annihilation operators are not actually required to serve the present purpose. 
Also $\Psi_{\bf NB}(v_{-\bf k},v_{\bf k},\tau)$ and $\Psi_{\bf B}(v_{-\bf k},v_{\bf k},\tau)$ are the associated wave functions for the scalar modes of the cosmological perturbation without and with having the effect of Bell's inequality violation.  Since the mathematical structures of the quantum mechanical Hamiltonian's are different with and without having the effect of Bell's inequality violation it is expected to be automatically reflected in the corresponding conformal time dependent wave functions as stated above. 

Now we are going to implement this well known Lewis Riesenfeld invariant trick in the above mentioned framework to explicitly determine the expressions for the non-trivial Hermitian operators, which we are going to compute in presence and absence of having the contributions from Bell's inequality violation within the framework of primordial quantum mechanical scalar fluctuations.  This can be easily being done by looking into the following the well known quantum Liouville equations explicitly written for the above mentioned two situations:
\bea && \underline{\rm Without~Bell's~inequality~violation:}~~~~~~~~~~~\frac{d \hat{\cal I}^{\bf LR}_{\bf NB,k}(\tau)}{d\tau}= \frac{\partial \hat{\cal I}^{\bf LR}_{\bf NB,k}(\tau)}{\partial\tau}+i\left[\hat{H}^{\bf NB}_{\bf k}(\tau),\hat{\cal I}^{\bf LR}_{\bf NB,k}(\tau)\right],~~~~~~~~~~~\\
&& \underline{\rm With~Bell's~inequality~violation:}~~~~~~~~~~\frac{d \hat{\cal I}^{\bf LR}_{\bf B,k}(\tau)}{d\tau}= \frac{\partial \hat{\cal I}^{\bf LR}_{\bf B,k}(\tau)}{\partial\tau}+i\left[\hat{H}^{\bf B}_{\bf k}(\tau),\hat{\cal I}^{\bf LR}_{\bf B,k}(\tau)\right],\eea
where,  $\hat{\cal I}^{\bf LR}_{\bf NB,k}(\tau)$ and $\hat{\cal I}^{\bf LR}_{\bf B,k}(\tau)$ are the Lewis Riesenfeld invariant quantum operators without and with having the effect of Bell's inequality violation in primordial cosmology.  It would be really interesting to find the underlying differences appearing in the above mentioned Lewis Riesenfeld invariant quantum operators computed without and with having the Bell's inequality violating effects. which will be helpful to explore various unknowns in quantum aspects of primordial cosmology. 

Now it is important to note that,  when  Lewis Riesenfeld invariant quantum operators not explicitly contain any conformal time derivatives one can write down the solution of the Schrödinger equation in the following simplified form:
\bea &&\Psi_{{\bf NB}}(v_{-\bf k},v_{\bf k},\tau)=\exp\bigg(i\gamma^{\bf LR, NB}_{{\bf k}}(\tau)\bigg)~\Phi_{{\bf NB}}(v_{-\bf k},v_{\bf k},\tau),\\
&&\Psi^{(n)}_{{\bf B}}(v_{-\bf k},v_{\bf k},\tau)=\exp\bigg(i\gamma^{\bf LR, B}_{n,{\bf k}}(\tau)\bigg)~\Phi_{{\bf B}}(v_{-\bf k},v_{\bf k},\tau).
\eea 
Here,  $\Phi_{{\bf NB}}(v_{-\bf k},v_{\bf k},\tau)$ and $\Phi_{{\bf B}}(v_{-\bf k},v_{\bf k},\tau)$ are corresponding eigenfunctions of Lewis Riesenfeld invariant quantum operators and most importantly,  $\gamma^{\bf LR, NB}_{{\bf k}}(\tau)$ and $\gamma^{\bf LR, NB}_{{\bf k}}(\tau)$ are identified to be the Lewis Riesenfeld phase factor without and with having the effect of Bell's inequality violation in primordial cosmology.  In the present context this phase can be computed from the following expressions for the two cases:
\bea \label{LRP1} \gamma^{\bf LR, NB}_{{\bf k}}(\tau):
&=&\gamma^{\bf PB, NB}_{{\bf k}}(\tau)+\delta^{\bf Dynamical,NB}_{{\bf k}}(\tau),\\
\label{LRP2} \gamma^{\bf LR, B}_{{\bf k}}(\tau):
&=&\gamma^{\bf PB, B}_{{\bf k}}(\tau)+\delta^{\bf Dynamical,B}_{{\bf k}}(\tau),\eea
where,  the {\it Pancharatnam-Berry phase} and the {\it dynamical phase} in this construction can be expressed as:
\bea \gamma^{\bf PB, NB}_{{\bf k}}(\tau):&=&i\int^{\tau}_{-\infty}d\tau^{\prime}~\langle\Phi_{\bf NB}(v_{-\bf k},v_{\bf k},\tau^{\prime})|\bigg(\frac{\partial}{\partial \tau^{\prime}}\bigg)|\Phi_{\bf NB}(v_{-\bf k},v_{\bf k},\tau^{\prime})\rangle,\\
\delta^{\bf Dynamical,NB}_{{\bf k}}(\tau):&=&-\int^{\tau}_{-\infty}d\tau^{\prime}~\langle\Phi_{\bf NB}(v_{-\bf k},v_{\bf k},\tau^{\prime})|\hat{H}^{\bf NB}_{\bf k}(\tau^{\prime})|\Phi_{\bf NB}(v_{-\bf k},v_{\bf k},\tau^{\prime})\rangle,~~~~~\eea
and 
\bea \gamma^{\bf PB, B}_{{\bf k}}(\tau):&=&i\int^{\tau}_{-\infty}d\tau^{\prime}~\langle\Phi_{\bf B}(v_{-\bf k},v_{\bf k},\tau^{\prime})|\bigg(\frac{\partial}{\partial \tau^{\prime}}\bigg)|\Phi_{\bf B}(v_{-\bf k},v_{\bf k},\tau^{\prime})\rangle,\\
\delta^{\bf Dynamical,B}_{{\bf k}}(\tau):&=&-\int^{\tau}_{-\infty}d\tau^{\prime}~\langle\Phi_{\bf B}(v_{-\bf k},v_{\bf k},\tau^{\prime})|\hat{H}^{\bf B}_{\bf k}(\tau^{\prime})|\Phi_{\bf B}(v_{-\bf k},v_{\bf k},\tau^{\prime})\rangle.~~~~~\eea
From this discussion it is also expected that:
\bea &&\gamma^{\bf PB, NB}_{{\bf k}}(\tau)\neq \gamma^{\bf PB, B}_{{\bf k}}(\tau),
\delta^{\bf Dynamical,NB}_{{\bf k}}(\tau)\neq \delta^{\bf Dynamical,B}_{{\bf k}}(\tau).\eea  
Now before going to the further details in this computation by seeing the mathematical structures of the Hamiltonians written in presence and absence of Bell's inequality violation it is quite justifiable to expect the following structure of the Lewis Riesenfeld invariant quantum operators for both the cases:
\bea && \underline{\rm Without~Bell's~inequality~violation:}~~~~~~~~~~~ \hat{\cal I}^{\bf LR}_{\bf NB,k}(\tau)=\hat{\cal I}^{\bf LR}_{\bf NB,k}(\hat{\Pi}_{v_{-\bf k}}(\tau)\hat{\Pi}_{v_{\bf k}}(\tau),\hat{v}_{-\bf k}(\tau))\hat{v}_{\bf k}(\tau)),~~~~~~\\
&& \underline{\rm With~Bell's~inequality~violation:}~~~~~~~~~~~  \hat{\cal I}^{\bf LR}_{\bf B,k}(\tau)=\hat{\cal I}^{\bf LR}_{\bf B,k}(\hat{\Pi}_{v_{-\bf k}}(\tau)\hat{\Pi}_{v_{\bf k}}(\tau),\hat{v}_{-\bf k}(\tau))\hat{v}_{\bf k}(\tau)).\eea

By following this trick one can construct the following {\it invariant operator} for the general time-dependent cosmological Hamiltonian constructed in the previous section:
\bea && \underline{\rm Without~Bell's~inequality~violation:}\nonumber\\
&& \underline{\rm A. ~ Before~integration~by-parts:}\nonumber\\
&&  \hat{\cal I}^{\bf LR}_{\bf NB,k}(\tau):
=\bigg[\left(\frac{\hat{v}_{-\bf k}(\tau)\hat{v}_{\bf k}(\tau)}{\left({\cal K}^{\bf NB}_{\bf k}(\tau)\right)^2}\right)+\bigg({\cal K}^{\bf NB}_{\bf k}(\tau)\hat{\Pi}_{v_{-\bf k}}(\tau)-\left(\frac{d {\cal K}^{\bf NB}_{\bf k}(\tau)}{d\tau}-\frac{1}{z(\tau)}\frac{d z(\tau)}{d\tau}{\cal K}^{\bf NB}_{\bf k}(\tau)\right)\hat{v}_{-\bf k}(\tau)\bigg)\nonumber\\
&&~~~~~~~~~~~~~~~~~~~~~~~~~~~~~~~~~~~~\times\bigg({\cal K}^{\bf NB}_{\bf k}(\tau)\hat{\Pi}_{v_{\bf k}}(\tau)-\left(\frac{d {\cal K}^{\bf NB}_{\bf k}(\tau)}{d\tau}-\frac{1}{z(\tau)}\frac{d z(\tau)}{d\tau}{\cal K}^{\bf NB}_{\bf k}(\tau)\right)\hat{v}_{\bf k}(\tau)\bigg)\bigg],~~~~~~~\\
&& \underline{\rm B. ~ After~integration~by-parts:}\nonumber\\
&&  \hat{\cal I}^{\bf LR}_{\bf NB,k}(\tau):
=\bigg[\left(\frac{\hat{v}_{-\bf k}(\tau)\hat{v}_{\bf k}(\tau)}{\left({\cal K}^{\bf NB}_{\bf k}(\tau)\right)^2}\right)+\bigg({\cal K}^{\bf NB}_{\bf k}(\tau)\hat{\Pi}_{v_{-\bf k}}(\tau)-\left(\frac{d {\cal K}^{\bf NB}_{\bf k}(\tau)}{d\tau}\right)\hat{v}_{-\bf k}(\tau)\bigg)\nonumber\\
&&~~~~~~~~~~~~~~~~~~~~~~~~~~~~~~\times\bigg({\cal K}^{\bf NB}_{\bf k}(\tau)\hat{\Pi}_{v_{\bf k}}(\tau)-\left(\frac{d {\cal K}^{\bf NB}_{\bf k}(\tau)}{d\tau}\right)\hat{v}_{\bf k}(\tau)\bigg)\bigg],\\ && \underline{\rm With~Bell's~inequality~violation:}\nonumber\\
&& \underline{\rm A. ~ Before~integration~by-parts:}\nonumber\\
&&  \hat{\cal I}^{\bf LR}_{\bf B,k}(\tau):
=\bigg[\left(\frac{\hat{v}_{-\bf k}(\tau)\hat{v}_{\bf k}(\tau)}{\left({\cal K}^{\bf B}_{\bf k}(\tau)\right)^2}\right)+\bigg({\cal K}^{\bf B}_{\bf k}(\tau)\hat{\Pi}_{v_{-\bf k}}(\tau)-\left(\frac{d {\cal K}^{\bf B}_{\bf k}(\tau)}{d\tau}-\frac{1}{z(\tau)}\frac{d z(\tau)}{d\tau}{\cal K}^{\bf B}_{\bf k}(\tau)\right)\hat{v}_{-\bf k}(\tau)\bigg)\nonumber\\
&&~~~~~~~~~~~~~~~~\times\bigg({\cal K}^{\bf B}_{\bf k}(\tau)\hat{\Pi}_{v_{\bf k}}(\tau)-\left(\frac{d {\cal K}^{\bf B}_{\bf k}(\tau)}{d\tau}-\frac{1}{z(\tau)}\frac{d z(\tau)}{d\tau}{\cal K}^{\bf B}_{\bf k}(\tau)\right)\hat{v}_{\bf k}(\tau)\bigg)+\frac{m(\tau)}{2{\cal H}}\frac{1}{z(\tau)}\hat{\Pi}_{v_{-\bf k}}(\tau)\bigg],\\
&& \underline{\rm B. ~ After~integration~by-parts:}\nonumber\\
&&  \hat{\cal I}^{\bf LR}_{\bf B,k}(\tau):
=\bigg[\left(\frac{\hat{v}_{-\bf k}(\tau)\hat{v}_{\bf k}(\tau)}{\left({\cal K}^{\bf B}_{\bf k}(\tau)\right)^2}\right)+\bigg({\cal K}^{\bf B}_{\bf k}(\tau)\hat{\Pi}_{v_{-\bf k}}(\tau)-\left(\frac{d {\cal K}^{\bf B}_{\bf k}(\tau)}{d\tau}\right)\hat{v}_{-\bf k}(\tau)\bigg)\nonumber\\
&&~~~~~~~~~~~~~~~~~~~\times\bigg({\cal K}^{\bf B}_{\bf k}(\tau)\hat{\Pi}_{v_{\bf k}}(\tau)-\left(\frac{d {\cal K}^{\bf B}_{\bf k}(\tau)}{d\tau}\right)\hat{v}_{\bf k}(\tau)\bigg)+\frac{m(\tau)}{2{\cal H}}\frac{1}{z(\tau)}\bigg(\hat{\Pi}_{v_{-\bf k}}(\tau)-\left(\frac{1}{z(\tau)}\frac{dz(\tau)}{d\tau}\right)\hat{v}_{\bf k}(\tau)\bigg)\bigg],~~~~~~~~~~~\eea
where the function ${\cal K}_{\bf k}(\tau)$ representing a {\it c-number} which satisfy the following quantum auxiliary equations which is commonly known the {\it Milne–Pinney equations} for the above mentioned two physical situations: 
\bea && \underline{\rm Without~Bell's~inequality~violation:}\nonumber
\\&&\bigg[\partial^2_{\tau}+\omega^2_{\rm eff}(k,\tau)\bigg]{\cal K}^{\bf NB}_{\bf k}(\tau)={\cal W}^{\bf NB}_{\bf k}(\tau)~~~~~~{\rm where}~~~~~{\cal W}^{\bf NB}_{\bf k}(\tau)=\frac{1}{\left({\cal K}^{\bf NB}_{\bf k}(\tau)\right)^3},\\
&& \underline{\rm With~Bell's~inequality~violation:}\nonumber\\&&\bigg[\partial^2_{\tau}+\Omega^2_{\rm eff}(k,\tau)\bigg]{\cal K}^{\bf B}_{\bf k}(\tau)={\cal W}^{\bf B}_{\bf k}(\tau)~~~~~~~~~~~~{\rm where}~~~~~{\cal W}^{\bf B}_{\bf k}(\tau)=\frac{1}{\left({\cal K}^{\bf B}_{\bf k}(\tau)\right)^3}.
\eea  
Here the effective-conformal time dependent frequency factors $\omega^2_{\rm eff}(k,\tau)$ and $\Omega^2_{\rm eff}(k,\tau)$ describing the non-Bell violating and Bell violating effects are explicitly defined in the earlier part of this paper.  Now it is important to note that,  the mathematical structure of these derived equations look like the equations of forced general conformal time-dependent parametric oscillator within the framework of primordial cosmology.  

Here it is important to note that,  if ($v^{(1),{\bf NB}}_{\bf k}(\tau)$,  $v^{(2),{\bf NB}}_{\bf k}(\tau)$) and ($v^{(1),{\bf B}}_{\bf k}(\tau)$,  $v^{(2),{\bf B}}_{\bf k}(\tau)$) are the two linearly independent solutions of the {\it Mukhanov-Sasaki equations} in absence and presence of Bell's inequality violation within the framework of primordial cosmology then the corresponding {\it Wronskian} for both the cases can be expressed as:
\bea && \underline{\rm Without~Bell's~inequality~violation:}\nonumber\\&&~~~~~~~~~~~~~{\bf W}_{\bf NB}\equiv\begin{vmatrix} v^{(1),{\bf NB}}_{\bf k}(\tau)~~ &~~ v^{(2),{\bf NB}}_{\bf k}(\tau) \\ 
& \\
\partial_{\tau}v^{(1),{\bf NB}}_{\bf k}(\tau)~~ &~~ \partial_{\tau}v^{(2),{\bf NB}}_{\bf k}(\tau) \end{vmatrix}={\cal A}={\rm Constant},\\
&& \underline{\rm With~Bell's~inequality~violation:}\nonumber\\&&~~~~~~~~~~~~~{\bf W}_{\bf B}\equiv\begin{vmatrix} v^{(1),{\bf B}}_{\bf k}(\tau)~~ &~~ v^{(2),{\bf B}}_{\bf k}(\tau) \\ 
& \\
\partial_{\tau}v^{(1),{\bf B}}_{\bf k}(\tau)~~ &~~ \partial_{\tau}v^{(2),{\bf B}}_{\bf k}(\tau) \end{vmatrix}={\cal B}={\rm Constant}.
 \eea
For both the cases the {\it Wronskian} turn out be constants ${\cal A}$ and ${\cal B}$ respectively.  In the above mentioned situations the most general solution of the quantum auxiliary equations or the {\it Milne–Pinney equations} can be expressed as:
 \bea && \underline{\rm Without~Bell's~inequality~violation:}\nonumber\\&&~~~~~~~~~~~~~{\cal K}^{\bf NB}_{\bf k}(\tau):=\sqrt{{\cal A}_1~v^{(1),{\bf NB}}_{\bf k}(\tau)+{\cal A}_2~v^{(2),{\bf NB}}_{\bf k}(\tau)+2{\cal A}_3~v^{(1),{\bf NB}}_{\bf k}(\tau)v^{(2),{\bf NB}}_{\bf k}(\tau)},~~~~~~~~~~~~\\
&& \underline{\rm Without~Bell's~inequality~violation:}\nonumber\\&&~~~~~~~~~~~~~{\cal K}^{\bf B}_{\bf k}(\tau):=\sqrt{{\cal B}_1~v^{(1),{\bf B}}_{\bf k}(\tau)+{\cal B}_2~v^{(2),{\bf B}}_{\bf k}(\tau)+2{\cal B}_3~v^{(1),{\bf B}}_{\bf k}(\tau)v^{(2),{\bf B}}_{\bf k}(\tau)}.~~~~~~~~~~~~ \eea
In terms of these above mentioned solutions of the functions ${\cal K}^{\bf NB}_{\bf k}(\tau)$ and ${\cal K}^{\bf B}_{\bf k}(\tau)$ the structure of the Lewis Riesenfeld invariant quantum operators are completely fixed for the above mentioned physical situations.
Here the constants ${\cal A}_i\forall i=1,2,3$ and ${\cal B}_i\forall i=1,2,3$ are determined by initial conditions,  which must satisfy the following constraint conditions for the above mentioned two physical situations:
 \bea && \underline{\rm Without~Bell's~inequality~violation:}~~~~~~~~~~{\cal A}=\pm \frac{1}{\sqrt{{\cal A}_1{\cal A}_2-{\cal A}^2_3}}={\rm Constant},~~~~~~\\
 && \underline{\rm With~Bell's~inequality~violation:}~~~~~~~~~~~~~~~{\cal B}=\pm \frac{1}{\sqrt{{\cal B}_1{\cal B}_2-{\cal B}^2_3}}={\rm Constant}.~~~~~~
 \eea
 For simplicity we will fix the initial condition as, 
 ${\cal A}_1={\cal C}_1={\cal B}_1, {\cal A}_2={\cal C}_2={\cal B}_2, {\cal A}_3={\cal C}_1{\cal C}_2={\cal B}_3,$ which further implies the following simplified constraint,
${\cal A}=\pm\frac{1}{\sqrt{{\cal C}_1{\cal C}_2}}~\frac{1}{\sqrt{1-{\cal C}_1{\cal C}_2}}={\cal B}.$
 Consequently,  the auxiliary c-function can be further simplified as:
  \bea && \underline{\rm Without~Bell's~inequality~violation:}\nonumber\\&&~~~~~~~~~~~~~{\cal K}^{\bf NB}_{\bf k}(\tau):=\bigg[{\cal C}_1~v^{(1),{\bf NB}}_{\bf k}(\tau)+{\cal C}_2~v^{(2),{\bf NB}}_{\bf k}(\tau)\bigg]=v^{{\bf NB}}_{\bf k}(\tau),~~~~~~~~~~~~\\
&& \underline{\rm Without~Bell's~inequality~violation:}\nonumber\\&&~~~~~~~~~~~~~{\cal K}^{\bf B}_{\bf k}(\tau):=\bigg[{\cal C}_1~v^{(1),{\bf B}}_{\bf k}(\tau)+{\cal C}_2~v^{(2),{\bf B}}_{\bf k}(\tau)\bigg]=v^{{\bf B}}_{\bf k}(\tau).~~~~~~~~~~~~ \eea
 In the present context of discussion,  in the asymptotic limit the two linearly independent solutions of the {\it Mukhanov-Sasaki equations} in absence and presence of Bell's inequality violation within the framework of primordial cosmology can be written as:
 \bea && \underline{\rm Without~Bell's~inequality~violation~(for~massless~case):}\nonumber\\
 &&v^{(1),{\bf NB}}_{\bf k}(\tau):=-\frac{2^{\nu_{\bf S}-\frac{3}{2}}~(-k\tau)^{\frac{3}{2}-\nu_{\bf S}}}{\sqrt{2k}} \biggl|\frac{\Gamma(\nu_{\bf S})}{\Gamma\left(\frac{3}{2}\right)}\biggr|\left(1-\frac{i}{k\tau}\right)\exp\biggl(-i\biggl\{k\tau+\frac{\pi}{2}\left(\nu_{\bf S}+\frac{1}{2}\right)\biggr\}\biggr),~~~~~~~~~~~~~\\
 &&v^{(2),{\bf NB}}_{\bf k}(\tau):=-\frac{2^{\nu_{\bf S}-\frac{3}{2}}~(-k\tau)^{\frac{3}{2}-\nu_{\bf S}}}{\sqrt{2k}} \biggl|\frac{\Gamma(\nu_{\bf S})}{\Gamma\left(\frac{3}{2}\right)}\biggr|\left(1+\frac{i}{k\tau}\right)\exp\biggl(i\biggl\{k\tau+\frac{\pi}{2}\left(\nu_{\bf S}+\frac{1}{2}\right)\biggr\}\biggr).~~~~~~~~~~~~~
 \\
  && \underline{\rm With~Bell's~inequality~violation:}\nonumber\\
  && \underline{\rm A. ~ For ~massive~case:}\nonumber\\
 &&v^{(1),{\bf B}}_{\bf k}(\tau):=(-\tau)^{\frac{3}{2}}~\exp\bigg(-\frac{1}{2}{\cal P}_1\tau\bigg)~\left({\cal P}_1\tau\right)^{{\cal P}_2+\frac{{\cal E}_2}{{\cal P}_1}}\nonumber\\
 &&~~~~~~~~~~~~~~~~~~~\times\bigg(1+\frac{{\cal P}_1{\cal P}_2}{{\cal P}_3}\tau+\frac{\Gamma\left({\cal P}_3\right)}{\Gamma\left({\cal P}_3-{\cal P}_2\right)}\left(-{\cal P}_1\tau\right)^{-{\cal P}_2}+\frac{\Gamma\left({\cal P}_3\right)}{\Gamma\left({\cal P}_2\right)}\left({\cal P}_1\tau\right)^{{\cal P}_2-{\cal P}_3}\exp\left({\cal P}_1\tau\right)\bigg),\\
 &&v^{(2),{\bf B}}_{\bf k}(\tau):=(-\tau)^{\frac{3}{2}}~\exp\bigg(-\frac{1}{2}{\cal P}_1\tau\bigg)~\left({\cal P}_1\tau\right)^{{\cal P}_2+\frac{{\cal E}_2}{{\cal P}_1}}\nonumber\\
 &&~~~~~~~~~~~~~~~~~~~\times\bigg(\frac{\Gamma\left(1-{\cal P}_3\right)}{\Gamma\left(1+{\cal P}_2-{\cal P}_3\right)}\bigg(1+\frac{{\cal P}_1{\cal P}_2}{{\cal P}_3}\tau\bigg)+\frac{\Gamma\left({\cal P}_3-1\right)}{\Gamma\left({\cal P}_2\right)}\left({\cal P}_1\tau\right)^{1-{\cal P}_3}+\left({\cal P}_1\tau\right)^{-{\cal P}_2}\bigg).~~~~~~~~~~~~~\\
 &&\underline{\rm B.~For~partially~massless~case:}\nonumber\\
&&v^{(1),{\bf NB}}_{\bf k}(\tau):=-\frac{2^{\sqrt{\nu^2_{\bf S}-1}-\frac{3}{2}}~(-k\tau)^{\frac{3}{2}-\sqrt{\nu^2_{\bf S}-1}}}{\sqrt{2k}} \biggl|\frac{\Gamma\left(\sqrt{\nu^2_{\bf S}-1}\right)}{\Gamma\left(\frac{3}{2}\right)}\biggr|\left(1-\frac{i}{k\tau}\right)\nonumber\\
&&~~~~~~~~~~~~~~~~~~~~~~~~~~~~~~~~~~~~~~~~~~~\times\exp\biggl(-i\biggl\{k\tau+\frac{\pi}{2}\left(\sqrt{\nu^2_{\bf S}-1}+\frac{1}{2}\right)\biggr\}\biggr),~~~~~~~~~~~~~\\
 &&v^{(2),{\bf NB}}_{\bf k}(\tau):=-\frac{2^{\sqrt{\nu^2_{\bf S}-1}-\frac{3}{2}}~(-k\tau)^{\frac{3}{2}-\sqrt{\nu^2_{\bf S}-1}}}{\sqrt{2k}} \biggl|\frac{\Gamma\left(\sqrt{\nu^2_{\bf S}-1}\right)}{\Gamma\left(\frac{3}{2}\right)}\biggr|\left(1+\frac{i}{k\tau}\right)\nonumber\\
&&~~~~~~~~~~~~~~~~~~~~~~~~~~~~~~~~~~~~~~~~~~~\times\exp\biggl(i\biggl\{k\tau+\frac{\pi}{2}\left(\sqrt{\nu^2_{\bf S}-1}+\frac{1}{2}\right)\biggr\}\biggr).~~~~~~~~~~~~~
\eea
All the representative quantities $\nu_{\bf S}$ and ${\cal P}_i\forall i=1,2,3$ are explicitly defined in the earlier part of the paper.  One more crucial thing we need to point out that,  the linearly independent solutions of the mode functions with Bell's inequality violation are obtained for a given specific conformal time dependent mass profile which we have quoted in the previous section of this paper. 

Our next goal is to determine the eigenfunction of the Lewis Riesenfeld invariant quantum operators in absence and presence of Bell's equality violation which satisfy the following eigenvalue equations respectively:
\bea && \underline{\rm Without~Bell's~inequality~violation:}~~~~~~~~~~~~~~\hat{\cal I}^{\bf LR}_{\bf NB,k}(\tau)\Phi_{{\bf NB}}(v_{-\bf k},v_{\bf k},\tau)=\lambda^{\bf NB}\Phi_{{\bf NB}}(v_{-\bf k},v_{\bf k},\tau),~~~~~~~~~\\
 && \underline{\rm With~Bell's~inequality~violation:} ~~~~~~~~~~~~~~\hat{\cal I}^{\bf LR}_{\bf B,k}(\tau)\Phi_{{\bf B}}(v_{-\bf k},v_{\bf k},\tau)=\lambda^{\bf B}\Phi_{{\bf B}}(v_{-\bf k},v_{\bf k},\tau),\eea
where $\lambda^{\bf NB}$ and $\lambda^{\bf B}$ are the corresponding eigenvalues in which we are particularly interested in the present context.  Now by seeing the previously mentioned mathematical structures of the Lewis Riesenfeld invariant quantum operators it is understandable that finding these eigen values are extremely difficult in the present context as in both of the physical situations the off-diagonal cross term exists in the matrix representation.  Like in the context of quantum mechanics, within the quantum description of primordial cosmology it is also advisable to introduce the following unitary transformations in the eigenfunctions:
\bea && \underline{\rm Without~Bell's~inequality~violation:}\nonumber\\
 &&\Phi_{{\bf NB}}(v_{-\bf k},v_{\bf k},\tau)~~~~~\longrightarrow ~~~~~\widetilde{\Phi_{{\bf NB}}(v_{-\bf k},v_{\bf k},\tau)}:=\hat{\cal U}^{\bf LR}_{\bf NB,k}(\tau)\Phi_{{\bf NB}}(v_{-\bf k},v_{\bf k},\tau),\\
 && \underline{\rm With~Bell's~inequality~violation:}\nonumber\\
 &&\Phi_{{\bf B}}(v_{-\bf k},v_{\bf k},\tau)~~~~~\longrightarrow ~~~~~\widetilde{\Phi_{{\bf B}}(v_{-\bf k},v_{\bf k},\tau)}:=\hat{\cal U}^{\bf LR}_{\bf B,k}(\tau)\Phi_{{\bf B}}(v_{-\bf k},v_{\bf k},\tau).\eea
 Here $\hat{\cal U}^{\bf LR}_{\bf NB,k}(\tau)$ and $\hat{\cal U}^{\bf LR}_{\bf B,k}(\tau)$ are the unitary operators in absence and presence of Bell's inequality violation,  which are given by the following expressions: 
 \bea  && \underline{\rm Without~Bell's~inequality~violation:}\nonumber\\
 && \underline{\rm A. ~ Before~integration~by-parts:}\nonumber\\
 &&\hat{\cal U}^{\bf LR}_{\bf NB,k}(\tau):
 =\exp\bigg(-\frac{i}{{\cal K}^{\bf NB}_{\bf k}(\tau)}\left(\frac{d {\cal K}^{\bf NB}_{\bf k}(\tau)}{d\tau}-\bigg(\frac{1}{z(\tau)}\frac{dz(\tau)}{d\tau}\bigg){\cal K}^{\bf NB}_{\bf k}(\tau)\right)\hat{v}_{-\bf k}(\tau)\hat{v}_{\bf k}(\tau)\bigg),~~~~~~~~~~~\\
 && \underline{\rm B. ~ After~integration~by-parts:}\nonumber\\
 &&\hat{\cal U}^{\bf LR}_{\bf NB,k}(\tau):
 =\exp\bigg(-\frac{i}{{\cal K}^{\bf NB}_{\bf k}(\tau)}\left(\frac{d {\cal K}^{\bf NB}_{\bf k}(\tau)}{d\tau}\right)\hat{v}_{-\bf k}(\tau)\hat{v}_{\bf k}(\tau)\bigg),~~~~~~~~\\
 && \underline{\rm With~Bell's~inequality~violation:}\nonumber\\ 
 && \underline{\rm A. ~ Before~integration~by-parts:}\nonumber\\
 &&\hat{\cal U}^{\bf LR}_{\bf B,k}(\tau):
=\exp\bigg(-i\bigg\{\frac{1}{{\cal K}^{\bf B}_{\bf k}(\tau)}\left(\frac{d {\cal K}^{\bf B}_{\bf k}(\tau)}{d\tau}-\bigg(\frac{1}{z(\tau)}\frac{dz(\tau)}{d\tau}\bigg){\cal K}^{\bf B}_{\bf k}(\tau)\right)\hat{v}_{-\bf k}(\tau)\hat{v}_{\bf k}(\tau)-\frac{m(\tau)}{2{\cal H}}\frac{1}{z(\tau)}\hat{\Pi}_{v_{-\bf k}}(\tau)\bigg\}\bigg),~~~~~~~~~~~~~\\
 && \underline{\rm B. ~ After~integration~by-parts:}\nonumber\\
 &&\hat{\cal U}^{\bf LR}_{\bf B,k}(\tau):
 =\exp\bigg(-i\bigg\{\frac{1}{{\cal K}^{\bf B}_{\bf k}(\tau)}\left(\frac{d {\cal K}^{\bf B}_{\bf k}(\tau)}{d\tau}\right)\hat{v}_{-\bf k}(\tau)\hat{v}_{\bf k}(\tau)-\frac{m(\tau)}{2{\cal H}}\frac{1}{z(\tau)}\bigg(\hat{\Pi}_{v_{-\bf k}}(\tau)-\left(\frac{1}{z(\tau)}\frac{dz(\tau)}{d\tau}\right)\hat{v}_{\bf k}(\tau)\bigg)\bigg\}\bigg),~~~~~~~~\eea
 
Consequently,  the diagonal simplified form of the {\it invariant operator} for the general conformal time-dependent cosmological Hamiltonian is given by:
\bea  && \underline{\rm Without~Bell's~inequality~violation:}\nonumber\\ &&\hat{\cal I}^{\bf LR}_{\bf NB,k}(\tau)~~~\longrightarrow~~~\widetilde{\hat{\cal I}^{\bf LR}_{\bf NB,k}(\tau)}=\hat{\cal U}^{\bf LR}_{\bf NB,k}(\tau)~\hat{\cal I}^{\bf LR}_{\bf NB,k}(\tau)~\left(\hat{\cal U}^{\bf LR}_{\bf NB,k}(\tau)\right)^{\dagger}=\bigg(\left(\frac{\hat{v}_{-\bf k}(\tau)\hat{v}_{\bf k}(\tau)}{\left({\cal K}^{\bf NB}_{\bf k}(\tau)\right)^2}\right)-\left({\cal K}^{\bf NB}_{\bf k}(\tau)\right)^2\hat{\partial}_{v_{-\bf k}}\hat{\partial}_{v_{\bf k}}\bigg),\\
&& \underline{\rm With~Bell's~inequality~violation:}\nonumber\\ &&\hat{\cal I}^{\bf LR}_{\bf B,k}(\tau)~~~\longrightarrow~~~\widetilde{\hat{\cal I}^{\bf LR}_{\bf B,k}(\tau)}=\hat{\cal U}^{\bf LR}_{\bf B,k}(\tau)~\hat{\cal I}^{\bf LR}_{\bf B,k}(\tau)~\left(\hat{\cal U}^{\bf LR}_{\bf B,k}(\tau)\right)^{\dagger}=\bigg(\left(\frac{\hat{v}_{-\bf k}(\tau)\hat{v}_{\bf k}(\tau)}{\left({\cal K}^{\bf B}_{\bf k}(\tau)\right)^2}\right)-\left({\cal K}^{\bf B}_{\bf k}(\tau)\right)^2\hat{\partial}_{v_{-\bf k}}\hat{\partial}_{v_{\bf k}}\bigg).~~~~~~~~~~~~~~~~~~~~~~~\eea

Consequently,  we get the following solutions of the eigenfunctions of the {\it invariant operator} after performing the unitary transformation:
\bea && \underline{\rm Without~Bell's~inequality~violation:}\widetilde{\Phi_{{\bf NB}}(v_{-\bf k},v_{\bf k},\tau)}:
 =\frac{1}{\sqrt{{\cal K}^{\bf NB}_{\bf k}(\tau)}}~{\cal P}\bigg(\frac{1}{2}\left(\lambda^{\bf NB}-1\right),\frac{\sqrt{2v_{-\bf k}(\tau)v_{\bf k}(\tau)}}{{\cal K}^{\bf NB}_{\bf k}(\tau)}\bigg),~~~~~~~~~\\ 
 && \underline{\rm With~Bell's~inequality~violation:}\widetilde{\Phi_{{\bf B}}(v_{-\bf k},v_{\bf k},\tau)}:
 =\frac{1}{\sqrt{{\cal K}^{\bf B}_{\bf k}(\tau)}}~{\cal P}\bigg(\frac{1}{2}\left(\lambda^{\bf B}-1\right),\frac{\sqrt{2v_{-\bf k}(\tau)v_{\bf k}(\tau)}}{{\cal K}^{\bf B}_{\bf k}(\tau)}\bigg).\eea
 Here it is important to note that $v^{\bf Sub}_{\bf k}(\tau)$ without and with having Bell's inequality violation can be expressed as:
 \bea && \underline{\rm Without~Bell's~inequality~violation:}~~~v^{\bf NB}_{\bf k}(\tau)=\footnotesize\bigg[{\cal C}_1~v^{(1),{\bf NB}}_{\bf k}(\tau)+{\cal C}_2~v^{(2),{\bf NB}}_{\bf k}(\tau)\bigg],~~~~~~~\\
 && \underline{\rm With~Bell's~inequality~violation:}~~~v^{\bf B}_{\bf k}(\tau)=\footnotesize\bigg[{\cal C}_1~v^{(1),{\bf B}}_{\bf k}(\tau)+{\cal C}_2~v^{(2),{\bf B}}_{\bf k}(\tau)\bigg],~~~~~~\eea
 where ${\cal C}_1$ and ${\cal C}_2$ are fixed by proper choice of the quantum initial conditions as mentioned earlier.  
 
 Further performing the inverse unitary transformation the eigenfunctions without and with having Bell's inequality violation can be computed as:
 \bea  && \underline{\rm Without~Bell's~inequality~violation:}\nonumber\\
 && \underline{\rm A. ~ Before~integration~by-parts:}\nonumber\\
 &&\footnotesize\Phi_{{\bf NB}}(v_{-\bf k},v_{\bf k},\tau):=\frac{1}{\sqrt{{\cal K}^{\bf NB}_{\bf k}(\tau)}}~\exp\bigg(\frac{i}{{\cal K}^{\bf NB}_{\bf k}(\tau)}\left(\frac{d {\cal K}^{\bf NB}_{\bf k}(\tau)}{d\tau}-\bigg(\frac{1}{z(\tau)}\frac{dz(\tau)}{d\tau}\bigg){\cal K}^{\bf NB}_{\bf k}(\tau)\right)|\hat{v}_{\bf k}(\tau)|^2\bigg)\nonumber\\
 &&~~~~~~~~~~~~~~~~~~~~~~~~~~~~~~~
 ~~~~~~~~~~~~~~~~~~~~~~~\footnotesize\times{\cal P}\bigg(\frac{1}{2}\left(\lambda^{\bf NB}-1\right),\frac{\sqrt{2|v_{\bf k}(\tau)|^2}}{{\cal K}^{\bf NB}_{\bf k}(\tau)}\bigg),~~~~~~~~\\
 && \underline{\rm B. ~ After~integration~by-parts:}\nonumber\\
 &&\footnotesize\Phi_{{\bf NB}}(v_{-\bf k},v_{\bf k},\tau):=\frac{1}{\sqrt{{\cal K}^{\bf B}_{\bf k}(\tau)}}~\exp\bigg(\frac{i}{{\cal K}^{\bf NB}_{\bf k}(\tau)}\left(\frac{d {\cal K}^{\bf NB}_{\bf k}(\tau)}{d\tau}\right)|\hat{v}_{\bf k}(\tau)|^2\bigg){\cal P}\bigg(\frac{1}{2}\left(\lambda^{\bf NB}-1\right),\frac{\sqrt{2|v_{\bf k}(\tau)|^2}}{{\cal K}^{\bf NB}_{\bf k}(\tau)}\bigg),\nonumber\\
 &&~~~~~~~~\\
 && \underline{\rm With~Bell's~inequality~violation:}\nonumber\\
 && \underline{\rm A. ~ Before~integration~by-parts:}\nonumber\\
 &&\footnotesize\Phi_{{\bf B}}(v_{-\bf k},v_{\bf k},\tau):=\frac{1}{\sqrt{{\cal K}^{\bf B}_{\bf k}(\tau)}}~\exp\bigg(i\bigg\{\frac{1}{{\cal K}^{\bf B}_{\bf k}(\tau)}\left(\frac{d {\cal K}^{\bf B}_{\bf k}(\tau)}{d\tau}-\bigg(\frac{1}{z(\tau)}\frac{dz(\tau)}{d\tau}\bigg){\cal K}^{\bf B}_{\bf k}(\tau)\right)|\hat{v}_{\bf k}(\tau)|^2\nonumber\\
 &&\footnotesize~~~~~~~~~~~~~~~~~~~~~~~~~~~~~
-\frac{m(\tau)}{2{\cal H}}\frac{1}{z(\tau)}\hat{\Pi}_{v_{-\bf k}}(\tau)\bigg\}\bigg)\times{\cal P}\bigg(\frac{1}{2}\left(\lambda^{\bf B}-1\right),\frac{\sqrt{2|v_{\bf k}(\tau)|^2}}{{\cal K}^{\bf B}_{\bf k}(\tau)}\bigg),~~~~~~~~\\
 && \underline{\rm B. ~ After~integration~by-parts:}\nonumber\\
 &&\footnotesize\Phi_{{\bf B}}(v_{-\bf k},v_{\bf k},\tau):=\frac{1}{\sqrt{{\cal K}^{\bf B}_{\bf k}(\tau)}}~\exp\bigg(i\bigg\{\frac{1}{{\cal K}^{\bf B}_{\bf k}(\tau)}\left(\frac{d {\cal K}^{\bf B}_{\bf k}(\tau)}{d\tau}\right)|\hat{v}_{\bf k}(\tau)|^2\nonumber\\
 &&~~~~~~
 \footnotesize~~~-\frac{m(\tau)}{2{\cal H}}\frac{1}{z(\tau)}\bigg(\hat{\Pi}_{v_{-\bf k}}(\tau)-\left(\frac{1}{z(\tau)}\frac{dz(\tau)}{d\tau}\right)\hat{v}_{\bf k}(\tau)\bigg)\bigg\}\bigg)\times{\cal P}\bigg(\frac{1}{2}\left(\lambda^{\bf B}-1\right),\frac{\sqrt{2|v_{\bf k}(\tau)|^2}}{{\cal K}^{\bf B}_{\bf k}(\tau)}\bigg),~~~~~~~~\eea
Here ${\cal P}(\nu,x)$ represents the {\it Parabolic Cylinder function} in this context.  In the next section we will explicitly compute the phase factor associated with wave function for both the physical situations without and with having Bell's inequality violation in primordial cosmology.  Before doing that explicit computation we will do the same analysis that we did in general here for the three specified region of interest,  which are sub-Hubble region,  super-Hubble region and from sub-Hubble to super-Hubble region horizon crossing point.  This will help us to understand the applicability of the obtained result in these three region of interested within the framework of primordial cosmology.  
\section{Geometric phase in Cosmology in different comoving scales}
\label{GP} 
In this section we explicitly compute the expression for the {\it Lewis Riesenfeld phase} factor within the framework of the quantum theory of primordial cosmology without and with having Bell's inequality violation.  To serve this purpose need to compute the following two quantities: 
\bea && \footnotesize\underline{\rm Without~Bell's~inequality~violation:}~~~~~\gamma^{\bf LR,NB}_{\bf k}(\tau):=-\lambda^{\bf NB}\int^{\tau}_{-\infty} d\tau^{\prime}~\frac{1}{\left({\cal K}^{\bf NB}_{\bf k}(\tau^{\prime})\right)^2},\\
&&\footnotesize \underline{\rm With~Bell's~inequality~violation:}~~~~~\gamma^{\bf LR,B}_{\bf k}(\tau):=-\lambda^{\bf B}\int^{\tau}_{-\infty} d\tau^{\prime}~\frac{1}{\left({\cal K}^{\bf B}_{\bf k}(\tau^{\prime})\right)^2}. \eea
 Now in the previous section we have explicitly obtained the solutions of the quantum auxiliary or the {\it Milne–Pinney equation} ${\cal K}^{\bf NB}_{\bf k}(\tau)$ and ${\cal K}^{\bf B}_{\bf k}(\tau)$ for a general conformal time and momentum dependent frequency,  which is valid for all cosmologically region of interest.  After substituting the derived expression for these auxiliary c-functions we get the following result for the {\it Lewis Riesenfeld phase} factor within the framework of the quantum theory of primordial cosmology:
 \bea && \underline{\rm Without~Bell's~inequality~violation:}\nonumber\\
 &&\footnotesize~~~~~\gamma^{\bf LR,NB}_{\bf k}(\tau):=-\lambda^{\bf NB}\int^{\tau}_{-\infty} d\tau^{\prime}~\frac{1}{\bigg[{\cal A}_1~v^{(1),{\bf NB}}_{\bf k}(\tau^{\prime})+{\cal A}_2~v^{(2),{\bf NB}}_{\bf k}(\tau^{\prime})+2{\cal A}_3~v^{(1),{\bf NB}}_{\bf k}(\tau^{\prime})v^{(2),{\bf NB}}_{\bf k}(\tau^{\prime})\bigg]},~~~~~~~~~\\
&& \underline{\rm With~Bell's~inequality~violation:}\nonumber\\
 &&\footnotesize~~~~~\gamma^{\bf LR,B}_{\bf k}(\tau):=-\lambda^{\bf B}\int^{\tau}_{-\infty} d\tau^{\prime}~\frac{1}{\bigg[{\cal B}_1~v^{(1),{\bf B}}_{\bf k}(\tau^{\prime})+{\cal B}_2~v^{(2),{\bf B}}_{\bf k}(\tau^{\prime})+2{\cal B}_3~v^{(1),{\bf B}}_{\bf k}(\tau^{\prime})v^{(2),{\bf B}}_{\bf k}(\tau^{\prime})\bigg]}.\eea
 Next for the further simplification and to make the consistency with the quantum initial conditions we choose the constants $\left({\cal A}_i,{\cal B}_i\right)\forall i=1,2,3$ as, 
${\cal A}_1={\cal C}_1={\cal B}_1,~~{\cal A}_2={\cal C}_2={\cal B}_2,~~{\cal A}_3={\cal C}_1{\cal C}_2={\cal B}_3.$ 
 Consequently,  we get the following simplified expression for the {\it Lewis Riesenfeld phase} factor within the framework of the quantum theory of primordial cosmology:
 \bea && \footnotesize\underline{\rm Without~Bell's~inequality~violation:}~~~~~\gamma^{\bf LR,NB}_{\bf k}(\tau):=-\lambda^{\bf NB}\int^{\tau}_{-\infty} d\tau^{\prime}~\frac{1}{\left(v^{\bf NB}_{\bf k}(\tau^{\prime})\right)^2},\\
&&\footnotesize \underline{\rm With~Bell's~inequality~violation:}~~~~~\gamma^{\bf LR,B}_{\bf k}(\tau):=-\lambda^{\bf B}\int^{\tau}_{-\infty} d\tau^{\prime}~\frac{1}{\left(v^{\bf B}_{\bf k}(\tau^{\prime})\right)^2}.\eea
 Now we will explicitly compute the expression for the dynamical phase factor in terms of the conformal time coordinate associated with the quantum theory of primordial cosmology without and with having the effect of Bell's inequality violation.  By doing explicit calculation in terms of the Hamiltonian derived in the earlier part of this paper we find the following result for the dynamical phase factor,  which is given by:
 \bea && \underline{\rm Without~Bell's~inequality~violation:}\nonumber\\
 &&\footnotesize~~~~~\delta^{\bf Dynamical,NB}_{{\bf k}}(\tau):=\lambda^{\bf NB}\int^{\tau}_{-\infty} d\tau^{\prime}~\bigg[\left({\cal K}^{\bf NB}_{\bf k}(\tau^{\prime})\right)^2\omega^2(k,\tau^{\prime})+\left(\frac{d {\cal K}^{\bf NB}_{\bf k}(\tau^{\prime})}{d\tau^{\prime}}\right)^2\bigg],\\
&& \underline{\rm With~Bell's~inequality~violation:}\nonumber\\
 &&\footnotesize~~~~~\delta^{\bf Dynamical,B}_{{\bf k}}(\tau):=\lambda^{\bf B}\int^{\tau}_{-\infty} d\tau^{\prime}~\bigg[\left({\cal K}^{\bf B}_{\bf k}(\tau^{\prime})\right)^2\Omega^2(k,\tau^{\prime})+\left(\frac{d {\cal K}^{\bf B}_{\bf k}(\tau^{\prime})}{d\tau^{\prime}}\right)^2\bigg]. \eea
 Further substituting the expression for the auxiliary c-functions which is consistent with the quantum initial condition we get:
  \bea && \underline{\rm Without~Bell's~inequality~violation:}\nonumber\\
 &&\footnotesize~~~~~\delta^{\bf Dynamical,NB}_{{\bf k}}(\tau):=\lambda^{\bf NB}\int^{\tau}_{-\infty} d\tau^{\prime}~\bigg[\left(v^{\bf NB}_{\bf k}(\tau^{\prime})\right)^2\omega^2(k,\tau^{\prime})+\left(\frac{d v^{\bf NB}_{\bf k}(\tau^{\prime})}{d\tau^{\prime}}\right)^2\bigg],\\
&& \underline{\rm With~Bell's~inequality~violation:}\nonumber\\
 &&\footnotesize~~~~~\delta^{\bf Dynamical,B}_{{\bf k}}(\tau):=\lambda^{\bf B}\int^{\tau}_{-\infty} d\tau^{\prime}~\bigg[\left(v^{\bf B}_{\bf k}(\tau^{\prime})\right)^2\Omega^2(k,\tau^{\prime})+\left(\frac{d v^{\bf B}_{\bf k}(\tau^{\prime})}{d\tau^{\prime}}\right)^2\bigg]. \eea
 Next using the {\it Lewis Riesenfeld phase} factor and the dynamical phase factor we get the following expression for the {\it Pancharatnam-Berry phase} within the framework of quantum description of primordial cosmology:
 \bea && \underline{\rm Without~Bell's~inequality~violation:}\nonumber\\
 &&\footnotesize\gamma^{\bf PB, NB}_{{\bf k}}(\tau):= -\lambda^{\bf NB}\int^{\tau}_{-\infty} d\tau^{\prime}~\bigg[{\cal K}^{\bf NB}_{\bf k}(\tau^{\prime})\bigg(\frac{d^2{\cal K}^{\bf NB}_{\bf k}(\tau^{\prime})}{d\tau^{\prime 2}}\bigg)-\bigg(\frac{d {\cal K}^{\bf NB}_{\bf k}(\tau^{\prime})}{d\tau^{\prime}}\bigg)^2\nonumber\\
 &&\footnotesize~~~~~~~~~~~~~~~~~~~~~~~~~~~~~~~~~~~~~~~~~~~~~~~~~~-\left( {\cal K}^{\bf NB}_{\bf k}(\tau^{\prime})\right)^2\bigg(\frac{1}{z(\tau^{\prime})}\frac{d^2z(\tau^{\prime})}{d\tau^{\prime 2}}-\bigg(\frac{1}{z(\tau^{\prime})}\frac{dz(\tau^{\prime})}{d\tau^{\prime }}\bigg)^2\bigg)\bigg],~~~~~~~~~~\\
&& \underline{\rm With~Bell's~inequality~violation:}\nonumber\\
 &&\footnotesize\gamma^{\bf PB, B}_{{\bf k}}(\tau):= -\lambda^{\bf B}\int^{\tau}_{-\infty} d\tau^{\prime}~\bigg[{\cal K}^{\bf B}_{\bf k}(\tau^{\prime})\bigg(\frac{d^2{\cal K}^{\bf B}_{\bf k}(\tau^{\prime})}{d\tau^{\prime 2}}\bigg)-\bigg(\frac{d {\cal K}^{\bf B}_{\bf k}(\tau^{\prime})}{d\tau^{\prime}}\bigg)^2\nonumber\\
 &&\footnotesize~~~~~~~~~~~~~~~~~~~~~~~~~~~~~~~~~~~~~~~~~~~~~~~~~~-\left( {\cal K}^{\bf B}_{\bf k}(\tau^{\prime})\right)^2\bigg(\frac{1}{z(\tau^{\prime})}\frac{d^2z(\tau^{\prime})}{d\tau^{\prime 2}}-\bigg(\frac{1}{z(\tau^{\prime})}\frac{dz(\tau^{\prime})}{d\tau^{\prime }}\bigg)^2\bigg)\bigg]. \eea 
 Here for both the physical situation for the simplification purpose we have explicitly used the quantum auxiliary equation or the {\it Milne Pinney equation} satisfied by the the auxiliary c-functions.  For the further simplification in the slowly conformal time varying regime one can introduce the following perturbative solution of the {\it Milne Pinney equation} which is given by:
 \bea && \underline{\rm Without~Bell's~inequality~violation:}\nonumber\\
 &&\footnotesize~~~~~{\cal K}^{\bf NB}_{\bf k}(\tau^{\prime}):=\underbrace{{\cal K}^{\bf NB,(0)}_{\bf k}(\tau^{\prime})}_{\rm Leading~contribution}+\underbrace{\sum^{\infty}_{n=0}\Delta^{n}{\cal K}^{\bf NB,(n)}_{\bf k}(\tau^{\prime})}_{\rm Perturbative~contributions}~~~~~~~{\rm where}~~~~~\Delta\ll 1,~~~~~~\\
 && \underline{\rm With~Bell's~inequality~violation:}\nonumber\\
 &&\footnotesize~~~~~{\cal K}^{\bf B}_{\bf k}(\tau^{\prime}):=\underbrace{{\cal K}^{\bf B,(0)}_{\bf k}(\tau^{\prime})}_{\rm Leading~contribution}+\underbrace{\sum^{\infty}_{n=0}\Delta^{n}{\cal K}^{\bf B,(n)}_{\bf k}(\tau^{\prime})}_{\rm Perturbative~contributions}~~~~~~~{\rm where}~~~~~\Delta\ll 1.~~~~~~\eea
 In this context $\Delta$ is the purterbative adiabatic slowly varying parameter which helps us to write the solution of {\it Milne Pinney equation} in an order by order expansion in a perturbative series in the slowly conformal time varying regime.  Also it is important to note that,  the leading order term in the above perturbative expansion is given by the following expression:
 \bea && \underline{\rm Without~Bell's~inequality~violation:}\nonumber\\
 &&\footnotesize~~~~~{\cal K}^{\bf NB,(0)}_{\bf k}(\tau^{\prime}):=\frac{1}{\bigg(\omega_0(k,\tau^{\prime})\bigg)^{\displaystyle 1/4}}~~~~{\rm where}~~~~\omega_0(k,\tau^{\prime}):=\sqrt{k^2-\bigg(\frac{1}{z(\tau^{\prime})}\frac{dz(\tau^{\prime})}{d\tau^{\prime }}\bigg)^2},~~~~~~~~~~~\\
 && \underline{\rm With~Bell's~inequality~violation:}\nonumber\\
 &&\footnotesize~~~~~{\cal K}^{\bf B,(0)}_{\bf k}(\tau^{\prime}):=\frac{1}{\bigg(\Omega_0(k,\tau^{\prime})\bigg)^{\displaystyle 1/4}}~~{\rm where}~~\Omega_0(k,\tau^{\prime}):=\sqrt{k^2+\frac{1}{\tau^2}\frac{m^2(\tau^{\prime})}{{\cal H}^2}-\bigg(\frac{1}{z(\tau^{\prime})}\frac{dz(\tau^{\prime})}{d\tau^{\prime }}\bigg)^2}.~~~~~~~~~\eea 
 
 Consequently we get the following structure of the {\it Milne Pinney equations} after substitution of these perturbative solutions:
  \bea && \underline{\rm Without~Bell's~inequality~violation:}\nonumber\\
 &&\footnotesize\Delta^2{\cal K}^{\bf NB,(0)}_{\bf k}(\tau^{\prime})\frac{d^2{\cal K}^{\bf NB,(0)}_{\bf k}(\tau^{\prime})}{d\tau^{\prime 2}}\nonumber\\
 &&\footnotesize~+\left({\cal K}^{\bf NB,(0)}_{\bf k}(\tau^{\prime})\right)^2\bigg[1+2{\cal K}^{\bf NB,(1)}_{\bf k}(\tau^{\prime})\Delta{\cal K}^{\bf NB,(0)}_{\bf k}(\tau^{\prime})\nonumber\\
 &&\footnotesize~+\Delta^2{\cal K}^{\bf NB,(1)}_{\bf k}(\tau^{\prime})+2{\cal K}^{\bf NB,(2)}_{\bf k}(\tau^{\prime})\Delta^2{\cal K}^{\bf NB,(0)}_{\bf k}(\tau^{\prime})\bigg]\overbrace{\bigg(\omega^2_0(k,\tau^{\prime})-\bigg(\frac{1}{z(\tau^{\prime})}\frac{d^2z(\tau^{\prime})}{d\tau^{\prime 2}}-\bigg(\frac{1}{z(\tau^{\prime})}\frac{dz(\tau^{\prime})}{d\tau^{\prime }}\bigg)^2\bigg)\bigg)}^{\displaystyle \equiv \omega^2(k,\tau^{\prime})}\nonumber\\
 &&\footnotesize~=\frac{1}{\bigg[\left({\cal K}^{\bf NB,(0)}_{\bf k}(\tau^{\prime})\right)^2+2{\cal K}^{\bf NB,(1)}_{\bf k}(\tau^{\prime})\Delta{\cal K}^{\bf NB,(0)}_{\bf k}(\tau^{\prime})+\Delta^2{\cal K}^{\bf NB,(1)}_{\bf k}(\tau^{\prime})+2{\cal K}^{\bf NB,(2)}_{\bf k}(\tau^{\prime})\Delta^2{\cal K}^{\bf NB,(0)}_{\bf k}(\tau^{\prime})\bigg]}+{\cal O}(\Delta^3)+\cdots,~~~~~~\\
 && \underline{\rm With~Bell's~inequality~violation:}\nonumber\\
 &&\footnotesize\Delta^2{\cal K}^{\bf B,(0)}_{\bf k}(\tau^{\prime})\frac{d^2{\cal K}^{\bf NB,(0)}_{\bf k}(\tau^{\prime})}{d\tau^{\prime 2}}\nonumber\\
 &&\footnotesize~+\left({\cal K}^{\bf B,(0)}_{\bf k}(\tau^{\prime})\right)^2\bigg[1+2{\cal K}^{\bf NB,(1)}_{\bf k}(\tau^{\prime})\Delta{\cal K}^{\bf B,(0)}_{\bf k}(\tau^{\prime})+\Delta^2{\cal K}^{\bf B,(1)}_{\bf k}(\tau^{\prime})\nonumber\\
 &&\footnotesize~+2{\cal K}^{\bf B,(2)}_{\bf k}(\tau^{\prime})\Delta^2{\cal K}^{\bf B,(0)}_{\bf k}(\tau^{\prime})\bigg]\overbrace{\bigg(\Omega^2_0(k,\tau^{\prime})+\frac{1}{\tau^{\prime 2}}\frac{m^2(\tau^{\prime})}{{\cal H}^2}-\bigg(\frac{1}{z(\tau^{\prime})}\frac{d^2z(\tau^{\prime})}{d\tau^{\prime 2}}-\bigg(\frac{1}{z(\tau^{\prime})}\frac{dz(\tau^{\prime})}{d\tau^{\prime }}\bigg)^2\bigg)\bigg)}^{\displaystyle \equiv \Omega^2(k,\tau^{\prime})}\nonumber\eea\bea
 &&\footnotesize~=\frac{1}{\bigg[\left({\cal K}^{\bf B,(0)}_{\bf k}(\tau^{\prime})\right)^2+2{\cal K}^{\bf B,(1)}_{\bf k}(\tau^{\prime})\Delta{\cal K}^{\bf B,(0)}_{\bf k}(\tau^{\prime})+\Delta^2{\cal K}^{\bf B,(1)}_{\bf k}(\tau^{\prime})+2{\cal K}^{\bf B,(2)}_{\bf k}(\tau^{\prime})\Delta^2{\cal K}^{\bf B,(0)}_{\bf k}(\tau^{\prime})\bigg]}+{\cal O}(\Delta^3)+\cdots.~~~~~~\eea
 Using the above mentioned details one can recast the {\it Pancharatnam-Berry phase} for the two frameworks of the quantum description of primordial cosmology in the following simplified fashion:
 \bea && \underline{\rm Without~Bell's~inequality~violation:}\nonumber\\
 &&\gamma^{\bf PB, NB}_{{\bf k}}(\tau):=-\lambda^{\bf NB}\int^{\tau}_{-\infty} d\tau^{\prime}~\frac{1}{\omega_0(k,\tau^{\prime})}~\bigg[\frac{1}{z(\tau^{\prime})}\frac{d^2z(\tau^{\prime})}{d\tau^{\prime 2}}-\bigg(\frac{1}{z(\tau^{\prime})}\frac{dz(\tau^{\prime})}{d\tau^{\prime }}\bigg)^2\bigg],~~~~~~~~~~\\
&& \underline{\rm With~Bell's~inequality~violation:}\nonumber\\
 &&\gamma^{\bf PB, B}_{{\bf k}}(\tau):=-\lambda^{\bf B}\int^{\tau}_{-\infty} d\tau^{\prime}~\frac{1}{\Omega_0(k,\tau^{\prime})}~\bigg[\frac{1}{z(\tau^{\prime})}\frac{d^2z(\tau^{\prime})}{d\tau^{\prime 2}}-\bigg(\frac{1}{z(\tau^{\prime})}\frac{dz(\tau^{\prime})}{d\tau^{\prime }}\bigg)^2\bigg]. \eea 
  Here in both the above mentioned expressions the common conformal time dependent factor appearing in the parenthesis bracket can be further simplified in terms of the slowly varying conformal time dependent parameters as:
\bea \bigg[\frac{1}{z(\tau^{\prime})}\frac{d^2z(\tau^{\prime})}{d\tau^{\prime 2}}- \left(\frac{1}{z(\tau^{\prime})}\frac{dz(\tau^{\prime})}{d\tau^{\prime}}\right)^2\bigg]
&\approx&\frac{1}{\tau^{\prime 2}}\bigg(1+5\epsilon(\tau^{\prime})-\eta(\tau^{\prime})\bigg)+{\cal O}(\epsilon^2(\tau^{\prime}),\eta^2(\tau^{\prime})).~~~~~~~\eea
Following the same logic one can further simplify the quantities appearing in the denominator of the above expressions in the slowly time varying limiting situation as:
 \bea && \underline{\rm Without~Bell's~inequality~violation:}\omega_0(k,\tau^{\prime})\approx\sqrt{k^2-\frac{1}{\tau^{\prime 2}}\bigg[1+2\bigg(2\epsilon(\tau^{\prime})-\eta(\tau^{\prime})\bigg)\bigg]},~~~~~~~~~~~\\
 && \underline{\rm With~Bell's~inequality~violation:}\Omega_0(k,\tau^{\prime})\approx \sqrt{k^2+\frac{1}{\tau^{\prime 2}}\bigg\{\frac{m^2(\tau^{\prime})}{{\cal H}^2}-\bigg[1+2\bigg(2\epsilon(\tau^{\prime})-\eta(\tau^{\prime})\bigg)\bigg]\bigg\}}.~~~~~~~~~\eea
Consequently,  in the slowly time varying limit we get the following simplified expression for the {\it Pancharatnam-Berry phase} for the two physical situations discussed in this paper:
 \bea && \underline{\rm Without~Bell's~inequality~violation:}\nonumber\\
 &&\footnotesize\gamma^{\bf PB, NB}_{{\bf k}}(\tau)\approx\lambda^{\bf NB}\int^{\displaystyle -\frac{1}{k}\bigg[1+2\bigg(2\epsilon(\tau)-\eta(\tau)\bigg)\bigg]}_{\tau} d\tau^{\prime}~\frac{\displaystyle \frac{1}{\tau^{\prime 2}}\bigg(1+5\epsilon(\tau^{\prime})-\eta(\tau^{\prime})\bigg)}{\displaystyle \sqrt{k^2-\frac{1}{\tau^{\prime 2}}\bigg[1+2\bigg(2\epsilon(\tau^{\prime})-\eta(\tau^{\prime})\bigg)\bigg]}}\nonumber\\
 &&\footnotesize~~~~~~~=\frac{\footnotesize\displaystyle \lambda^{\bf NB}\bigg(1+5\epsilon(\tau)-\eta(\tau)\bigg)}{\displaystyle\sqrt{\bigg[1+2\bigg(2\epsilon(\tau)-\eta(\tau)\bigg)\bigg]}}\Bigg[{\rm tan}^{-1}\bigg(\sqrt{2\bigg(2\epsilon(\tau)-\eta(\tau)\bigg)}\bigg)-{\rm tan}^{-1}\left(\frac{\displaystyle \tau~\sqrt{k^2-\frac{1}{\tau^{2}}\bigg[1+2\bigg(2\epsilon(\tau)-\eta(\tau)\bigg)\bigg]}}{\displaystyle\sqrt{\bigg[1+2\bigg(2\epsilon(\tau)-\eta(\tau)\bigg)\bigg]}}\right)\Bigg],~~~~~~~~~~\\
&& \underline{\rm With~Bell's~inequality~violation:}\nonumber\\
 &&\gamma^{\bf PB, B}_{{\bf k}}(\tau)\approx\lambda^{\bf B}\int^{\displaystyle -\frac{1}{k}\bigg\{\bigg[1+2\bigg(2\epsilon(\tau)-\eta(\tau)\bigg)\bigg]-\frac{m^2(\tau)}{{\cal H}^2}\bigg\}}_{\tau} d\tau^{\prime}~\frac{\displaystyle \frac{1}{\tau^{\prime 2}}\bigg(1+5\epsilon(\tau^{\prime})-\eta(\tau^{\prime})\bigg)}{\displaystyle \sqrt{k^2+\frac{1}{\tau^{\prime 2}}\bigg\{\frac{m^2(\tau^{\prime})}{{\cal H}^2}-\bigg[1+2\bigg(2\epsilon(\tau^{\prime})-\eta(\tau^{\prime})\bigg)\bigg]\bigg\}}},\nonumber\\
 &&~~~~~~~~~~~~~=\frac{\displaystyle \lambda^{\bf B}\bigg(1+5\epsilon(\tau)-\eta(\tau)\bigg)}{\displaystyle\sqrt{\bigg\{\bigg[1+2\bigg(2\epsilon(\tau)-\eta(\tau)\bigg)\bigg]-\frac{m^2(\tau)}{{\cal H}^2}\bigg\}}}\Bigg[{\rm tan}^{-1}\bigg(\sqrt{2\bigg(2\epsilon(\tau)-\eta(\tau)\bigg)-\frac{m^2(\tau)}{{\cal H}^2}}\bigg)\nonumber\\
 &&~~~~~~~~~~~~~~~~~~~~~~~-{\rm tan}^{-1}\left(\frac{\displaystyle \tau~\sqrt{k^2-\frac{1}{\tau^{2}}\bigg\{\bigg[1+2\bigg(2\epsilon(\tau)-\eta(\tau)\bigg)\bigg]-\frac{m^2(\tau)}{{\cal H}^2}\bigg\}}}{\displaystyle\sqrt{\bigg\{\bigg[1+2\bigg(2\epsilon(\tau)-\eta(\tau)\bigg)\bigg]-\frac{m^2(\tau)}{{\cal H}^2}\bigg\}}}\right)\Bigg].\nonumber\\
 && \eea
 where it is important to note that the upper limits of the integrations appearing in the expression for the {\it Pancharatnam-Berry phase} are fixed in the conformal time scale from the poles of the above mentioned results.           

\section{Observational consequences of Cosmological Geometric Phase}
\label{OC}
\subsection{Connecting Pancharatnam-Berry phase with cosmological observables}
In this section our prime objective is to make connection between the computed {\it Pancharatnam-Berry phase} from the quantum description of primordial cosmology without and with having Bell's inequality violation with the cosmological observation to address the very relevant question that {\it whether or not one can able to detect this in cosmological observation?} To address this question first of all one need to connect our computed results for the {\it Pancharatnam-Berry phase} with the perturbation theory originated cosmologically relevant quantities which are probed in various observations related to early universe cosmology.  These cosmologically relevant quantities related to the scalar and tensor perturbations in the Planckian unit ($M_p=1$) are given by \cite{Akrami:2018odb}:
\bea &&{\rm Amplitude ~of ~the~power~spectrum:}~~~~~~P_{\zeta}(k)=\left(\frac{k^3}{2\pi^2}~\frac{|v_{\bf k}(\tau)|^2}{z^2(\tau)}\right),\\
&&{\rm Spectral~tilt/Spectral~index:}~~~~~~~~~~~~~~~n_{\zeta}(k)=\left(\frac{d\ln P_{\zeta}(k)}{d\ln k}\right)=1+2\eta(\tau)-4\epsilon(\tau),~~~~~~~~\\
&&{\rm Tensor-to-scalar~ratio:}~~~~~~~~~~~~~~~~~~r(k)=16\epsilon(\tau). \eea 
Here one crucial point we have to mention that,  the above mentioned cosmological consistency relations are written by assuming the well known Bunch Davies vacuum state as the initial condition by fixing ${\cal C}_1=1$ and ${\cal C}_2=0$ in the solution of the scalar modes which we have obtained by solving the {\it Mukhanov Sasaki equation}.  The prime reason for this is because the observational probes only look into the Bunch Davies vacuum state to compare the outcomes from different models \cite{Akrami:2018odb}.  If this is not fully convincing then one can interpret this in a different way.  Once the observational probes give numerical constraints on the above mentioned observations via cosmological parameter estimation using likelihood analysis,  the corresponding contributions in the above mentioned observables for other non Bunch Davies initial conditions are adjusted in such a way that it confronts well with the outcome of the observation.  In presence of non-Bunch Davies initial it is expected to have the non-trivial contributions from both ${\cal C}_1$ and ${\cal C}_2$ as appearing in the solution of scalar modes.  The determining relation for the power-spectrum in terms of the scalar modes will be exactly same for the non-Bunch Davies states,  but the changes will be reflected in the expression once we substitute the expression for the modes that we have computed for in absence and in presence of Bell's inequality violation in this paper.  Once can explicitly show that the overall co-efficient,  which is usually identified as the amplitude of the scalar power spectrum,  will exactly turn out be the same as obtained from the Bunch Davies quantum initial condition for both Bell violating and non Bell violating cases.  The changes will appear for both the cases in the additional multiplicative factor which are the functions of both ${\cal C}_1$ and ${\cal C}_2$ and additionally the comoving scale $k$ and the conformal time $\tau$.  Since the contribution from this multiplicative factor is different for different choices of non Bunch Davies vacuum states,  it is further expected to have changes in the other observables,  which are the scalar spectral tilt/index and in the tensor-to-scalar ratio.  For the non Bunch Davies initial condition for both the observables we get the contribution from the usual Bunch Davies part and then some additive sub-leading correction terms which turns out to be small once we impose the constraints from observation.  Though the overall structure becomes same.  So from this discussion it appears to that,  at the level of determining and constrain the {\it Pancharatnam-Berry phase} in absence and in presence of Bell's inequality violation,  it is not possible to distinguish the contributions from different quantum initial conditions. 

In the slowly time varying limit we get the following simplified expression for the {\it Pancharatnam-Berry phase} for the two physical situations:
 \bea && \underline{\rm Without~Bell's~inequality~violation:}\nonumber\\
 &&\gamma^{\bf PB, NB}_{{\bf k}}(\tau)\approx\frac{\displaystyle \lambda^{\bf NB}\bigg(1+\frac{3}{16}r(k)-\frac{(1-n_{\zeta}(k))}{2}\bigg)}{\displaystyle\sqrt{2-n_{\zeta}(k)}}{\rm tan}^{-1}\left(\frac{\displaystyle \sqrt{1-n_{\zeta}(k)}-\tau~\sqrt{k^2-\frac{1}{\tau^{2}}\bigg[2-n_{\zeta}(k)\bigg]}}{\displaystyle1+\sqrt{1-n_{\zeta}(k)}\sqrt{2-n_{\zeta}(k)}}\right),\\
&& \underline{\rm With~Bell's~inequality~violation:}\nonumber\\
 &&\footnotesize\gamma^{\bf PB, B}_{{\bf k}}(\tau)\approx\frac{\displaystyle \lambda^{\bf B}\bigg(1+\frac{3}{16}r(k)-\frac{(1-n_{\zeta}(k))}{2}\bigg)}{\displaystyle\sqrt{\bigg\{2-n_{\zeta}(k)-\frac{m^2(\tau)}{{\cal H}^2}\bigg\}}}{\rm tan}^{-1}\left(\frac{\displaystyle \sqrt{1-n_{\zeta}(k)-\frac{m^2(\tau)}{{\cal H}^2}}-\tau~\sqrt{k^2-\frac{1}{\tau^{2}}\bigg\{\bigg[2-n_{\zeta}(k)\bigg]-\frac{m^2(\tau)}{{\cal H}^2}\bigg\}}}{\displaystyle 1+\sqrt{1-n_{\zeta}(k)-\frac{m^2(\tau)}{{\cal H}^2}}\sqrt{2-n_{\zeta}(k)-\frac{m^2(\tau)}{{\cal H}^2}}}\right). ~~~~~\eea 
 For the partially massless case we have to fix $m/{\cal H}\sim 1$ in the massive case result in the region,  sub Hubble super Hubble region and at the horizon crossing point.
\begin{figure}[htb!]
	\centering
	\subfigure[Normalized~Pancharatnam~Berry~phase~
	vs~scalar~spectral~tilt~without~Bell's~inequality~violation~in~sub-Hubble~region.]{
		\includegraphics[width=7cm,height=6.5cm] {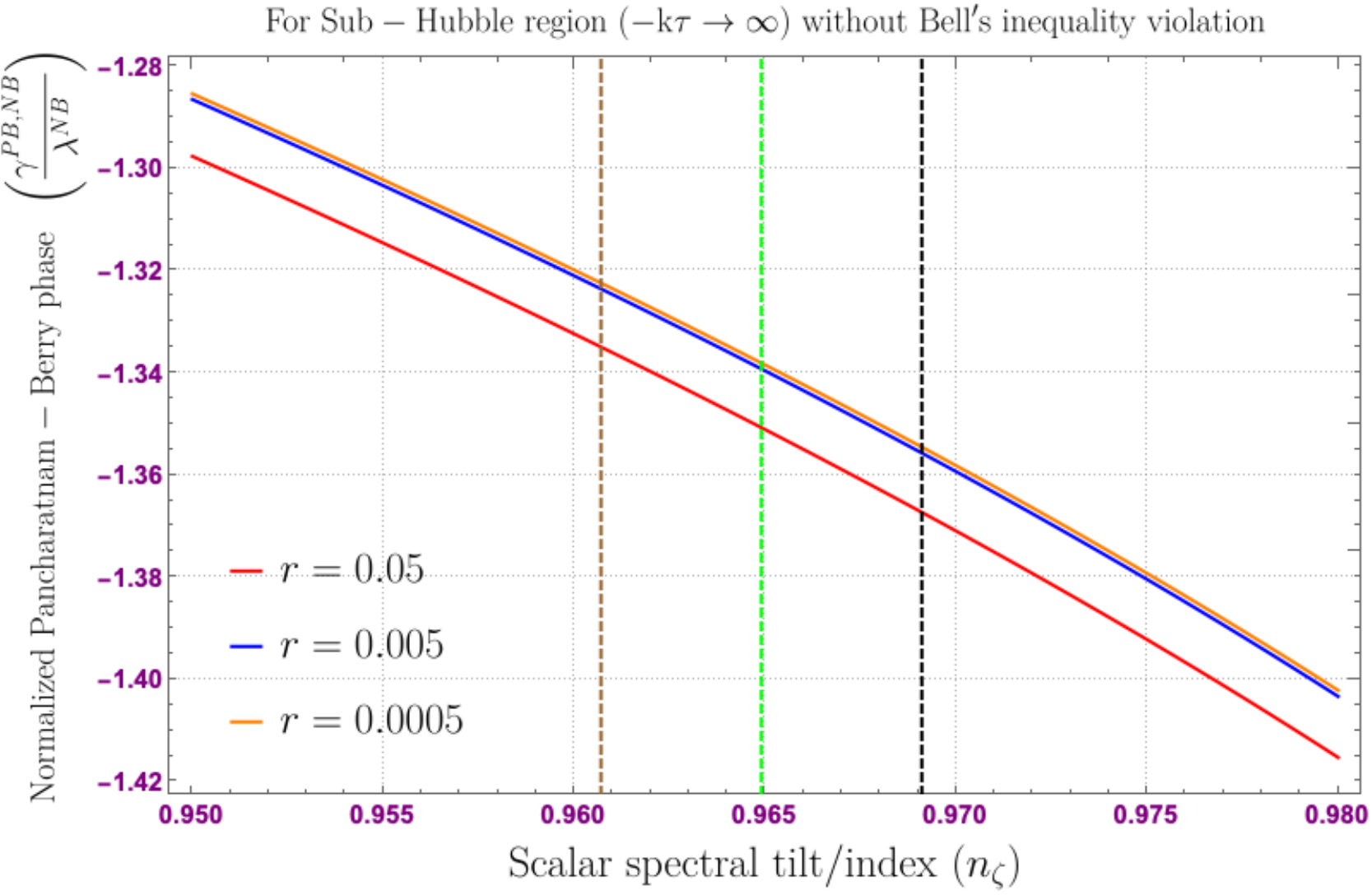}
	}
	\subfigure[Normalized~Pancharatnam~Berry~phase~
	vs~scalar~spectral~tilt~without~Bell's~inequality~violation~in~super-Hubble~region.]{
		\includegraphics[width=7cm,height=6.5cm] {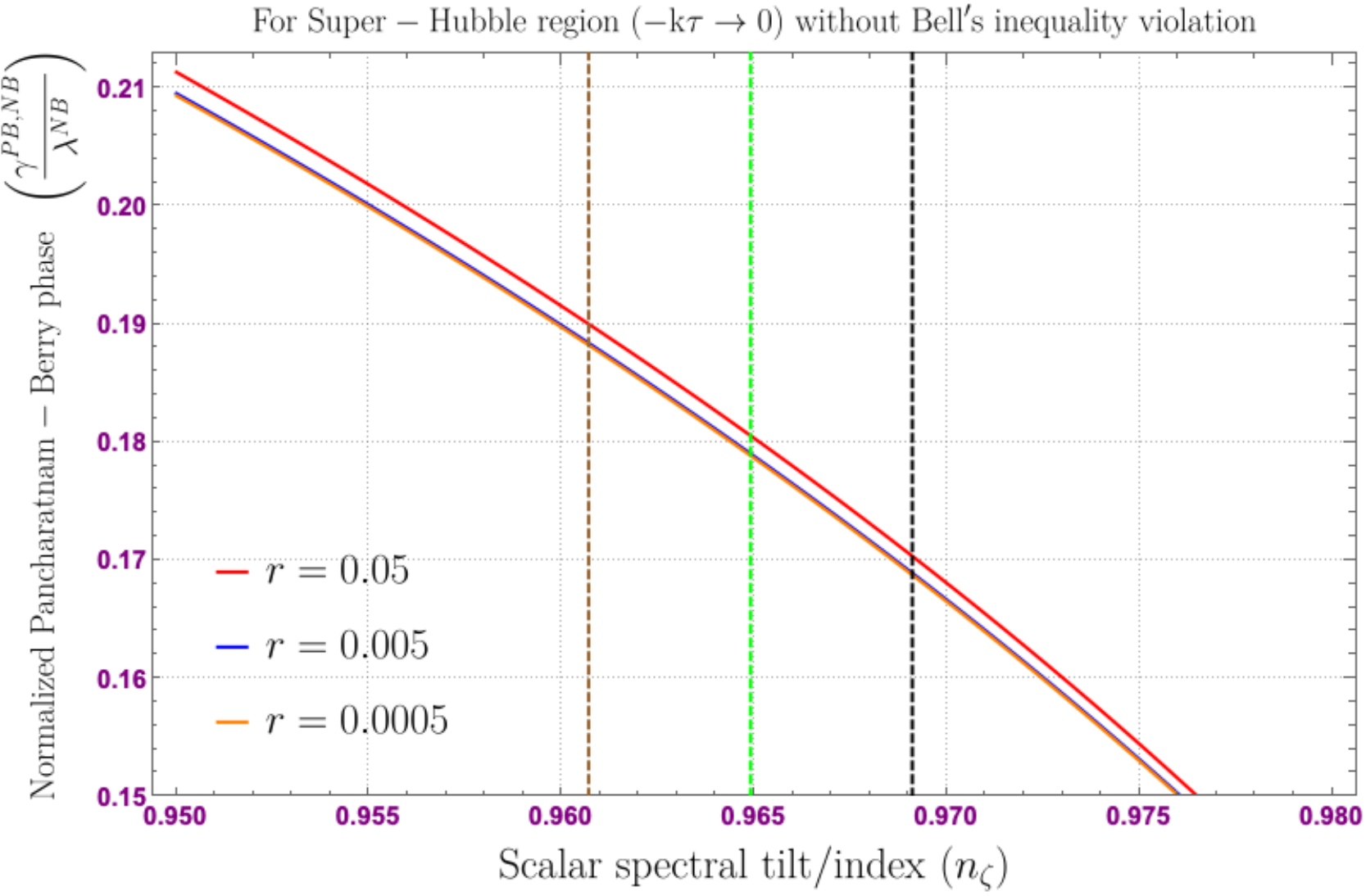}
	}
	\caption{Behavior of the Normalized Pancharatnam Berry phase with scalar spectral tilt without having Bell's inequality violation in sub and super Hubble regime.}
	\label{fig:1}
\end{figure}  
\begin{figure}[htb!]
	\centering
	\subfigure[Real~ part~ of~ the~ Normalized~Pancharatnam~Berry~phase~vs~scalar~spectral~tilt~without~Bell's~inequality~violation~at~horizon~exit~point]{
		\includegraphics[width=7cm,height=6.5cm] {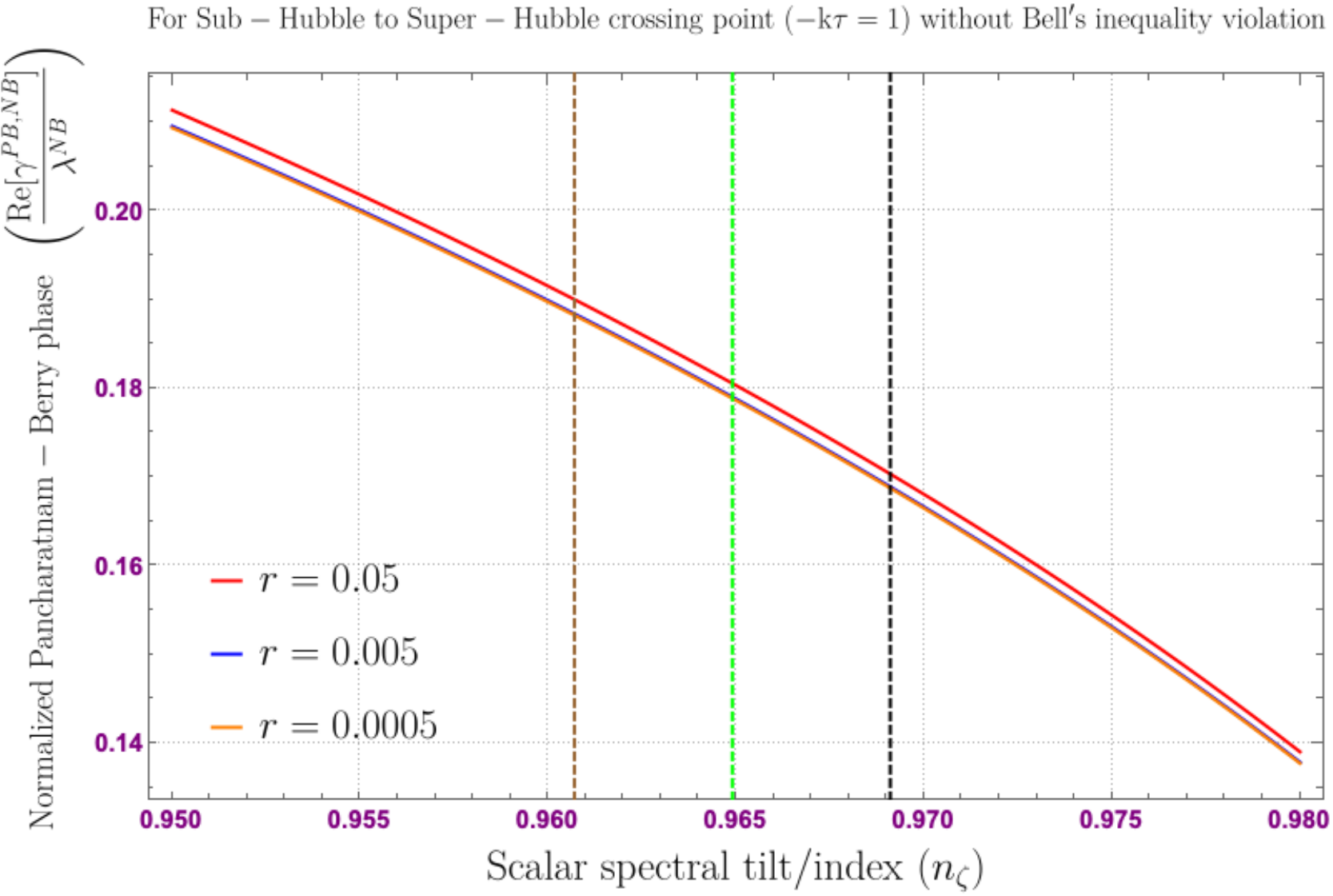}
	}
	\subfigure[Imaginary~ part~ of~ the~ Normalized~Pancharatnam~Berry~phase~vs~
scalar~spectral~tilt~without~Bell's~inequality~violation~at~horizon~
exit~point.]{
		\includegraphics[width=7cm,height=6.5cm] {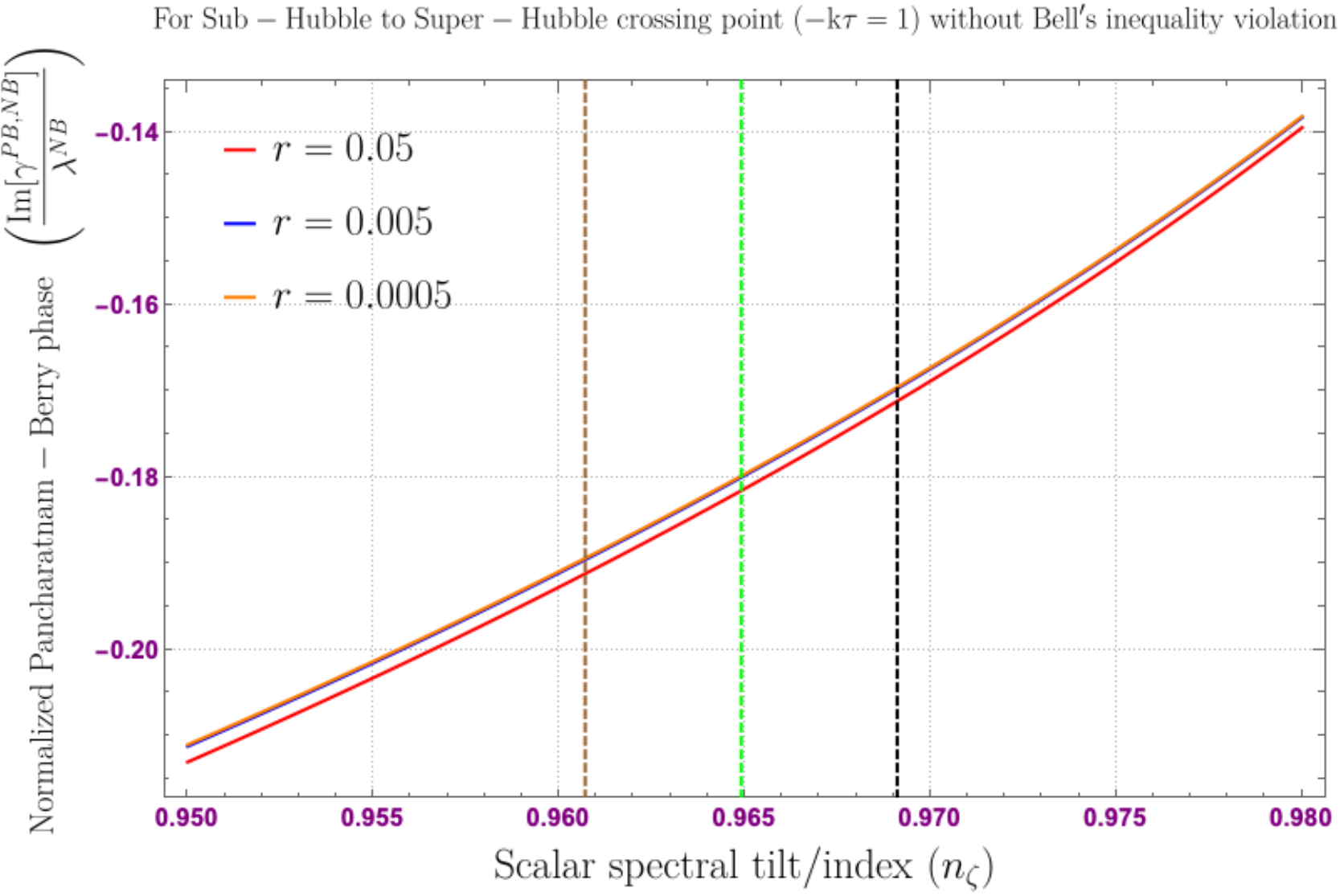}
	}
	\caption{Behavior of the complex Normalized Pancharatnam Berry phase with scalar spectral tilt without having Bell's inequality violation at horizon exit.}
	\label{fig:2}
\end{figure} 

\begin{figure}[htb!]
	\centering
	{
		\includegraphics[width=10cm,height=6.5cm] {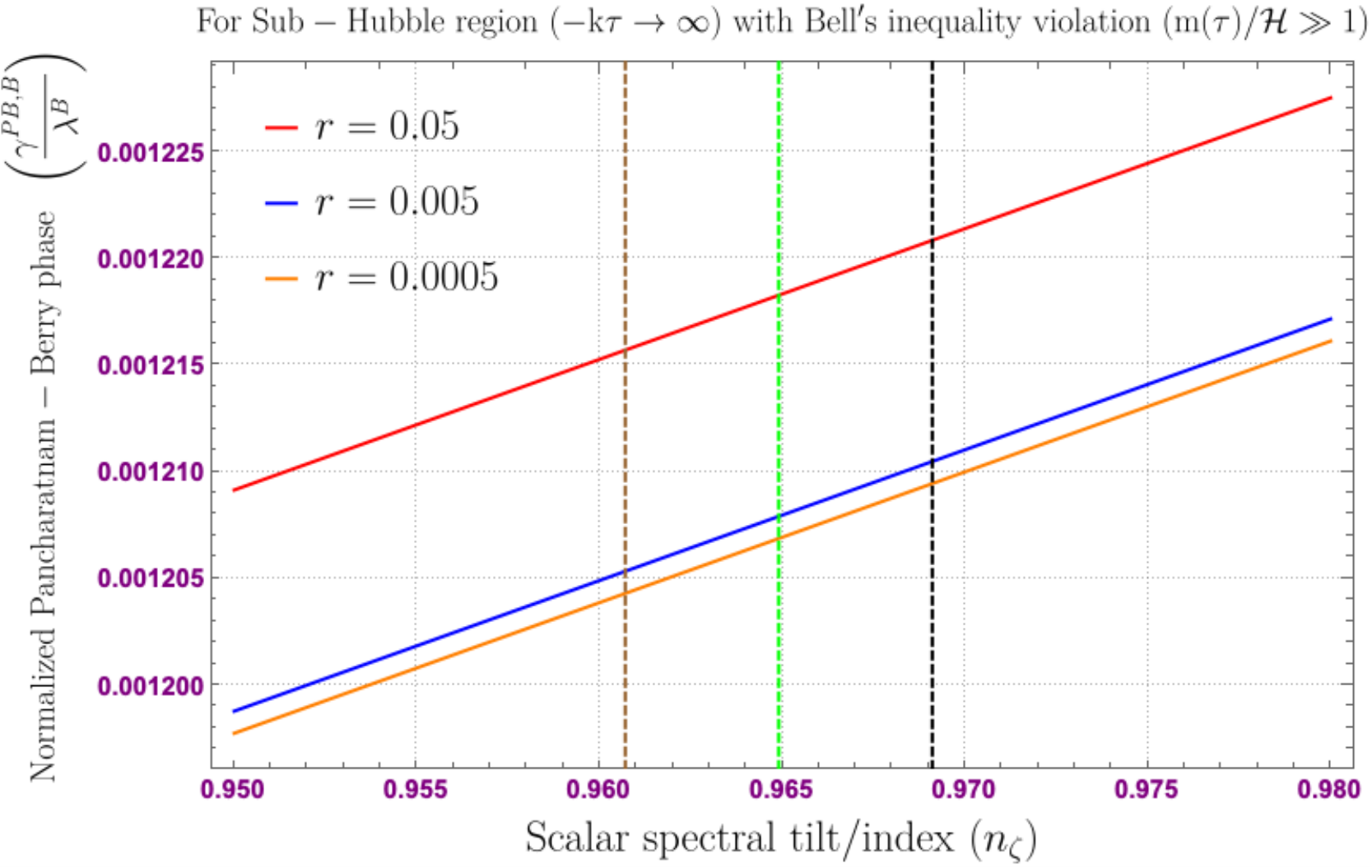}
	}
	\caption{Behavior of the Normalized Pancharatnam Berry phase with scalar spectral tilt with having Bell's inequality violation in sub-Hubble region for massive field.}
	\label{fig:3}
\end{figure} 

\begin{figure}[htb!]
	\centering
	\subfigure[Real~ part~ of~ the~ Normalized~Pancharatnam~Berry~phase~vs~scalar~spectral~tilt~with
	~Bell's~inequality~violation~in~the~sub-Hubble~region.]{
		\includegraphics[width=7cm,height=6.5cm] {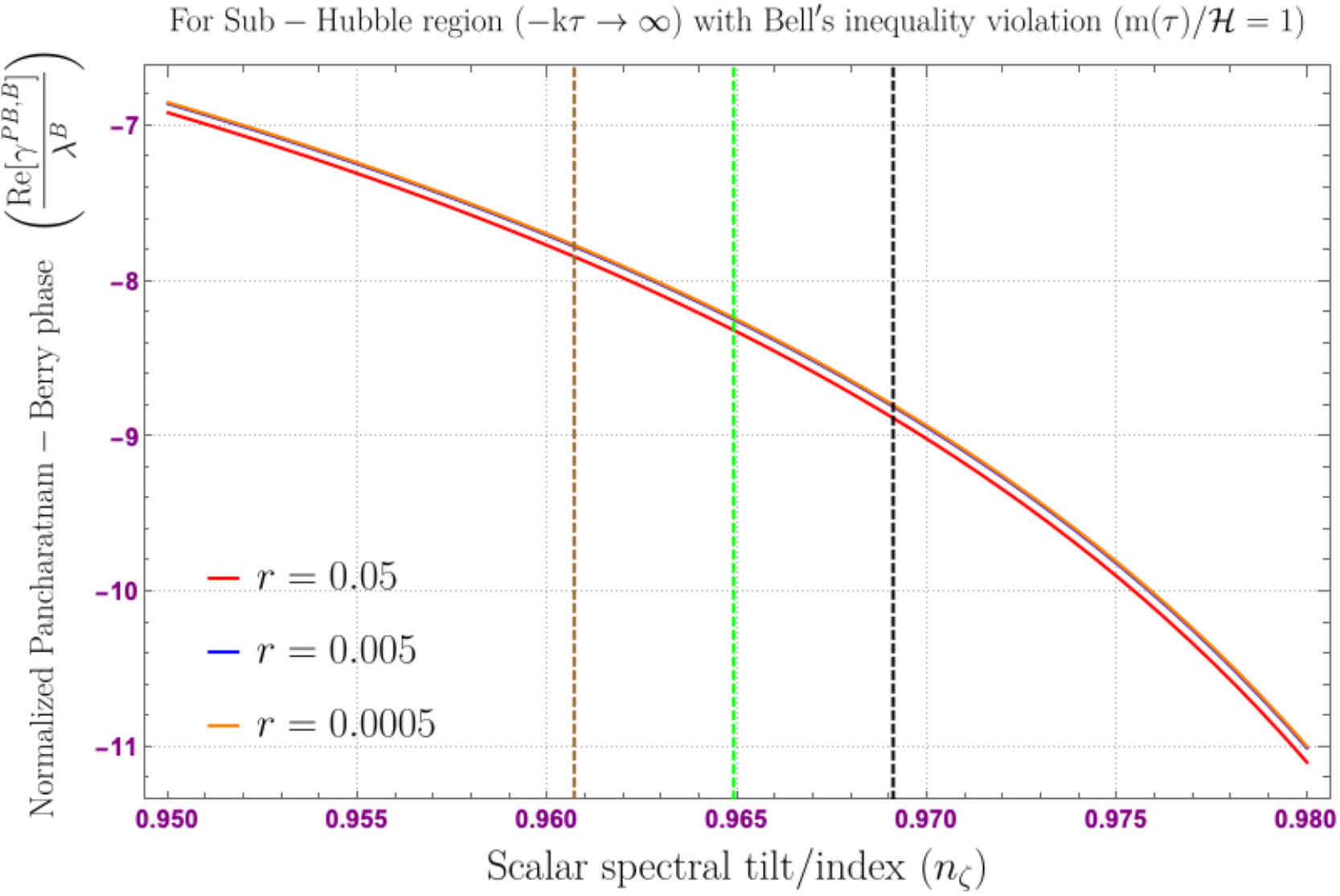}
	}
	\subfigure[Imaginary~ part~ of~ the~ Normalized~Pancharatnam~Berry~phase~vs~scalar~spectral~tilt~without~Bell's~inequality~violation~in the sub-Hubble region.]{
		\includegraphics[width=7cm,height=6.5cm] {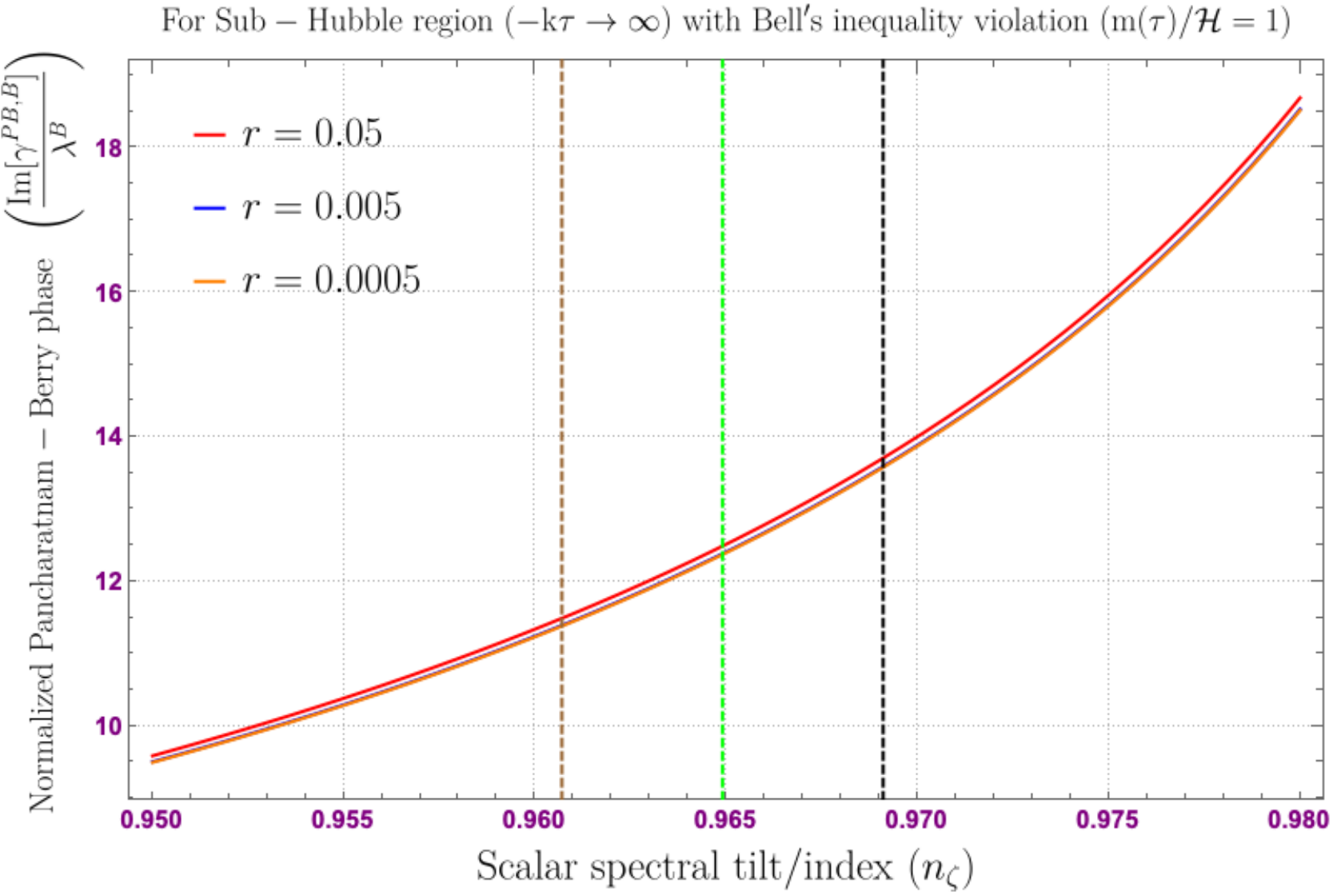}
	}
	\caption{Behavior of the complex Normalized Pancharatnam Berry phase with scalar spectral tilt with having Bell's inequality violation in sub-Hubble region for partially massless field.}
	\label{fig:4}
\end{figure} 

\begin{figure}[htb!]
	\centering
	\subfigure[Real~ part~ of~ the~ Normalized~Pancharatnam~Berry~phase~vs~scalar~spectral~tilt~with
	~Bell's~inequality~violation~in~the~super-Hubble~region.]{
		\includegraphics[width=7cm,height=6.5cm] {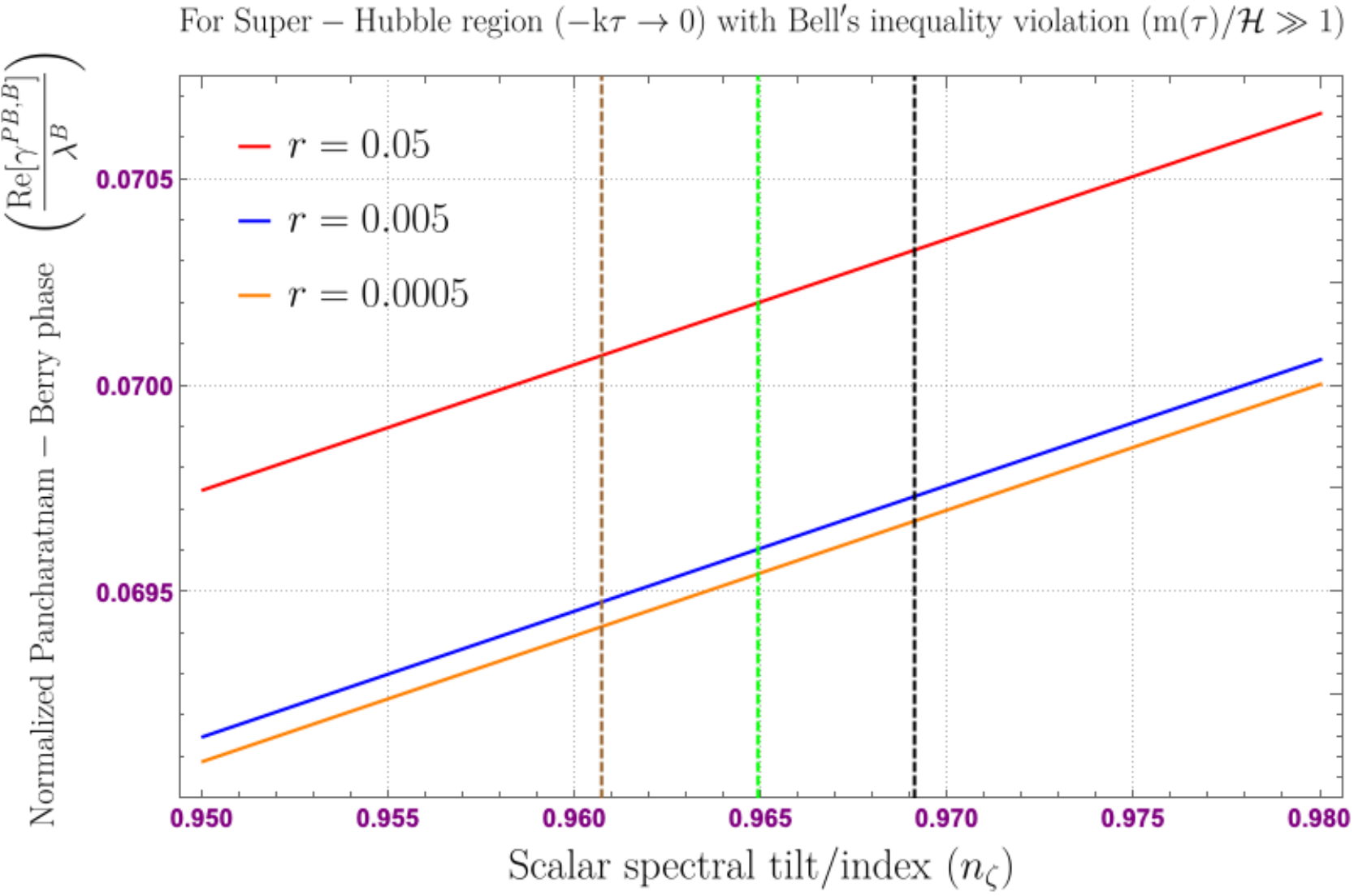}
	}
	\subfigure[Imaginary~ part~ of~ the~ Normalized~Pancharatnam~Berry~phase~vs~scalar~spectral~tilt~with
	~Bell's~inequality~violation~in~the~super-Hubble~region.]{
		\includegraphics[width=7cm,height=6.5cm] {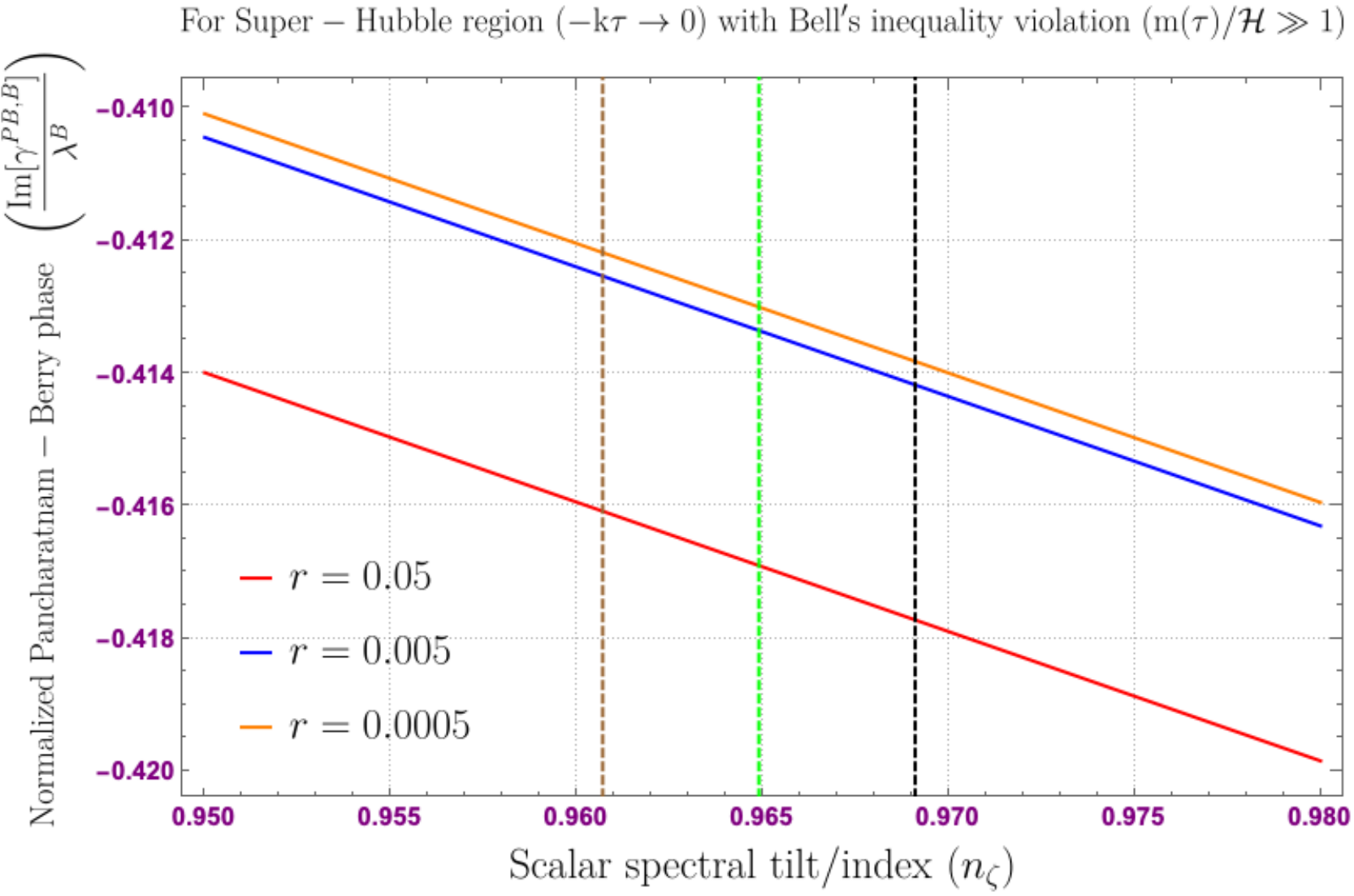}
	}
	\caption{Behavior of the complex Normalized Pancharatnam Berry phase with scalar spectral tilt with having Bell's inequality violation in sub-Hubble region for massive field.}
	\label{fig:5}
\end{figure} 

\begin{figure}[htb!]
	\centering
		\includegraphics[width=10cm,height=6.5cm] {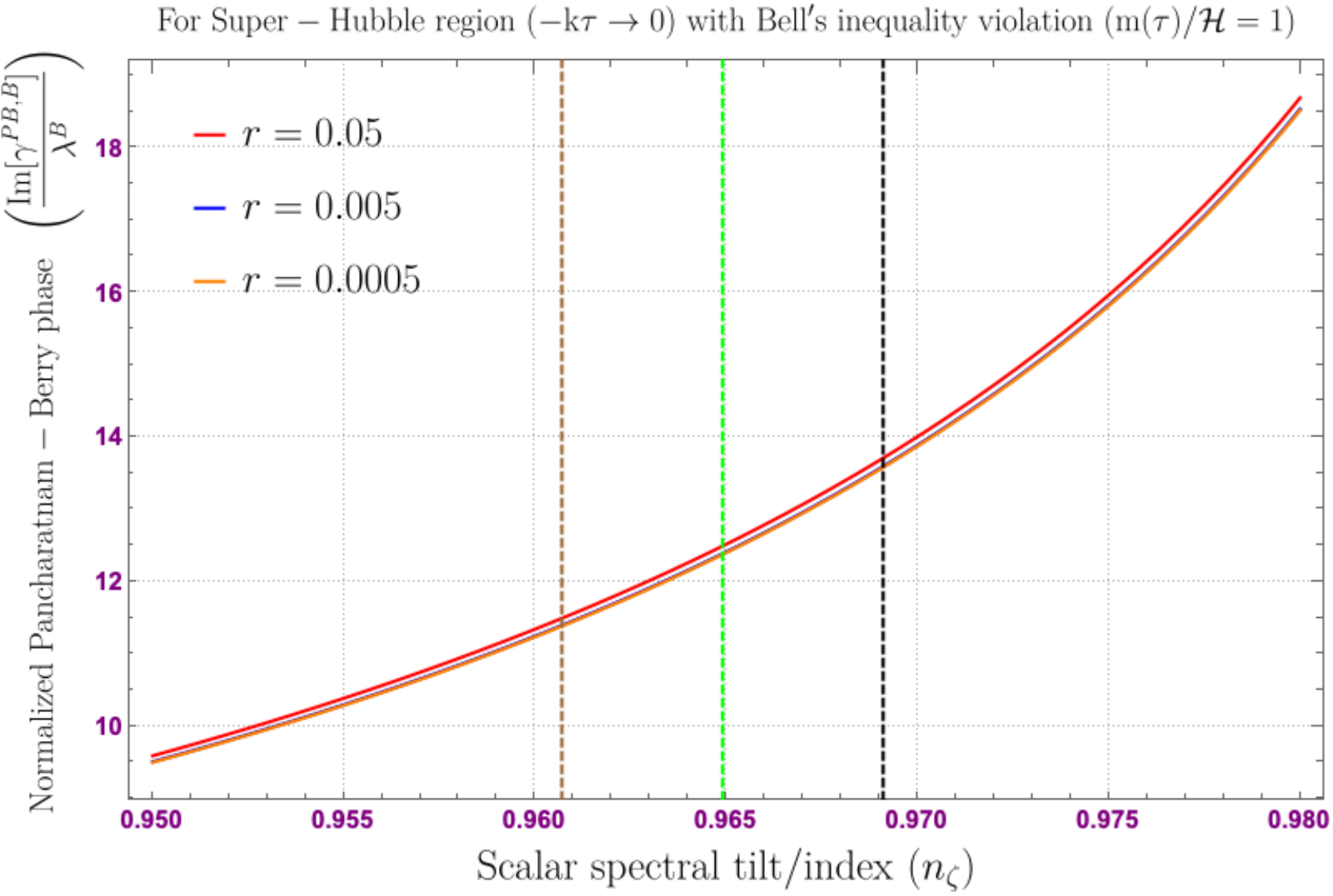}
	\caption{Behavior of the purely imaginary Normalized Pancharatnam Berry phase with scalar spectral tilt with having Bell's inequality violation in super-Hubble region for partially massless field.}
	\label{fig:6}
\end{figure} 

\begin{figure}[htb!]
	\centering
	\subfigure[Real~ part~ of~ the~ Normalized~Pancharatnam~Berry~phase~vs~scalar~spectral~tilt~with
	~Bell's~inequality~violation~at~horizon~crossing.]{
		\includegraphics[width=7cm,height=6.5cm] {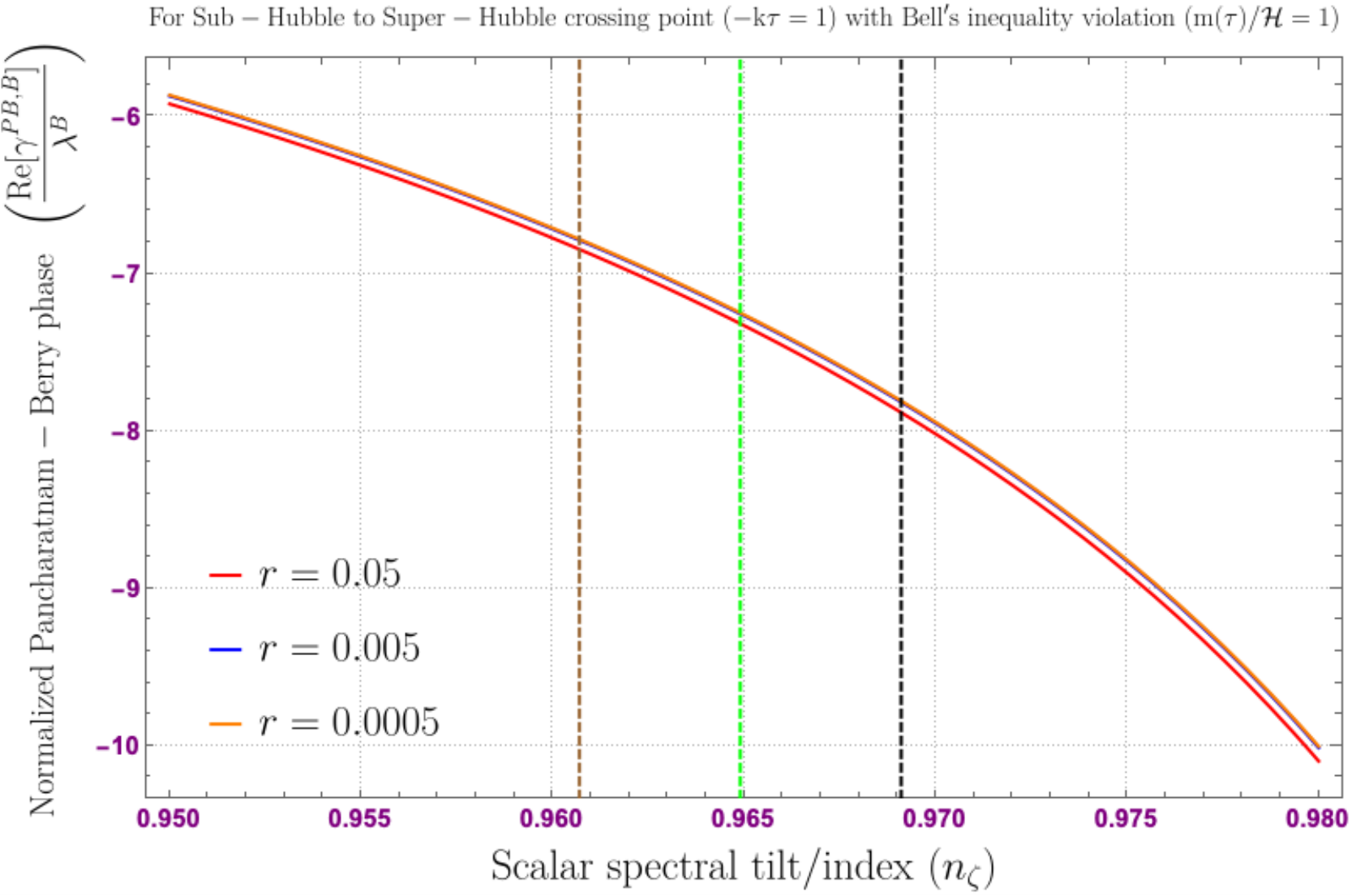}
	}
	\subfigure[Imaginary~ part~ of~ the~ Normalized~Pancharatnam~Berry~phase~vs~scalar~spectral~tilt~with
	~Bell's~inequality~violation~at~horizon~crossing.]{
		\includegraphics[width=7cm,height=6.5cm] {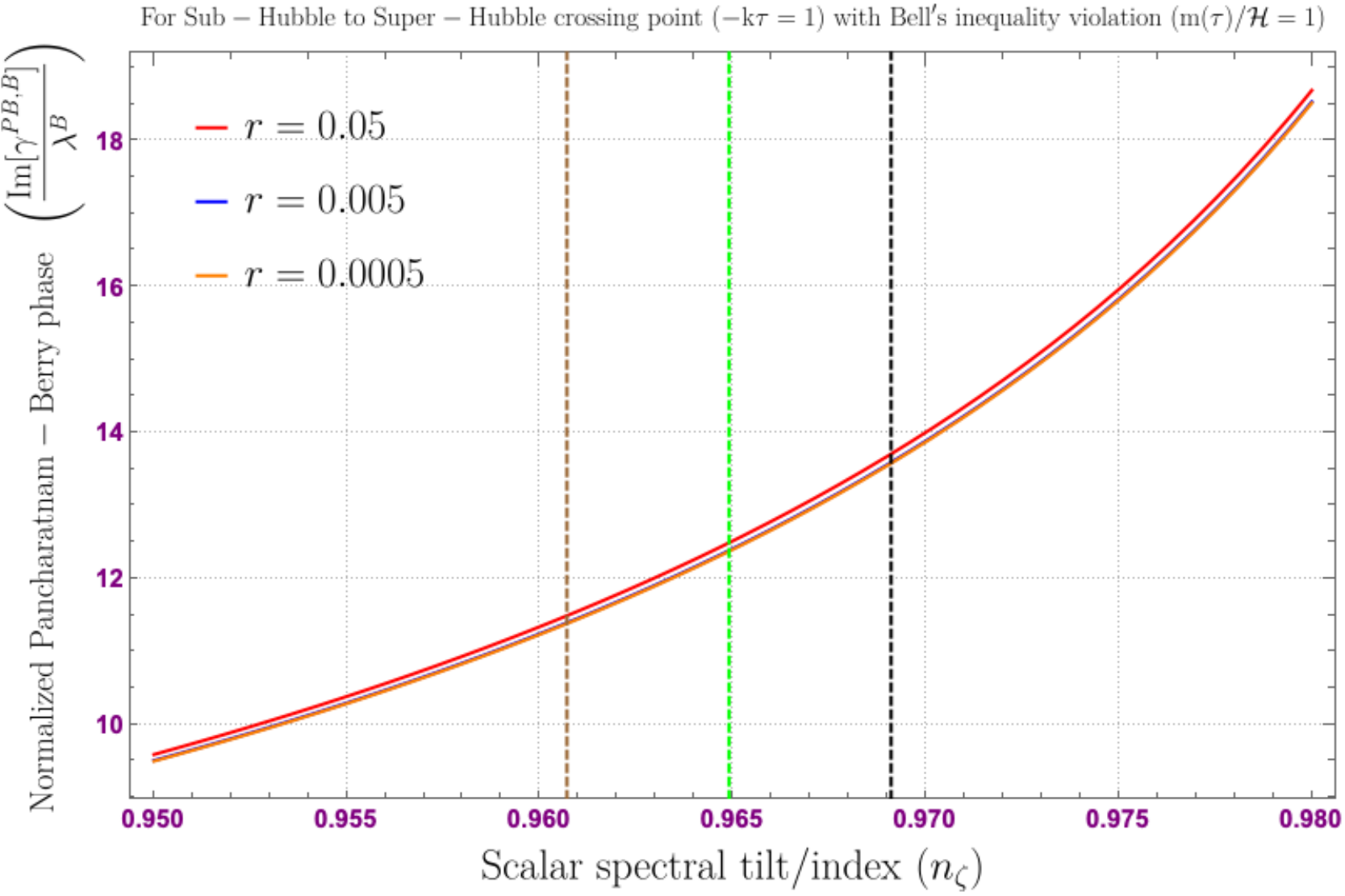}
	}
	\caption{Behavior of the complex Normalized Pancharatnam Berry phase with scalar spectral tilt with having Bell's inequality violation at horizon crossing for partially massless field.}
	\label{fig:7}
\end{figure}

\begin{figure}[htb!]
	\centering
	\subfigure[Real~ part~ of~ the~ Normalized~Pancharatnam~Berry~phase~vs~scalar~spectral~tilt~with
	~Bell's~inequality~violation~at~horizon~crossing.]{
		\includegraphics[width=7cm,height=6.5cm] {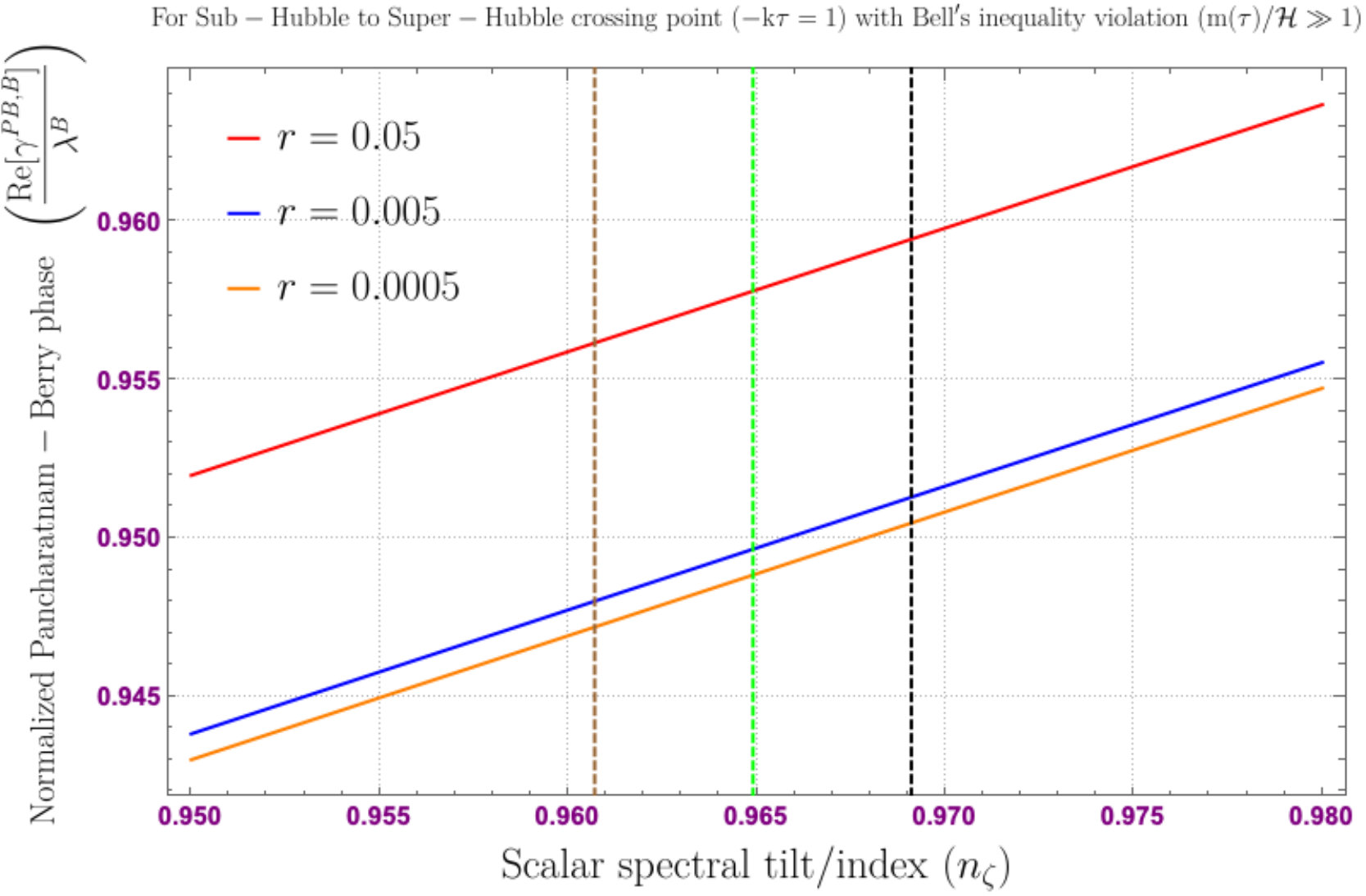}
	}
	\subfigure[Imaginary~ part~ of~ the~ Normalized~Pancharatnam~Berry~phase~vs~scalar~spectral~tilt~with
	~Bell's~inequality~violation~at~horizon~crossing.]{
		\includegraphics[width=7cm,height=6.5cm] {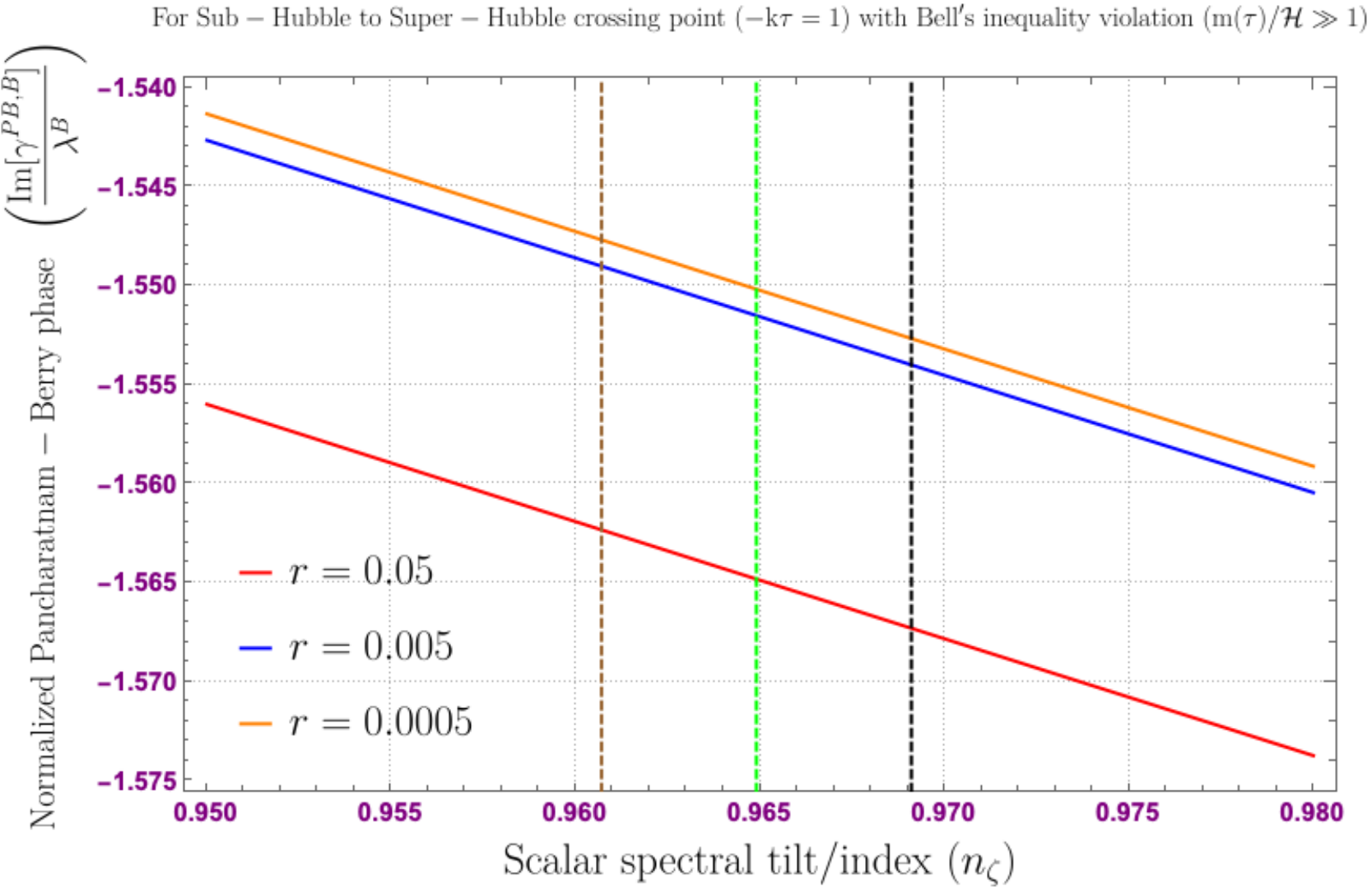}
	}
	\caption{Behavior of the complex Normalized Pancharatnam Berry phase with scalar spectral tilt with having Bell's inequality violation at horizon crossing for massive field.}
	\label{fig:8}
\end{figure} 
\begin{figure}[htb!]
	\centering
	\subfigure[Normalized ~Pancharatnam~ Berry~ phase ~vs~ tensor-to-scalar~ ratio~ without~ Bell's~ inequality~ violation~in~sub-Hubble~region.]{
		\includegraphics[width=7cm,height=6.5cm] {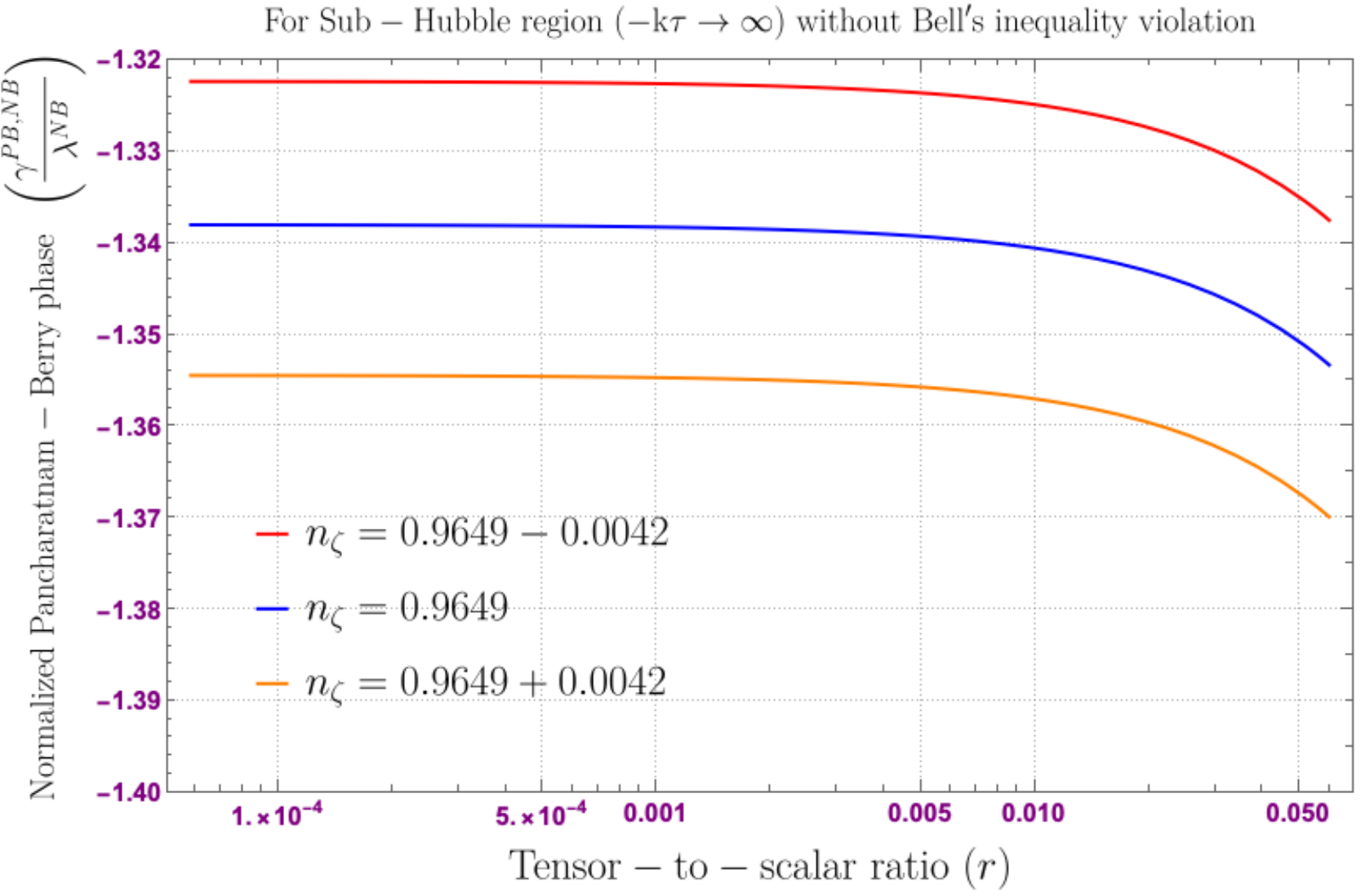}
	}
	\subfigure[Normalized ~Pancharatnam~ Berry~ phase ~vs~ tensor-to-scalar~ ratio~ without~ Bell's~ inequality~ violation~in~super-Hubble~region.]{
		\includegraphics[width=7cm,height=6.5cm] {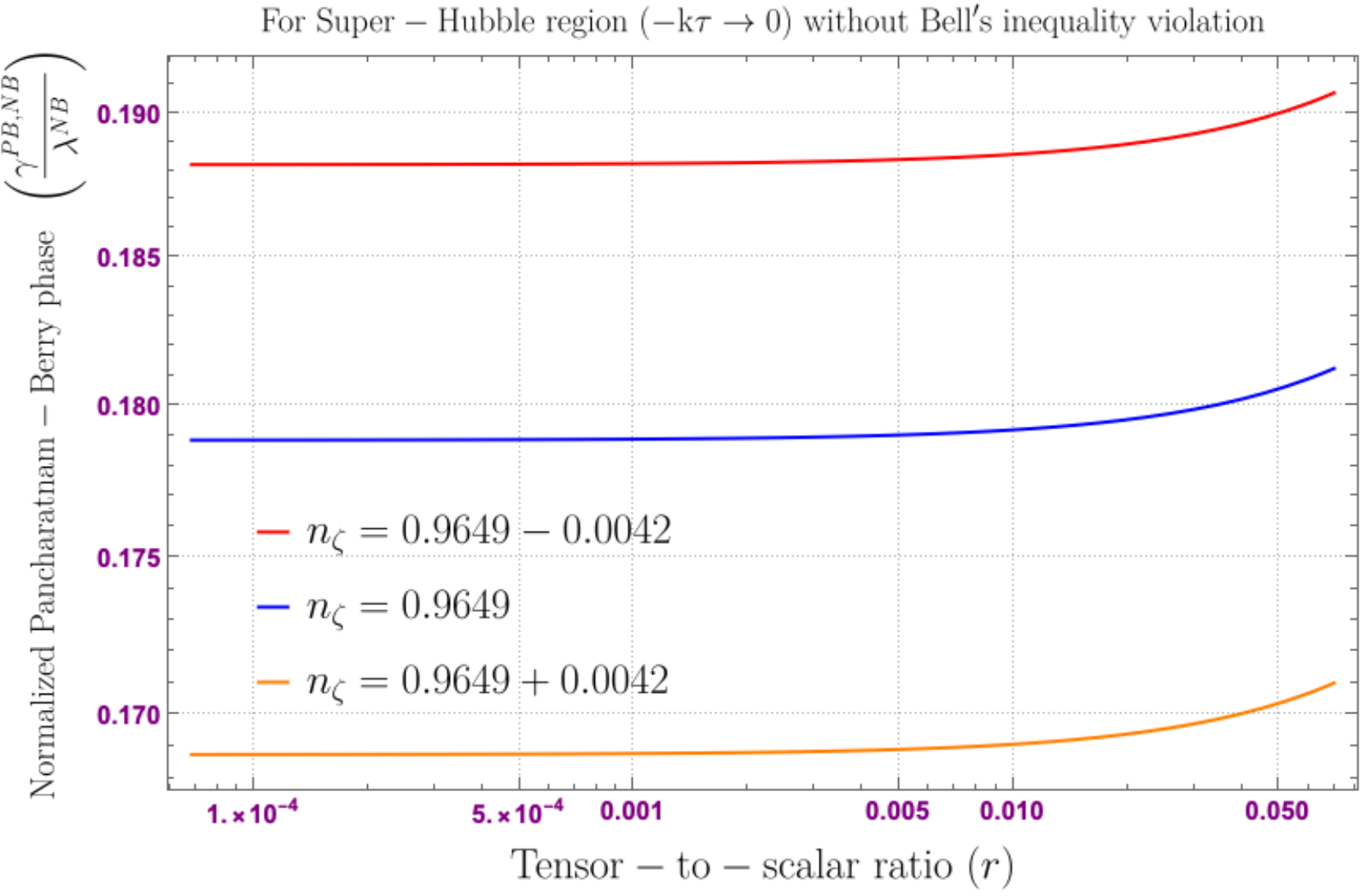}
	}
	\caption{Behavior of the Normalized Pancharatnam Berry phase with tensor-to-scalar ratio without having Bell's inequality violation in sub and super Hubble regime.}
	\label{fig:9}
\end{figure} 
\begin{figure}[htb!]
	\centering
	\subfigure[Real~part~of~the~Normalized ~Pancharatnam~ Berry~ phase ~vs~ tensor-to-scalar~ ratio~ without~ Bell's~ inequality~ violation~at~horizon~crossing~point.]{
		\includegraphics[width=7cm,height=6.5cm] {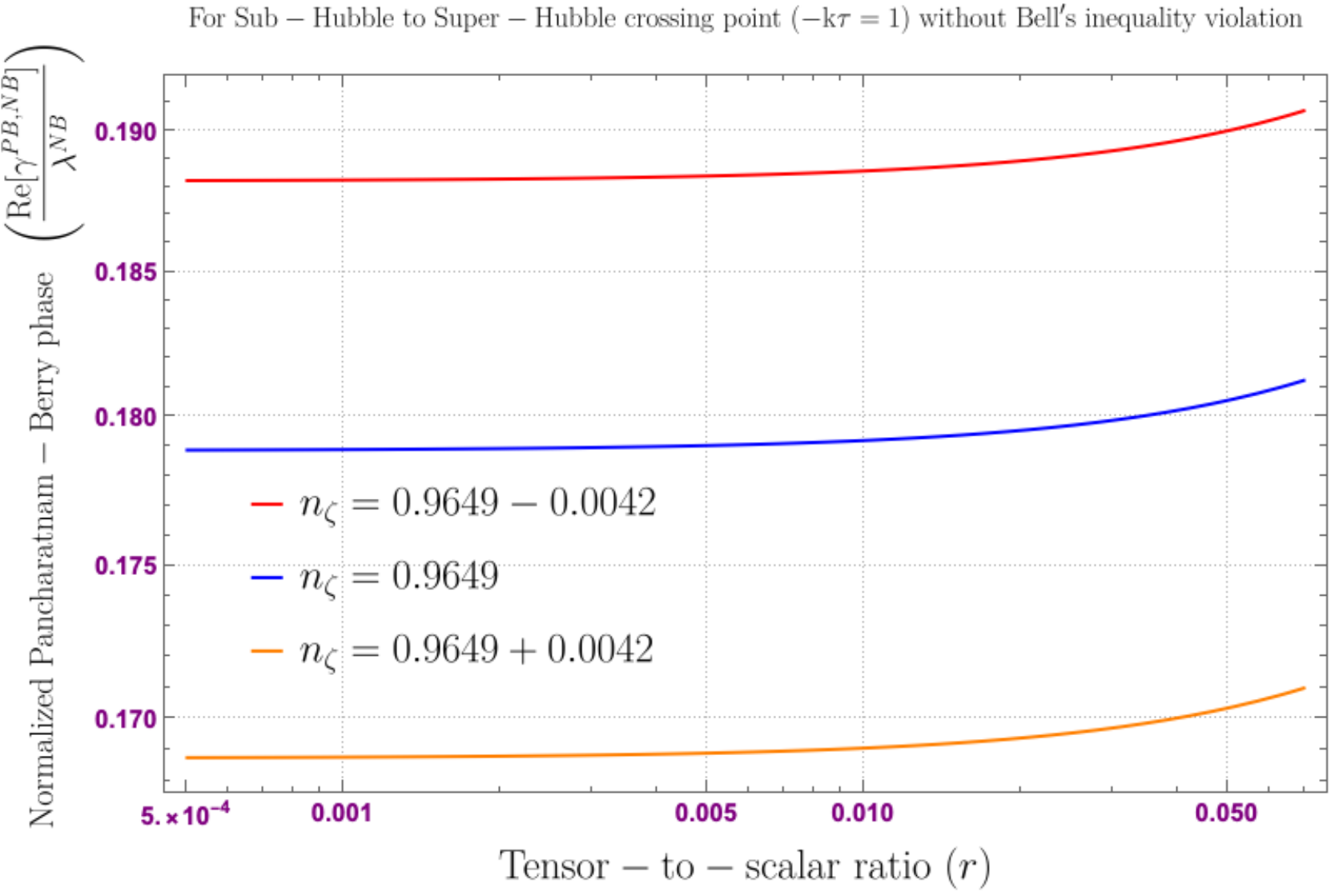}
	}
	\subfigure[Imaginary~part~of~the~Normalized ~Pancharatnam~ Berry~ phase ~vs~ tensor-to-scalar~ ratio~ without~ Bell's~ inequality~ violation~at~horizon~crossing~point.]{
		\includegraphics[width=7cm,height=6.5cm] {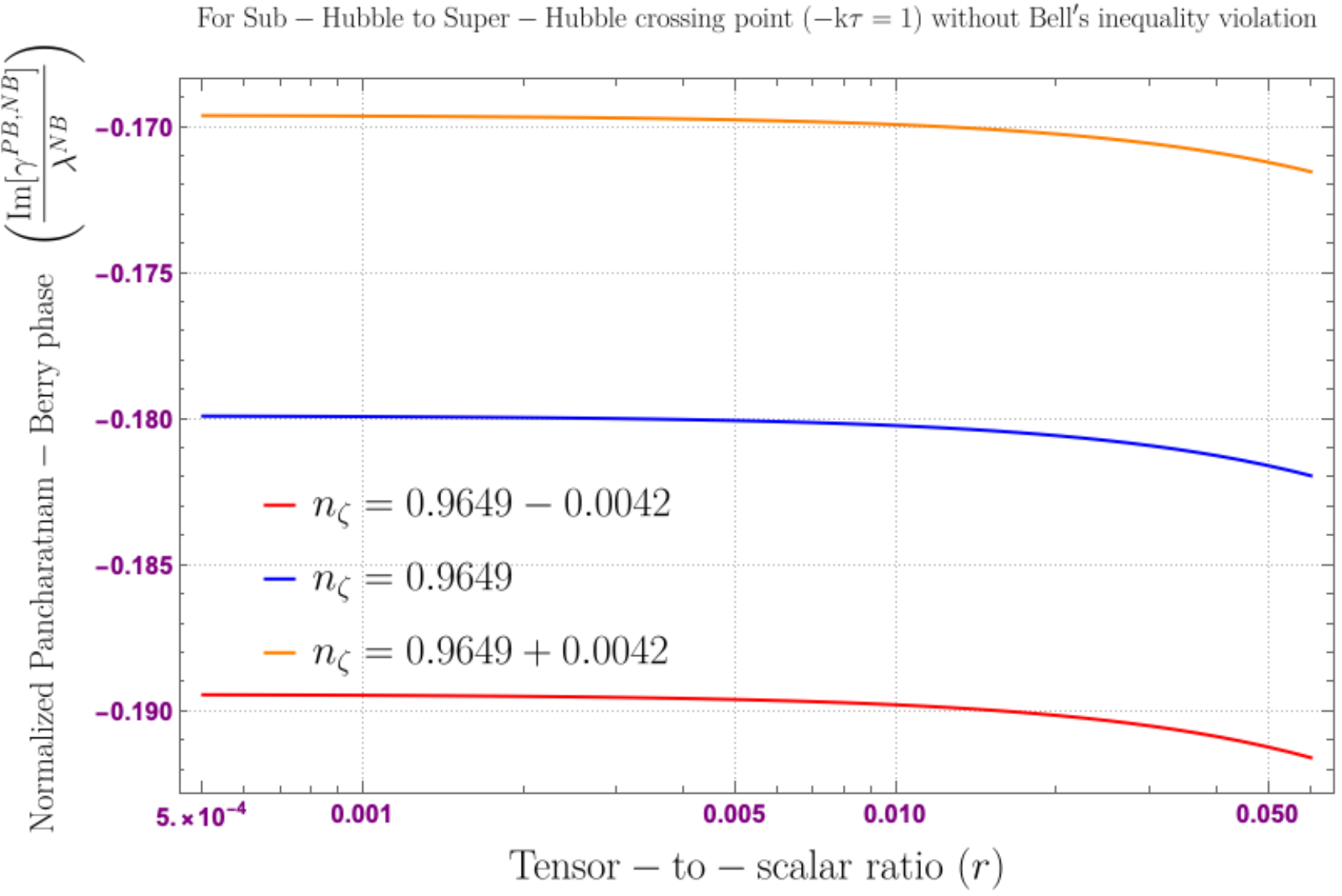}
	}
	\caption{Behavior of the Normalized Pancharatnam Berry phase with tensor-to-scalar ratio without having Bell's inequality at the horizon crossing point.}
	\label{fig:10}
\end{figure} 
\begin{figure}[htb!]
	\centering
		\includegraphics[width=10cm,height=6.5cm] {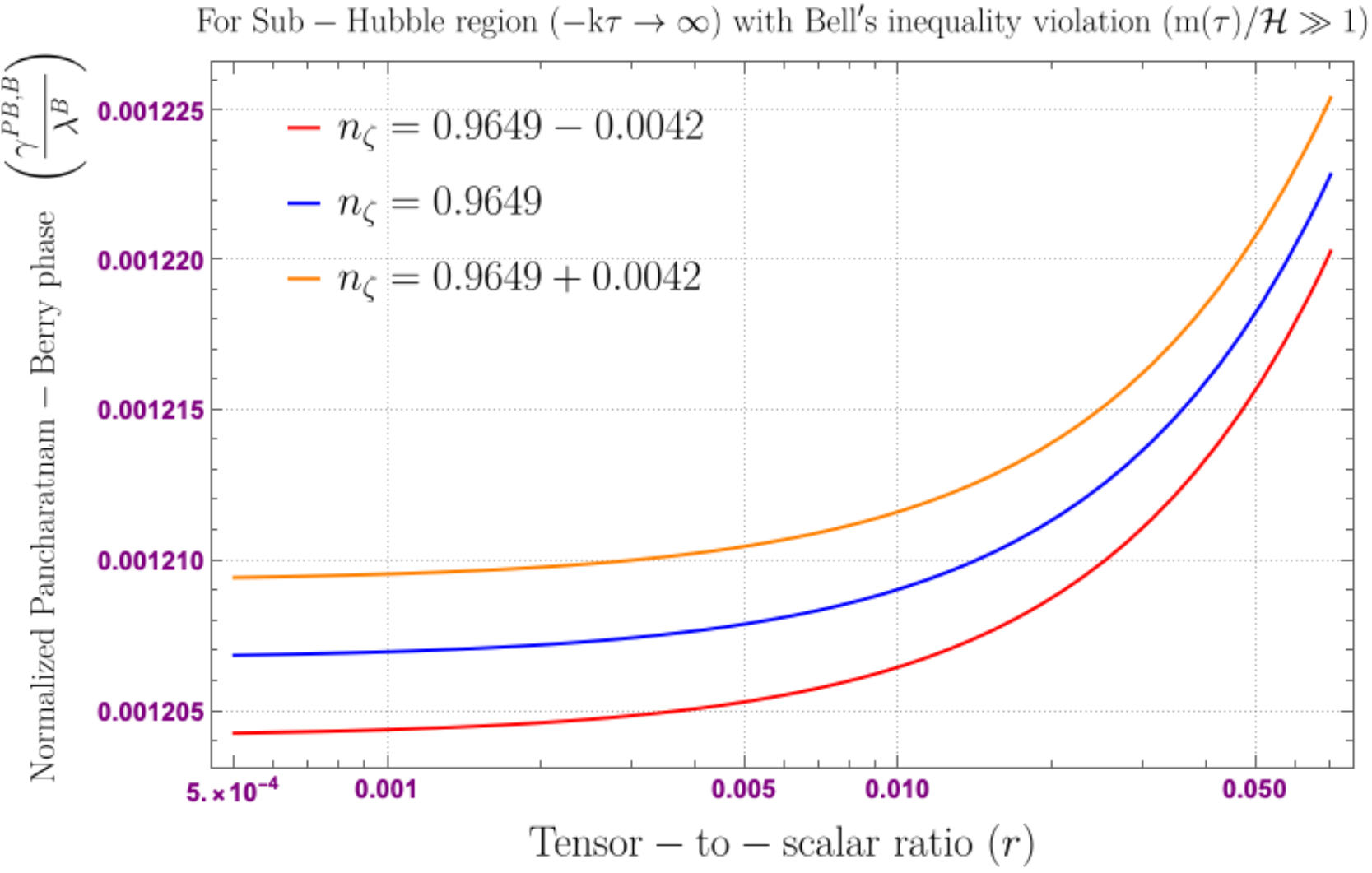}
	\caption{Behavior of the Normalized Pancharatnam Berry phase with tensor-to-scalar ratio with having Bell's inequality in the sub-Hubble region for heavy field.}
	\label{fig:11}
\end{figure} 
\begin{figure}[htb!]
	\centering
	\subfigure[Real~part~of~the~Normalized ~Pancharatnam~ Berry~ phase ~vs~ tensor-to-scalar~ ratio~ with~ Bell's~ inequality~ violation~in~super-Hubble~region.]{
		\includegraphics[width=7cm,height=6.5cm] {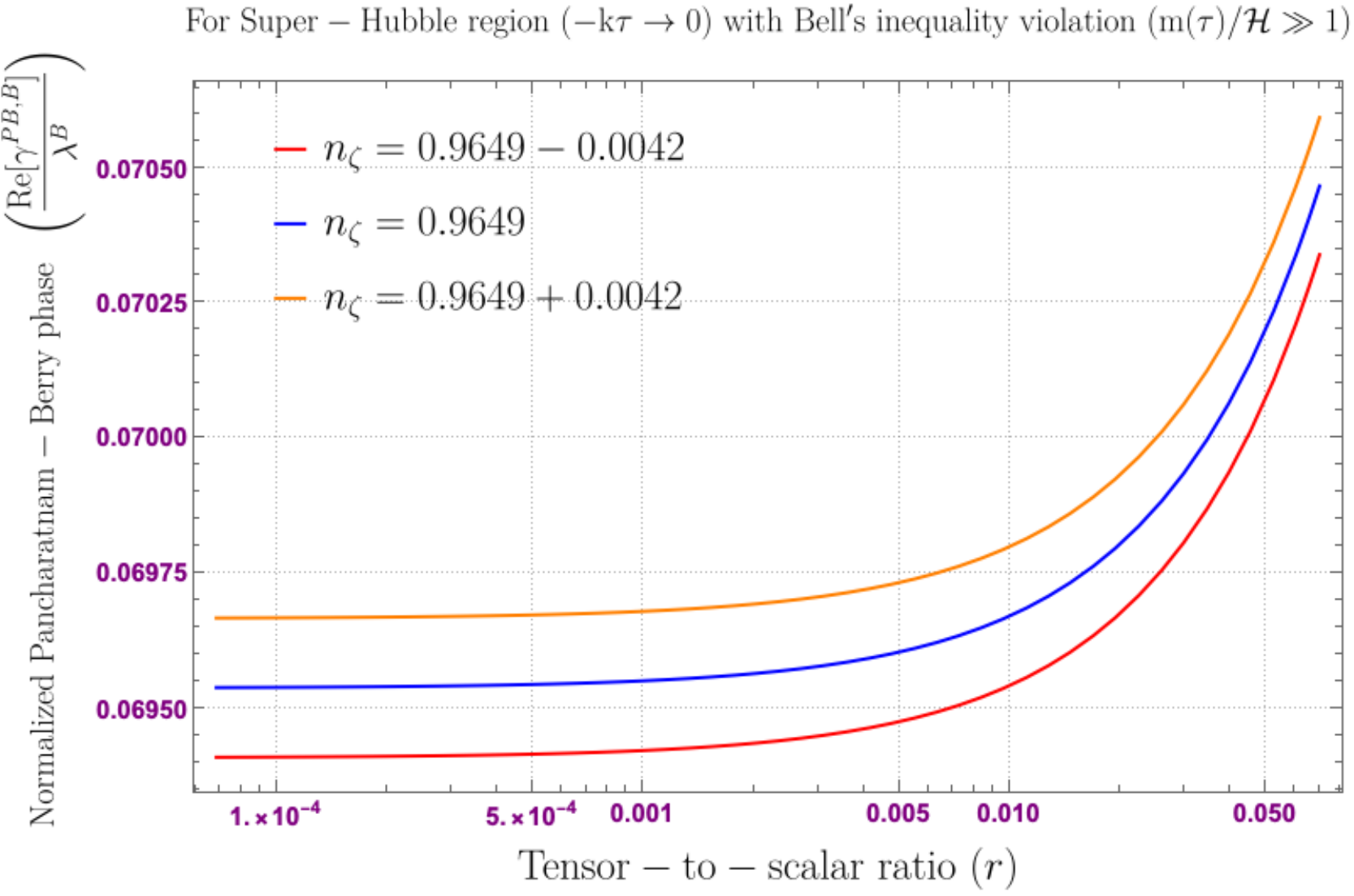}
	}
	\subfigure[Imaginary~part~of~the~Normalized ~Pancharatnam~ Berry~ phase ~vs~ tensor-to-scalar~ ratio~ with~ Bell's~ inequality~ violation~in~super-Hubble~region.]{
		\includegraphics[width=7cm,height=6.5cm] {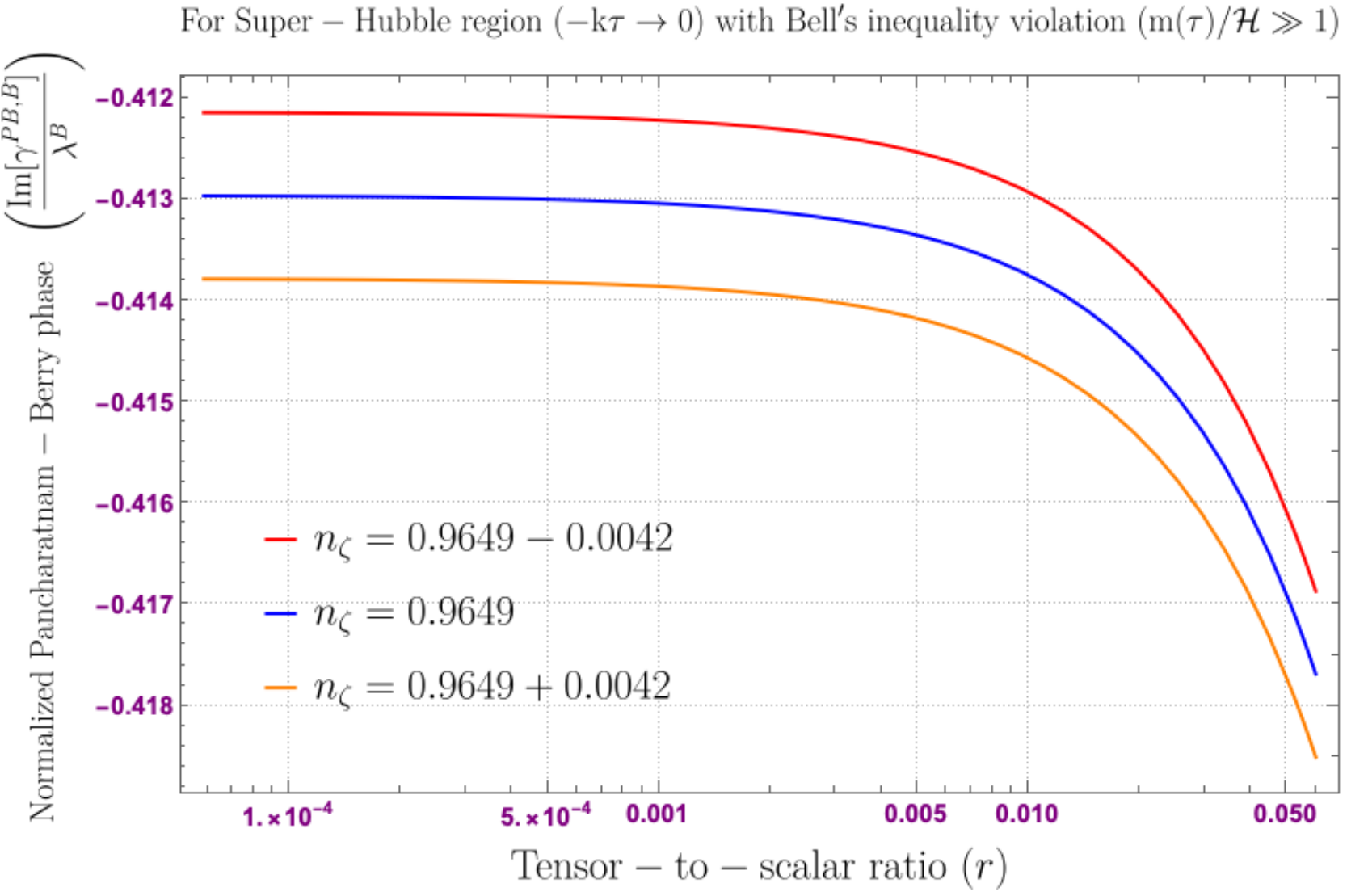}
	}
	\caption{Behavior of the complex Normalized Pancharatnam Berry phase with tensor-to-scalar ratio with having Bell's inequality in the super-Hubble region for heavy field.}
	\label{fig:12}
\end{figure} 
\begin{figure}[htb!]
	\centering
	\subfigure[Real~part~of~the~Normalized ~Pancharatnam~ Berry~ phase ~vs~ tensor-to-scalar~ ratio~ with~ Bell's~ inequality~ violation~in~sub-Hubble~region.]{
		\includegraphics[width=7cm,height=6.5cm] {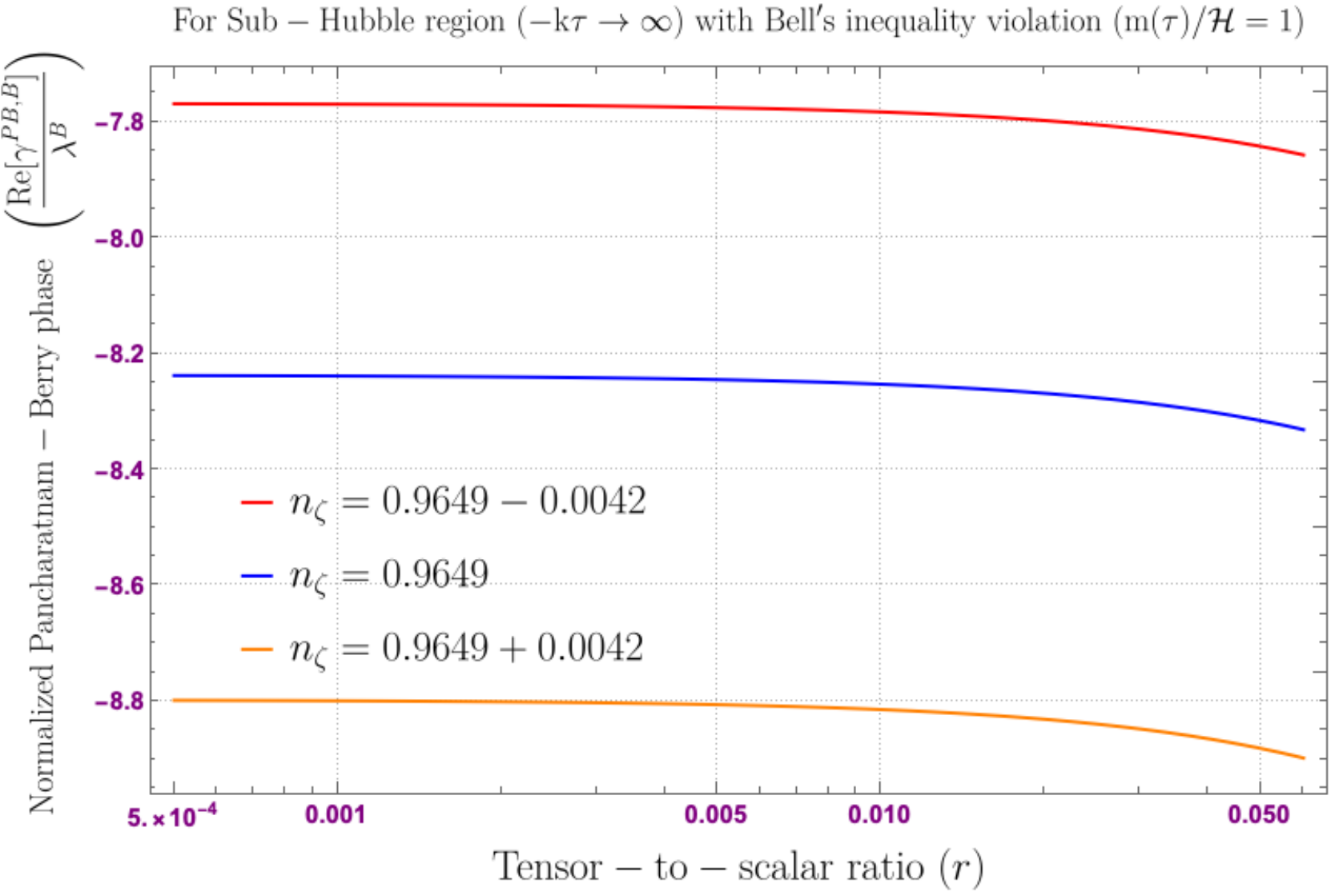}
	}
	\subfigure[Imaginary~part~of~the~Normalized ~Pancharatnam~ Berry~ phase ~vs~ tensor-to-scalar~ ratio~ with~ Bell's~ inequality~ violation~in~sub-Hubble~region.]{
		\includegraphics[width=7cm,height=6.5cm] {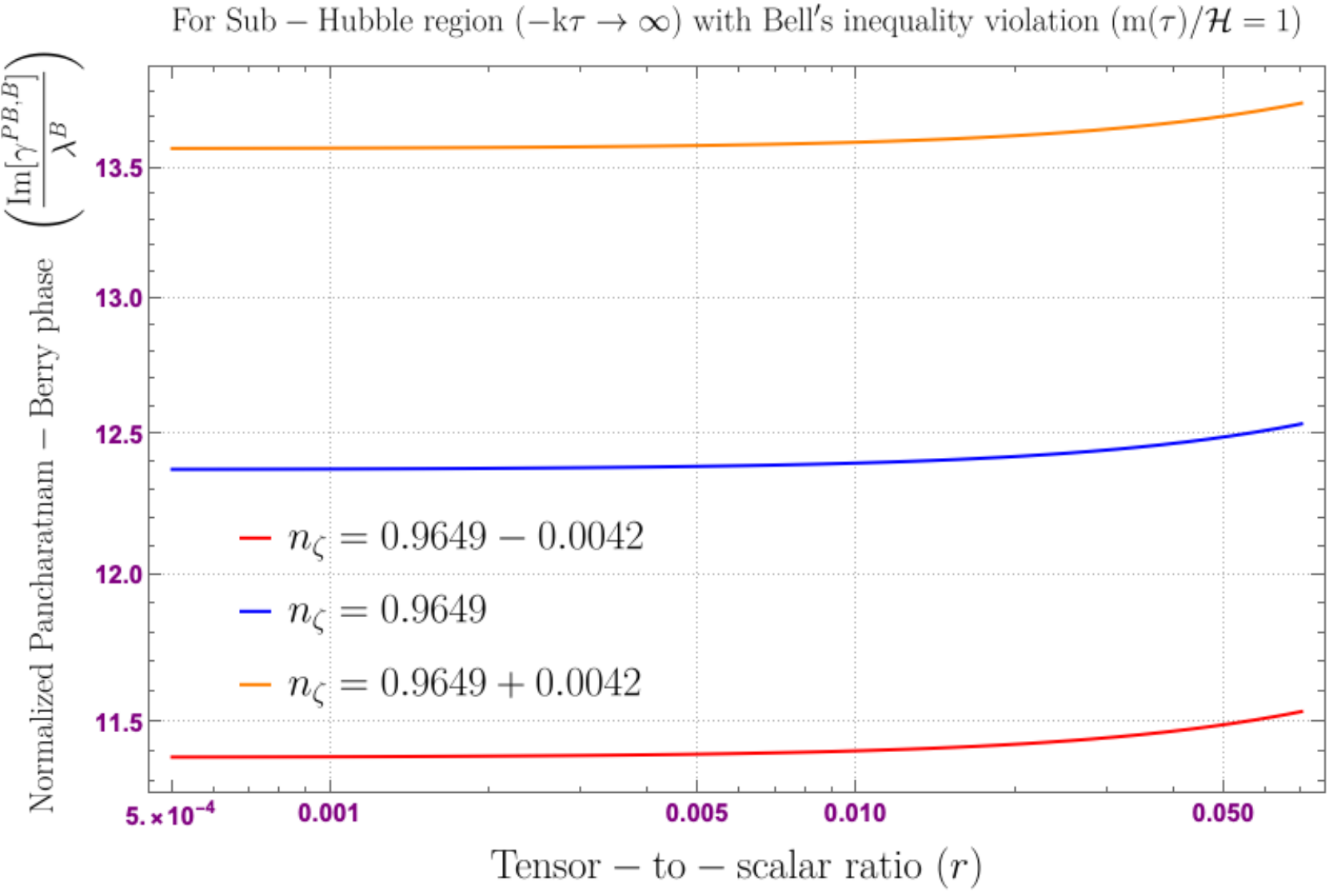}
	}
	\caption{Behavior of the complex Normalized Pancharatnam Berry phase with tensor-to-scalar ratio with having Bell's inequality in the sub-Hubble region for partially massless field.}
	\label{fig:13}
\end{figure} 

\begin{figure}[htb!]
	\centering
	{
		\includegraphics[width=10cm,height=6.5cm] {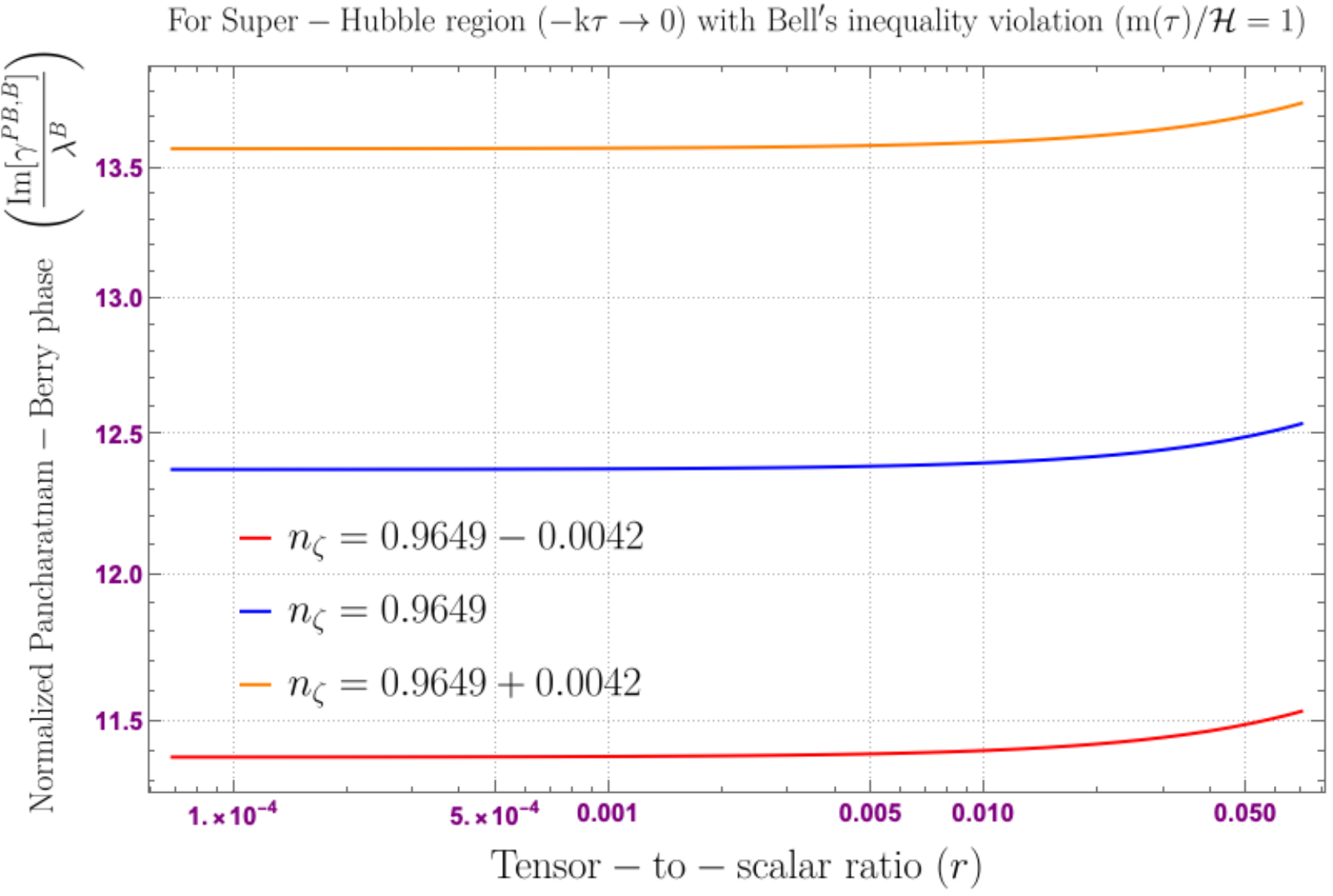}
	}
	\caption{Behavior of the complex Normalized Pancharatnam Berry phase with tensor-to-scalar ratio with having Bell's inequality in the super-Hubble region for partially massless field.}
	\label{fig:14}
\end{figure} 

\begin{figure}[htb!]
	\centering
	\subfigure[Real~part~of~the~Normalized ~Pancharatnam~ Berry~ phase ~vs~ tensor-to-scalar~ ratio~ with~ Bell's~ inequality~ violation~at~horizon~exit.]{
		\includegraphics[width=7cm,height=6.5cm] {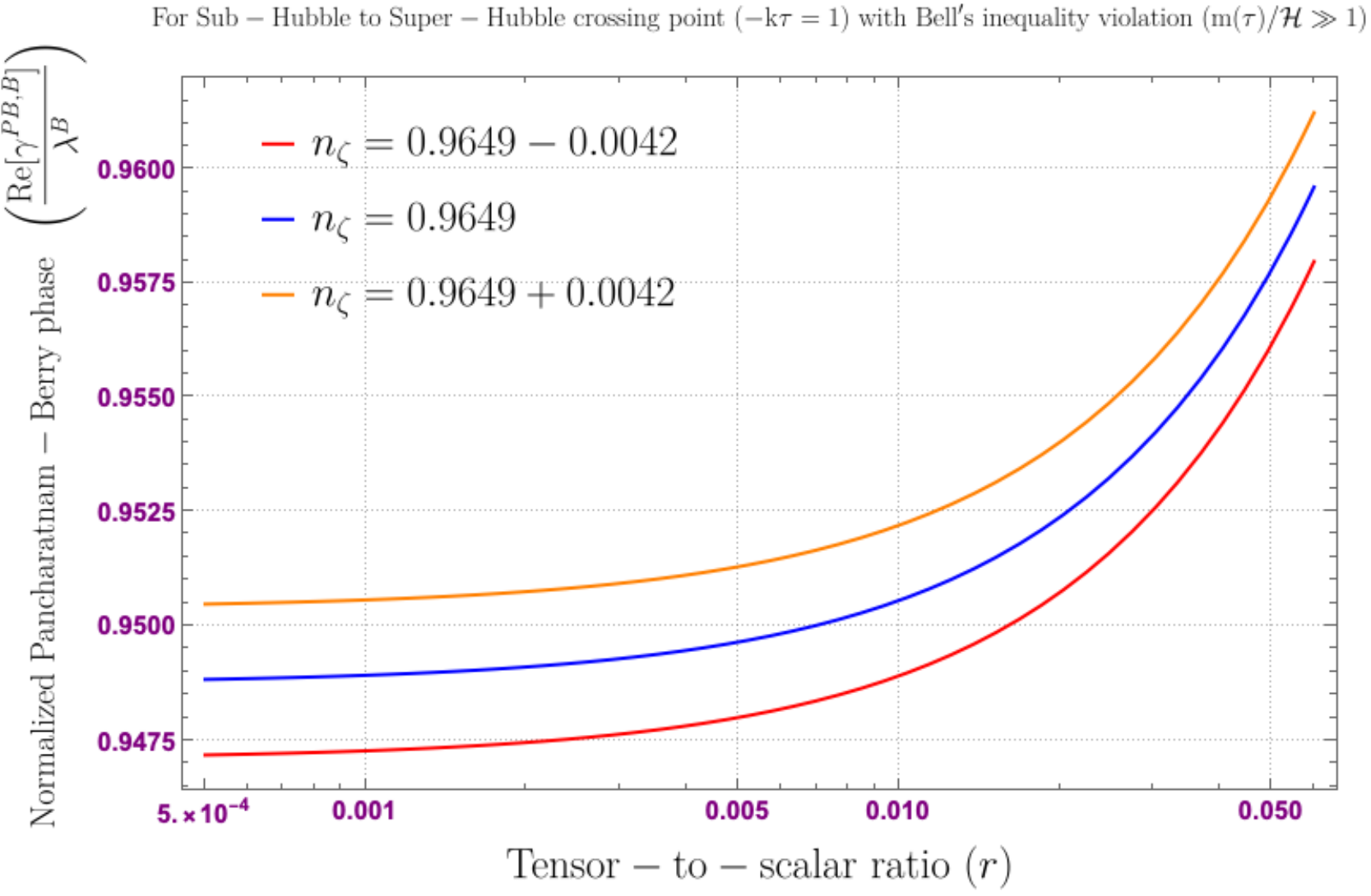}
	}
	\subfigure[Imaginary~part~of~the~Normalized ~Pancharatnam~ Berry~ phase ~vs~ tensor-to-scalar~ ratio~ with~ Bell's~ inequality~ violation~at~horizon~exit.]{
		\includegraphics[width=7cm,height=6.5cm] {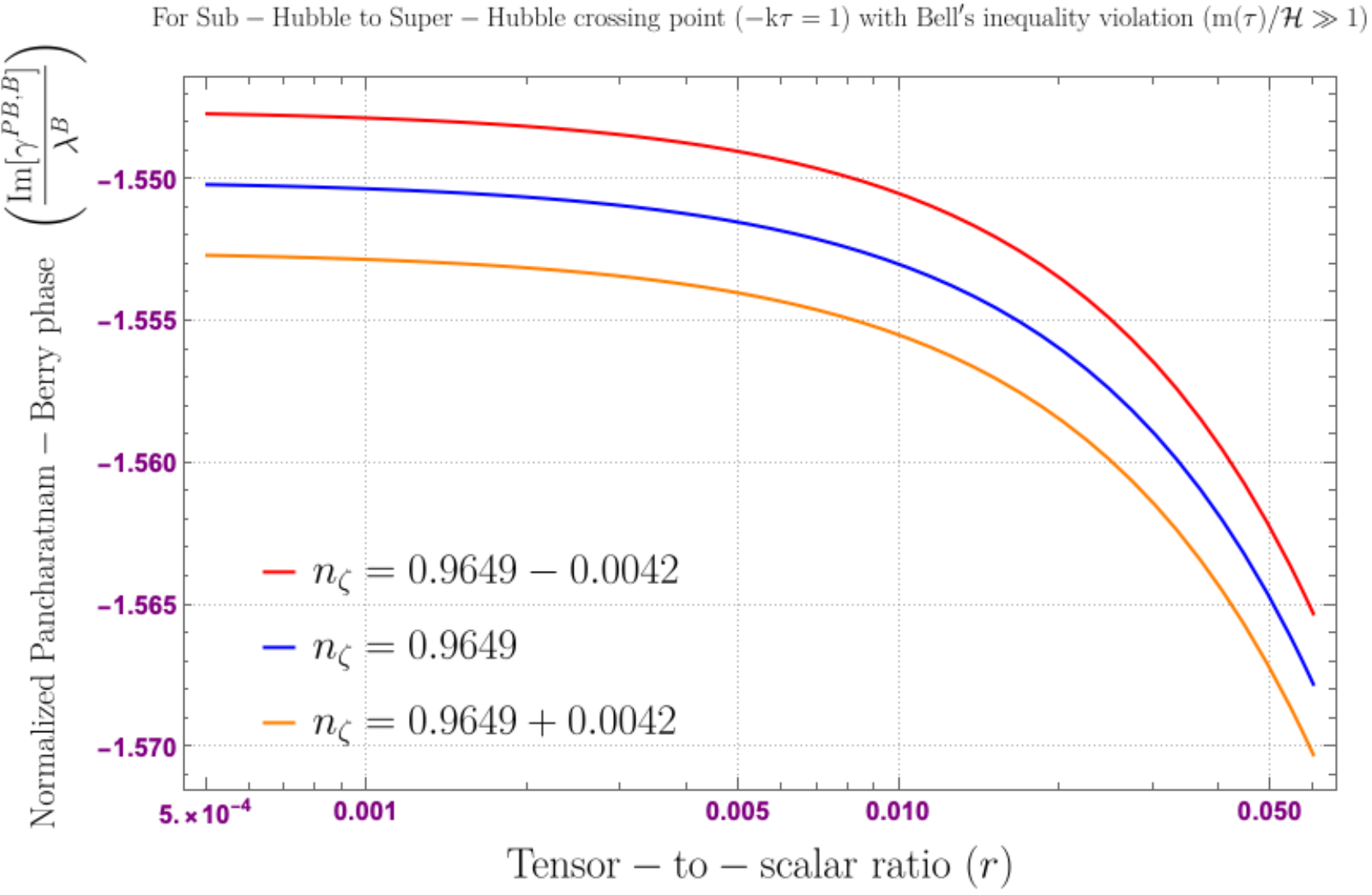}
	}
	\caption{Behavior of the complex Normalized Pancharatnam Berry phase with tensor-to-scalar ratio with having Bell's inequality at the horizon exit point for massive field.}
	\label{fig:15}
\end{figure} 

\begin{figure}[htb!]
	\centering
	\subfigure[Real~part~of~the~Normalized ~Pancharatnam~ Berry~ phase ~vs~ tensor-to-scalar~ ratio~ with~ Bell's~ inequality~ violation~at~horizon~exit.]{
		\includegraphics[width=7cm,height=6.5cm] {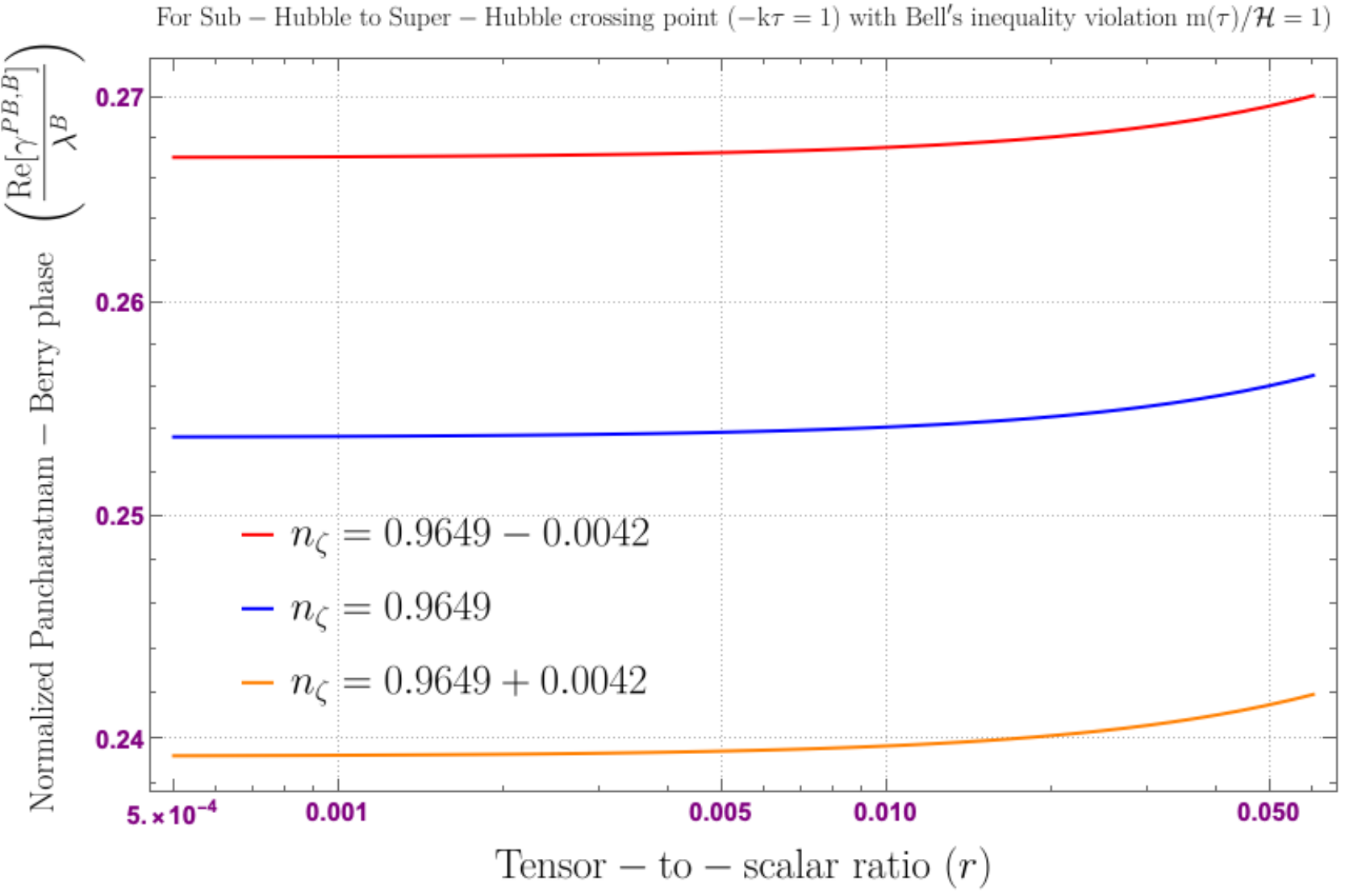}
	}
	\subfigure[Imaginary~part~of~the~Normalized ~Pancharatnam~ Berry~ phase ~vs~ tensor-to-scalar~ ratio~ with~ Bell's~ inequality~ violation~at~horizon~exit.]{
		\includegraphics[width=7cm,height=6.5cm] {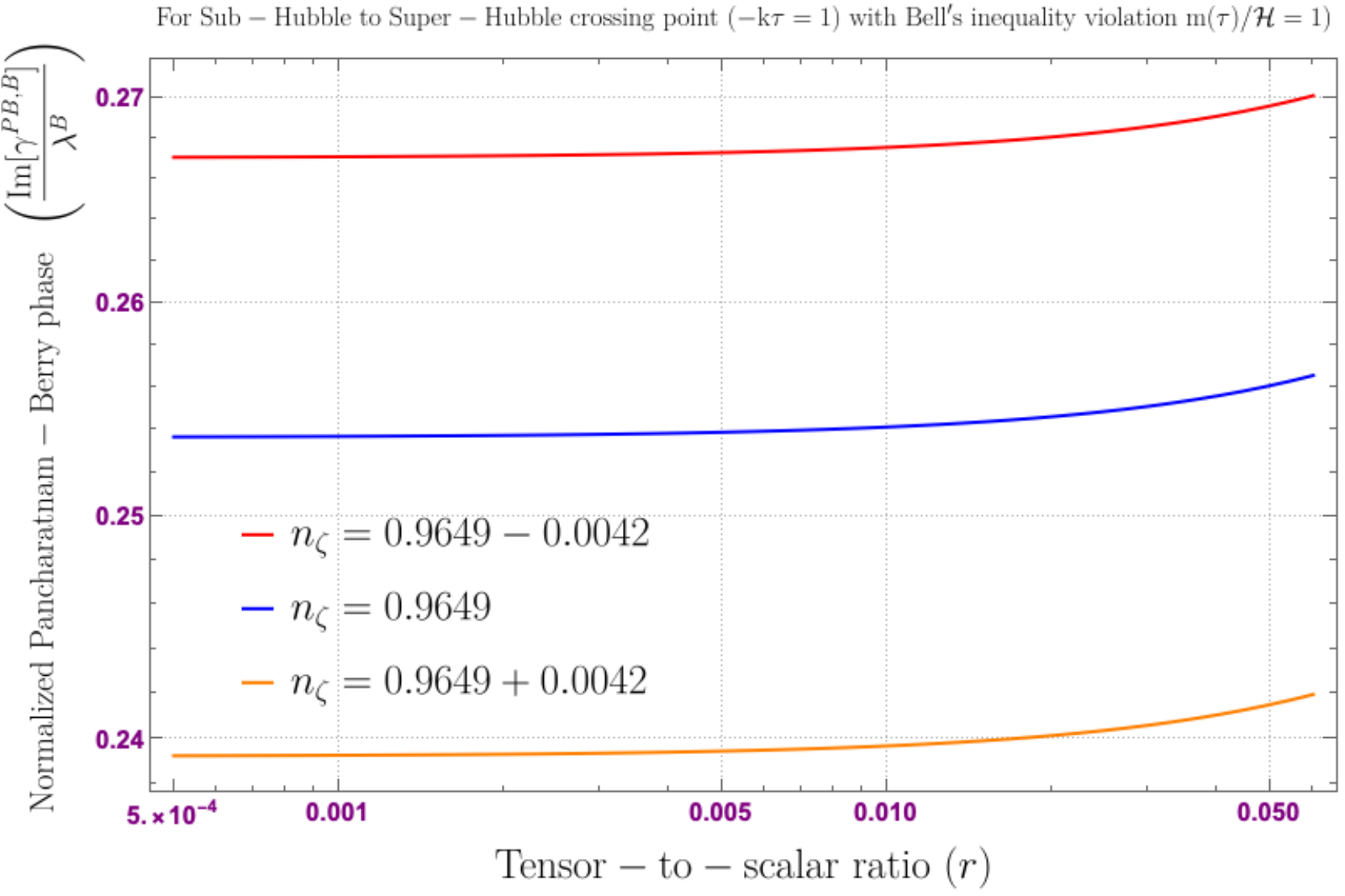}
	}
	\caption{Behavior of the complex Normalized Pancharatnam Berry phase with tensor-to-scalar ratio with having Bell's inequality at the horizon exit point for partially massless field.}
	\label{fig:16}
\end{figure} 
\subsection{Things to remember for physical interpretation} 
 To interpret these obtained results physically and to connect with cosmological observation one need to point the following crucial facts,  which are appended below point-wise:
 \begin{enumerate}
 \item From the Planck 2018 combined with the BICEP2/Keck Array BK15 datasets (Planck TT,TE,EE
+lowE+lensing+BK15  likelihood data) gives the following combined constraints on the amplitude of the power spectrum, spectral tilt and tensor-to-scalar-ratio \cite{Akrami:2018odb}:
 \bea &&{\rm Amplitude ~of ~the~power~spectrum:}~~~~~~~P_{\zeta}(k_{*})=\left(2.975\pm 0.056\right)\times 10^{-10}~~~68\%~{\rm CL},~~~~~~~~\\
&&{\rm Spectral~tilt/Spectral~index:}~~~~~~~~~~~~~~~n_{\zeta}(k_{*})=0.9649\pm 0.0042~~~~~~~~~~~~~68\%~{\rm CL},~~~~~~~~\\
&&{\rm Tensor-to-scalar~ratio:}~~~~~~~~~~~~~~~~~~r_{\zeta}(k_{*})<0.056~~~~~~~~~~~~~~~~~~~~~~~~~~~95\%~{\rm CL}, \eea
where $k_{*}$ is identified to be pivot or the normalization scale in the following parametrization,  
$P_{\zeta}(k)=P_{\zeta}(k_{*})~(k/k_{*})^{n_{\zeta}(k_{*})-1}.$
In this computation it is fixed at $k_{*}=0.002~{\rm Mpc}^{-1}$.  Using the above mentioned constraints our plan is to give a observationally viable tight constraint on the  {\it Pancharatnam-Berry phase} computed from scalar perturbation in the quantum set up for primordial cosmology in absence and in the presence of Bell's inequality violating effects.   

 \item If we consider the results without having any Bell's inequality violating effect then the previously mentioned three constraints are sufficient enough to put tighter constraint on the  {\it Pancharatnam-Berry phase}.  But if we allow the Bell's inequality violating effects in the primordial scalar cosmological perturbation set up then one need to supply an additional information regarding the conformal time dependent mass profile $m^2(\tau)/{\cal H}^2$,  which we have taken a particular profile for the purpose of the computation in this paper \cite{Maldacena:2015bha,Choudhury:2016cso,Choudhury:2016pfr}.  Apart from using this particular profile one may choose various other profiles to study the effect in Bell's inequality violation from a broad range of possibilities.  Here the information regarding the quantum fluctuation in presence of Bell's inequality violation is extremely model dependent which rely on a particular type of or the class of conformal time dependent profile which can finally give rise to desirable analytical result for the scalar perturbed modes.  Though we need to point here that,  apart from having model or profile dependence on the analytical expression for the quantum fluctuated scalar modes the final expression for the {\it Pancharatnam-Berry phase} can be directly expressed in terms of the scalar spectral tilt $n_{\zeta}$ and in terms of the ratio $m^2(\tau)/{\cal H}^2$.  In this final expression now if we supply the form of this profile and compute the spectral tilt from the quantum fluctuated scalar modes in presence of the given profile then one can fix the {\it Pancharatnam-Berry phase} in presence of Bell's inequality violation.  
In refs~\cite{Maldacena:2015bha,Choudhury:2016cso,Choudhury:2016pfr} it was explicitly pointed that the above mentioned profile for the conformal time dependent mass is very successful to describe the effects and impacts of Bell's inequality violation within the set up of quantum theory of primordial cosmological perturbation theory.  It was also pointed in refs.~\cite{Maldacena:2015bha,Choudhury:2016cso,Choudhury:2016pfr} that in this type of set up one can have non vanishing one-point function of co-moving scalar curvature perturbation field variable:
\bea && \underline{\rm A. ~ Sub~Hubble~region:}\nonumber\\
&&\langle \hat{\zeta}_{\bf k}(\tau)\rangle^{\bf Sub}=i\lim_{\tau\rightarrow -\infty}\int^{\tau_0}_{\tau}~\frac{d\tau^{\prime}}{\tau^{\prime}}~\frac{a(\tau^{\prime})}{z^2(\tau^{\prime})}~\frac{m(\tau^{\prime})}{{\cal H}}~\bigg(v_{\bf k}(\tau_0)\frac{dv_{-\bf k}(\tau^{\prime})}{d\tau^{\prime}}-v_{-\bf k}(\tau_0)\frac{dv_{\bf k}(\tau^{\prime})}{d\tau^{\prime}}\bigg)\neq 0,~~~~~~~~~\\
&& \underline{\rm B. ~ Super~Hubble~region:}\nonumber\\
&&\langle \hat{\zeta}_{\bf k}(\tau)\rangle^{\bf Super}=i\lim_{\tau\rightarrow 0}\int^{\tau_0}_{\tau}~\frac{d\tau^{\prime}}{\tau^{\prime}}~\frac{a(\tau^{\prime})}{z^2(\tau^{\prime})}~\frac{m(\tau^{\prime})}{{\cal H}}~\bigg(v_{\bf k}(\tau_0)\frac{dv_{-\bf k}(\tau^{\prime})}{d\tau^{\prime}}-v_{-\bf k}(\tau_0)\frac{dv_{\bf k}(\tau^{\prime})}{d\tau^{\prime}}\bigg)\neq 0,~~~~~~~~~\\
 && \underline{\rm C. ~ Sub~Hubble~to~Super~Hubble~crossing~point:}\nonumber\\
&&\langle\hat{\zeta}_{\bf k}(\tau)\rangle^{\bf Cross}=i\lim_{\tau\rightarrow -\frac{1}{k}}\int^{\tau_0}_{\tau}~\frac{d\tau^{\prime}}{\tau^{\prime}}~\frac{a(\tau^{\prime})}{z^2(\tau^{\prime})}~\frac{m(\tau^{\prime})}{{\cal H}}~\bigg(v_{\bf k}(\tau_0)\frac{dv_{-\bf k}(\tau^{\prime})}{d\tau^{\prime}}-v_{-\bf k}(\tau_0)\frac{dv_{\bf k}(\tau^{\prime})}{d\tau^{\prime}}\bigg)\neq 0.~~~~~~~~~
\eea
 In this one-point function the upper limit of the integration in all the primordial cosmological scales is given by the following expression:
 \bea \tau_0&=&\underbrace{-\frac{1}{k}\bigg[2-n_{\zeta}(k)\bigg]}_{\rm Non~Bell~violating~contribution}-\underbrace{\frac{1}{k}\frac{m^2(\tau)}{{\cal H}^2}}_{\rm Bell~violating~contribution} \eea 
 the lower limit of the integration is fixed from the the cosmological scale of interest,  sub Hubble region ($\tau\rightarrow -\infty$),  super Hubble region ($\tau\rightarrow 0$) and sub Hubble to super Hubble crossing point ($\tau\rightarrow -1/k$) respectively.  This is the direct outcome of having Bell's inequality violation as the following statement is true for all cosmological scales \cite{Maldacena:2015bha,Choudhury:2016cso,Choudhury:2016pfr}:
\bea \sqrt{\langle \left[\hat{\zeta}_{\bf k}(\tau),\hat{\Pi}_{\zeta, {\bf k}}(\tau)\right]\rangle}\propto\langle \hat{\zeta}_{\bf k}(\tau)\rangle \neq0 ~~~{\rm because}~~~~\frac{m(\tau)}{{\cal H}}\neq 0~~~~\&~~~~\left|\frac{m(\tau)}{{\cal H}}\right|\gg 1. ~~~~~~~~~\eea
Without Bell's inequality violation the above quantity is explicitly zero as $m(\tau)/{\cal H}=0$ in that case.  Here the quantum operator $\hat{\zeta}_{\bf k}(\tau)$ is defined as,  $\hat{\zeta}_{\bf k}(\tau)=\hat{v}_{\bf k}(\tau)/z(\tau)$ in the Planckian unit ($M_p=1$).  The classical counterpart ${v}_{\bf k}(\tau)$ we have computed by solving the {\it Mukhanov Sasaki equation} for a given conformal time dependent profile for $m(\tau)/{\cal H}$ which satisfy the constraint $|m(\tau)/{\cal H}|\gg 1$.  From this discussion it implies that if in near future observations one can able to detect such heavy mass quantum fluctuations,  then the existence of one-point function and the related square-root of the expectation value of the commutator bracket (which is commonly known as spread) can be confirmed,  which will be treated to be the direct observational confirmation of Bell's inequality violation in the primordial cosmology.  Since without having Bell's inequality violation in cosmology we mostly deal with Gaussian random variables it is expected to have zero one-point function.  Here in the present set up the one-point function is significantly non vanishing and finite due to having heavy mass profile $|m(\tau)/{\cal H}|\gg 1$,  it is in turn expected that the amount of primordial non-Gaussianity will be large compared to the result obtained without Bell's inequality violation from single field cosmological perturbation theory.  So if in near future the observational probe can able to detect large value of the cosmological three and four point functions then that can also be treated to be second confirmation of having Bell's inequality violation in the primordial cosmological set up.  So to give a justifiable and physically consistent prediction for {\it Pancharatnam-Berry phase} computed from the Bell's inequality violating set up.

 \item To give a as much as possible a model independent numerical estimation of the theoretically computed expression for the {\it Pancharatnam-Berry phase} in the primordial cosmological set up we are going to follow the mentioned steps as appended below:
 \begin{enumerate}
 \item \underline{In absence of Bell's inequality violation:}\\
 In this case we are going to first of all fix the value of the tensor-to-scalar ratio in five different range of values in the decimal places.  This is because of the fact that except the upper bound the exact value with statistical error bar (CL) is not yet measured from the observational probes.  For this reason we have to choose different prior values of the tensor-to-scalar ratio.  Once we choose these five different prior value that will automatically fix the energy scale of inflation associated with this scenario and the corresponding underlying physical framework responsible for generating inflationary effective potential which will appear in the expression for the Hubble parameter in the slowly time varying region through the Einstein's equation written in spatially flat FLRW background (Friedmann equantion) with quasi de Sitter solution.  With having this information in our hand we will vary the scalar spectral tilt within the previously mentioned observational window obtained from Planck 2018 combined with the BICEP2/Keck Array BK15 datasets (Planck TT,TE,EE
+lowE+lensing+BK15  likelihood data) \cite{Akrami:2018odb} as appearing in the theoretically computed expression for the {\it Pancharatnam-Berry phase}.  This will serve the purpose and finally it is possible to obtain the numerically predicted value of the {\it Pancharatnam-Berry phase} which is consistent with the above mentioned combined observational constraints for other observables.
 \item \underline{In presence of Bell's inequality violation:}\\
 In this case we need to follow the same steps as stated in the previous point.  Along with that we need to supply the the conformal time dependent heavy mass profile which we have stated before.  Using this information along with five different prior values of the tensor-to-scalar ratio we will vary the scalar spectral tilt within the previously mentioned observational window obtained from Planck 2018 combined with the BICEP2/Keck Array BK15 datasets (Planck TT,TE,EE
+lowE+lensing+BK15  likelihood data) \cite{Akrami:2018odb} as appearing in the theoretically computed expression for the {\it Pancharatnam-Berry phase} in presence of Bell's inequality violation.  Though the conformal time dependent behaviour of the {\it Pancharatnam-Berry phase} in presence of Bell's inequality violation in this case will be dependent on the chosen profile structure of the heavy mass fluctuation,  though it is expected that the numerically predicted value of this phase will not change drastically for other choice of the heavy mass profile.     
 \end{enumerate}
 
 \item Next to give a model independent numerical estimation of the theoretically computed expression for the {\it Pancharatnam-Berry phase} in the primordial cosmological set up we are going to follow some different mentioned steps as appended below:
 \begin{enumerate}
 \item \underline{In absence of Bell's inequality violation:}\\
 In this case we are going to first of all fix the value of the scalar spectral tilt in five different range of values within the observed window predicted from from Planck 2018 combined with the BICEP2/Keck Array BK15 datasets (Planck TT,TE,EE
+lowE+lensing+BK15  likelihood data) \cite{Akrami:2018odb}.  With having this information in our hand we will vary the tensor-to-scalar ratio as appearing in the theoretically computed expression for the {\it Pancharatnam-Berry phase} where we will consider the range of variation below the predicted upper bound from observation.  This will serve the purpose and finally it is possible to obtain the numerically predicted value of the {\it Pancharatnam-Berry phase} which is consistent with the above mentioned combined observational constraints for other observables.
 \item \underline{In presence of Bell's inequality violation:}\\
 In this case we need to follow the same steps as stated in the previous point.  Along with that we need to supply the the conformal time dependent heavy mass profile which we have stated before.  Using this information along with five different prior values of the scalar spectral tilt within the observed window we will vary the tensor-to-scalar ratio below the previously mentioned observational upper bound.  Additionally we need provide the conformal time dependent profile of the heavy mass fluctuation,  though as we have mentioned earlier it is similarly expected here as well that the numerically predicted value of this phase will not change drastically for other choice of the profile.     
 \end{enumerate}
 \item Combination of the last two performed steps for non Bell violating and Bell violating case in primordial cosmological set up will fix the numerically predicted range of the {\it Pancharatnam-Berry phase} which confronts well with the previously mentioned observational constraints.  
 
 \item During performing the numerics and implement that to get relevant plots we have actually used the ratio of the {\it Pancharatnam-Berry phase} and the corresponding eigenvalue of the the {\it Lewis Riesenfeld} quantum operator i.e.  $\gamma^{\bf PB,NB}/\lambda^{\bf NB}$ for the non Bell violating case and $\gamma^{\bf PB,B}/\lambda^{\bf B}$.   For simplicity we have identified both of them to be the {\it Normalized Pancharatnam-Berry phase},  this terminology we are going to further use throughout our analysis performed in this section.
 
 \item Using the present analysis it is possible to put stringent constraint on the {\it Normalized Pancharatnam-Berry phase} without and with having Bell's inequality violation within the framework of primordial cosmology.  Since in this construction in the sub-Hubble region ($-k\tau\rightarrow \infty$ or $-k\tau\gg 1$) the Hamiltonian of the primordial cosmological perturbation can be expressed in terms of quantum mechanical Harmonic oscillator then in this regime the eigenvalue of the {\it Lewis Riesenfeld} quantum operator in absence of Bell's inequality violation can be expressed in terms discrete positive integers as:
 \bea \lim_{-k\tau\rightarrow \infty}\lambda^{\bf NB}:=\bigg(n+\frac{1}{2}\bigg)~~~~{\rm where}~~~n=\mathbb{Z}^{+}.\eea
 Here by fixing the values of $n$ one can further able to obtain simplified constraint on the {\it Pancharatnam-Berry phase} $\gamma^{\bf PB,NB}$ instead of fixing the ratio $\gamma^{\bf PB,NB}/\lambda^{\bf NB}$ for the non-Bell violating case.  On the other hand in the sub-Hubble region due to having additional contributions in the case for Bell's inequality violation it is not at all possible to treat the corresponding Hamiltonian of the primordial cosmological perturbation for the scalar modes as quantum mechanical Harmonic oscillator.  Consequently,  in this case the eigenvalue of  the {\it Lewis Riesenfeld} quantum operator cannot be represented in terms of discrete positive integers.  Instead of that it turns out to be that in this case the eigenvalue $\lambda^{\bf B}$ becomes continuous and for this reason in the Bell violating case the ratio $\gamma^{\bf PB,B}/\lambda^{\bf B}$ can only be constrained.  Now further if we want to remove the information regarding this eigenvalue and just only constrain the {\it Pancharatnam-Berry phase} $\gamma^{\bf PB,B}$,  then we need to integrate the over all possible eigenvalues values to get the following normalised version of the average value of the phase, which is given by:
 \bea \overline{\lim_{-k\tau\rightarrow \infty}\gamma^{\bf PB,B}(\tau)}:=\frac{\displaystyle \int d\lambda^{\bf B}~{\cal V}(\lambda^{\bf B})~\gamma^{\bf PB,B}(\tau)}{\displaystyle \int d\lambda^{\bf B}~{\cal V}(\lambda^{\bf B})},\eea
 where ${\cal V}(\lambda^{\bf B})$ is the distribution function of the continuous eigenvalue $\lambda^{\bf B}$ which one need to consider to compute this average value.

 \end{enumerate}  
 \subsection{Numerical results: Physical interpretation and numerical constraints }
The detailed interpretation of the plots are appended below point-wise:
\begin{itemize}
\item In fig.~(\ref{fig:1}\textcolor{Sepia}{(a)}) and fig.~(\ref{fig:1}\textcolor{Sepia}{(b)}),  we have plotted the behavior of the {\it Normalized Pancharatnam Berry phase} with scalar spectral tilt without having Bell's inequality violation in sub and super Hubble regime for massless case respectively.  For each cases we have fixed the value of tensor-to-scalar ratio at $r=0.05, 0.005,0.0005$ values.  We have drawn three vertical lines at $n_{\zeta}(k_{*})=0.9649- 0.0042$, $n_{\zeta}(k_{*})=0.9649$ and $n_{\zeta}(k_{*})=0.9649+ 0.0042$ (from left to right) which are drawn with 68\%~{\rm CL} observed value \cite{Akrami:2018odb}.  From this analysis we found for the massless field case without having the Bell's inequality violation following constraints from the plots appear in the sub and super Hubble region:
\bea && \underline{\rm Sub-Hubble~ region~with ~massless~field~(without~Bell's~inequality~violation):}~~~~~~\nonumber\\
&&\underline{{\rm For}~~r=0.05}~~\Longrightarrow~~ ~~~~~~~~-1.368 \le \frac{\gamma^{PB,NB}}{\lambda^{NB}}\le -1.337,\\
&&\underline{{\rm For}~~r=0.005}~~\Longrightarrow~~ ~~~~~~~-1.358 \le \frac{\gamma^{PB,NB}}{\lambda^{NB}}\le -1.327,\\
&&\underline{{\rm For}~~r=0.0005}~~\Longrightarrow~~ ~~~~~~-1.356 \le \frac{\gamma^{PB,NB}}{\lambda^{NB}}\le -1.325,\\ 
&&  \underline{\rm Super-Hubble~ region~with ~massless~field~(without~Bell's~inequality~violation):}~~~~~~\nonumber\\
&&\underline{{\rm For}~~r=0.05}~~\Longrightarrow~~~~~~~~~~~~~0.172 \le \frac{\gamma^{PB,NB}}{\lambda^{NB}}\le 0.191,\\
&&\underline{{\rm For}~~r=0.005}~~\Longrightarrow~~~~~~~~~~~0.168 \le \frac{\gamma^{PB,NB}}{\lambda^{NB}}\le 0.187,\\
&&\underline{{\rm For}~~r=0.0005}~~\Longrightarrow~~~~~~~~~~0.167 \le \frac{\gamma^{PB,NB}}{\lambda^{NB}}\le 0.186.\eea

\item In fig.~(\ref{fig:2}\textcolor{Sepia}{(a)}) and fig.~(\ref{fig:2}\textcolor{Sepia}{(b)}),  we have plotted the behavior of the real and imaginary part of complex {\it Normalized Pancharatnam Berry phase} with scalar spectral tilt without having Bell's inequality violation at the sub to super Hubble crossing point for massless case respectively.  From this analysis we found for the massless field case without having the Bell's inequality violation following constraints from the plots appear at the sub Hubble to super Hubble crossing point:
\bea && \underline{\rm Horizon-crossing~point~with ~massless~field~(without~Bell's~inequality~violation):}~~~~~~\nonumber\\
&&\underline{\rm From~real~part~(Oscillatory~amplitude~contribution):}~~~~~~\nonumber\\
&&\underline{{\rm For}~~r=0.05}~~\Longrightarrow~~ ~~~~~~~~0.168 \le \frac{{\rm Re}\left[\gamma^{PB,NB}\right]}{\lambda^{NB}}\le 0.192,\\
&&\underline{{\rm For}~~r=0.005}~~\Longrightarrow~~ ~~~~~~0.165 \le \frac{{\rm Re}\left[\gamma^{PB,NB}\right]}{\lambda^{NB}}\le 0.190,\\
&&\underline{{\rm For}~~r=0.0005}~~\Longrightarrow~~ ~~~~~0.163 \le \frac{{\rm Re}\left[\gamma^{PB,NB}\right]}{\lambda^{NB}}\le 0.188,\\
&&\underline{\rm From~imaginary~part~(Growing~amplitude~contribution):}~~~~~~\nonumber\\
&&\underline{{\rm For}~~r=0.05}~~\Longrightarrow~~ ~~~~~~~~-0.195 \le \frac{{\rm Im}\left[\gamma^{PB,NB}\right]}{\lambda^{NB}}\le -0.176,\\
&&\underline{{\rm For}~~r=0.005}~~\Longrightarrow~~ ~~~~~~~-0.194 \le \frac{{\rm Im}\left[\gamma^{PB,NB}\right]}{\lambda^{NB}}\le -0.174,\\
&&\underline{{\rm For}~~r=0.0005}~~\Longrightarrow~~ ~~~~~~-0.193 \le \frac{{\rm Im}\left[\gamma^{PB,NB}\right]}{\lambda^{NB}}\le -0.172.\eea

\item In fig.~(\ref{fig:3}),  we have plotted the behavior of the {\it Normalized Pancharatnam Berry phase} with scalar spectral tilt with having Bell's inequality violation at the sub Hubble region for massive/heavy field case.   From this analysis we found for the massive/heavy field case with having the Bell's inequality violation following constraints from the plots appear in the sub-Hubble region:
\bea && \underline{\rm Sub-Hubble~region~with ~massive~field~(with~Bell's~inequality~violation):}~~~~~~\nonumber\\
&&\underline{{\rm For}~~r=0.05}~~\Longrightarrow~~ ~~~~~~~~1.216\times 10^{-3} \le \frac{\gamma^{PB,B}}{\lambda^{B}}\le 1.221\times 10^{-3},\\
&&\underline{{\rm For}~~r=0.005}~~\Longrightarrow~~ ~~~~~~1.206\times 10^{-3} \le \frac{\gamma^{PB,B}}{\lambda^{B}}\le 1.211\times 10^{-3},\\
&&\underline{{\rm For}~~r=0.0005}~~\Longrightarrow~~ ~~~~~1.204\times 10^{-3} \le \frac{\gamma^{PB,B}}{\lambda^{B}}\le 1.209\times 10^{-3}.\eea

\item In fig.~(\ref{fig:4}\textcolor{Sepia}{(a)}) and fig.~(\ref{fig:4}\textcolor{Sepia}{(b)}),  we have plotted the behavior of the real and imaginary part of complex {\it Normalized Pancharatnam Berry phase} with scalar spectral tilt with having Bell's inequality violation in the sub Hubble region for partially massless case respectively.  From this analysis we found for the partially massless feild case with having the Bell's inequality violation following constraints from the plots appear in the sub Hubble region:
\bea && \underline{\rm Sub-Hubble~region~with ~partially~massless~field~(with~Bell's~inequality~violation):}~~~~~~\nonumber\\
&&\underline{\rm From~real~part~(Oscillatory~amplitude~contribution):}~~~~~~\nonumber\\
&&\underline{{\rm For}~~r=0.05}~~\Longrightarrow~~ ~~~~~~~~-8.8 \le \frac{{\rm Re}\left[\gamma^{PB,NB}\right]}{\lambda^{NB}}\le -7.8,\\
&&\underline{{\rm For}~~r=0.005}~~\Longrightarrow~~ ~~~~~~-8.7 \le \frac{{\rm Re}\left[\gamma^{PB,NB}\right]}{\lambda^{NB}}\le -7.7,\\
&&\underline{{\rm For}~~r=0.0005}~~\Longrightarrow~~ ~~~~~8.6 \le \frac{{\rm Re}\left[\gamma^{PB,NB}\right]}{\lambda^{NB}}\le -7.6,\\
&&\underline{\rm From~imaginary~part~(Decaying~amplitude~contribution):}~~~~~~\nonumber\\
&&\underline{{\rm For}~~r=0.05}~~\Longrightarrow~~ ~~~~~~~~11.5 \le \frac{{\rm Im}\left[\gamma^{PB,NB}\right]}{\lambda^{NB}}\le 13.5,\\
&&\underline{{\rm For}~~r=0.005}~~\Longrightarrow~~ ~~~~~~~11.3 \le \frac{{\rm Im}\left[\gamma^{PB,NB}\right]}{\lambda^{NB}}\le 13.3,\\
&&\underline{{\rm For}~~r=0.0005}~~\Longrightarrow~~ ~~~~~~11.2 \le \frac{{\rm Im}\left[\gamma^{PB,NB}\right]}{\lambda^{NB}}\le 13.2.\eea

\item In fig.~(\ref{fig:5}\textcolor{Sepia}{(a)}) and fig.~(\ref{fig:5}\textcolor{Sepia}{(b)}),  we have plotted the behavior of the real and imaginary part of complex {\it Normalized Pancharatnam Berry phase} with scalar spectral tilt with having Bell's inequality violation in the super Hubble region for massive field case respectively.  From this analysis we found for the partially massive field case with having the Bell's inequality violation following constraints from the plots appear in the super Hubble region:
\bea && \underline{\rm Super-Hubble~region~with ~massive~field~(with~Bell's~inequality~violation):}~~~~~~\nonumber\\
&&\underline{\rm From~real~part~(Oscillatory~amplitude~contribution):}~~~~~~\nonumber\\
&&\underline{{\rm For}~~r=0.05}~~\Longrightarrow~~ ~~~~~~~~7.01\times 10^{-2} \le \frac{{\rm Re}\left[\gamma^{PB,NB}\right]}{\lambda^{NB}}\le 7.03\times 10^{-2},\\
&&\underline{{\rm For}~~r=0.005}~~\Longrightarrow~~ ~~~~~~6.95\times 10^{-2} \le \frac{{\rm Re}\left[\gamma^{PB,NB}\right]}{\lambda^{NB}}\le 6.97\times 10^{-2},\\
&&\underline{{\rm For}~~r=0.0005}~~\Longrightarrow~~ ~~~~~6.94\times 10^{-2} \le \frac{{\rm Re}\left[\gamma^{PB,NB}\right]}{\lambda^{NB}}\le 6.96\times 10^{-2},\\
&&\underline{\rm From~imaginary~part~(Growing~amplitude~contribution):}~~~~~~\nonumber\\
&&\underline{{\rm For}~~r=0.05}~~\Longrightarrow~~ ~~~~~~~~-4.175\times 10^{-1} \le \frac{{\rm Im}\left[\gamma^{PB,NB}\right]}{\lambda^{NB}}\le -4.16\times 10^{-1},\\
&&\underline{{\rm For}~~r=0.005}~~\Longrightarrow~~ ~~~~~~~-4.141\times 10^{-1} \le \frac{{\rm Im}\left[\gamma^{PB,NB}\right]}{\lambda^{NB}}\le -4.125\times 10^{-1},\\
&&\underline{{\rm For}~~r=0.0005}~~\Longrightarrow~~ ~~~~~~-4.137\times 10^{-1} \le \frac{{\rm Im}\left[\gamma^{PB,NB}\right]}{\lambda^{NB}}\le -4.121\times 10^{-1}.~~~~~~~~~~\eea 

\item In fig.~(\ref{fig:6}),  we have plotted the behavior of the purely imaginary part of the {\it Normalized Pancharatnam Berry phase} with scalar spectral tilt with having Bell's inequality violation at the super Hubble region for partially massless field case.   From this analysis we found for the partially massless field case with having the Bell's inequality violation following constraints from the plots appear in the super-Hubble region:
\bea && \underline{\rm Super-Hubble~region~with ~partially~massless~field~(with~Bell's~inequality~violation):}~~~~~~\nonumber\\
&&\underline{{\rm For}~~r=0.05}~~\Longrightarrow~~ ~~~~~~~~11.5 \le \frac{\gamma^{PB,B}}{\lambda^{B}}\le 13.5,\\
&&\underline{{\rm For}~~r=0.005}~~\Longrightarrow~~ ~~~~~~11.2 \le \frac{\gamma^{PB,B}}{\lambda^{B}}\le 13.4,\\
&&\underline{{\rm For}~~r=0.0005}~~\Longrightarrow~~ ~~~~~11.1 \le \frac{\gamma^{PB,B}}{\lambda^{B}}\le 13.3.\eea

\item In fig.~(\ref{fig:7}\textcolor{Sepia}{(a)}) and fig.~(\ref{fig:7}\textcolor{Sepia}{(b)}),  we have plotted the behavior of the real and imaginary part of complex {\it Normalized Pancharatnam Berry phase} with scalar spectral tilt with having Bell's inequality violation at the sub Hubble to super Hubble crossing point for partially massless field case respectively.   From this analysis we found for the partially massless field case with having the Bell's inequality violation following constraints from the plots appear at the sub Hubble to super Hubble crossing point:
\bea && \underline{\rm Horizon-crossing~point~with ~partially~massless~field~(with~Bell's~inequality~violation):}~~~~~~\nonumber\\
&&\underline{\rm From~real~part~(Oscillatory~amplitude~contribution):}~~~~~~\nonumber\\
&&\underline{{\rm For}~~r=0.05}~~\Longrightarrow~~ ~~~~~~~~-7.2 \le \frac{{\rm Re}\left[\gamma^{PB,NB}\right]}{\lambda^{NB}}\le -8.2,\\
&&\underline{{\rm For}~~r=0.005}~~\Longrightarrow~~ ~~~~~~-7.3 \le \frac{{\rm Re}\left[\gamma^{PB,NB}\right]}{\lambda^{NB}}\le 8.3,\\
&&\underline{{\rm For}~~r=0.0005}~~\Longrightarrow~~ ~~~~~-7.35 \le \frac{{\rm Re}\left[\gamma^{PB,NB}\right]}{\lambda^{NB}}\le -8.35,\\
&&\underline{\rm From~imaginary~part~(Decaying~amplitude~contribution):}~~~~~~\nonumber\\
&&\underline{{\rm For}~~r=0.05}~~\Longrightarrow~~ ~~~~~~~~11.5 \le \frac{{\rm Im}\left[\gamma^{PB,NB}\right]}{\lambda^{NB}}\le 13.5,\\
&&\underline{{\rm For}~~r=0.005}~~\Longrightarrow~~ ~~~~~~~11.4\le \frac{{\rm Im}\left[\gamma^{PB,NB}\right]}{\lambda^{NB}}\le 13.4,\\
&&\underline{{\rm For}~~r=0.0005}~~\Longrightarrow~~ ~~~~~~11.3 \le \frac{{\rm Im}\left[\gamma^{PB,NB}\right]}{\lambda^{NB}}\le 13.3.~~~~~~~~~~\eea   

\item In fig.~(\ref{fig:8}\textcolor{Sepia}{(a)}) and fig.~(\ref{fig:8}\textcolor{Sepia}{(b)}),  we have plotted the behavior of the real and imaginary part of complex {\it Normalized Pancharatnam Berry phase} with scalar spectral tilt with having Bell's inequality violation at the sub Hubble to super Hubble crossing point for massive field case respectively.   From this analysis we found for the massive field case with having the Bell's inequality violation following constraints from the plots appear at the sub Hubble to super Hubble crossing point:
\bea && \underline{\rm Horizon-crossing~point~with ~massive~field~(with~Bell's~inequality~violation):}~~~~~~\nonumber\\
&&\underline{\rm From~real~part~(Oscillatory~amplitude~contribution):}~~~~~~\nonumber\\
&&\underline{{\rm For}~~r=0.05}~~\Longrightarrow~~ ~~~~~~~~0.956 \le \frac{{\rm Re}\left[\gamma^{PB,NB}\right]}{\lambda^{NB}}\le 0.959,\\
&&\underline{{\rm For}~~r=0.005}~~\Longrightarrow~~ ~~~~~~0.947 \le \frac{{\rm Re}\left[\gamma^{PB,NB}\right]}{\lambda^{NB}}\le 0.951,\\
&&\underline{{\rm For}~~r=0.0005}~~\Longrightarrow~~ ~~~~~0.946 \le \frac{{\rm Re}\left[\gamma^{PB,NB}\right]}{\lambda^{NB}}\le 0.950,\\
&&\underline{\rm From~imaginary~part~(Growing~amplitude~contribution):}~~~~~~\nonumber\\
&&\underline{{\rm For}~~r=0.05}~~\Longrightarrow~~ ~~~~~~~~-1.567 \le \frac{{\rm Im}\left[\gamma^{PB,NB}\right]}{\lambda^{NB}}\le -1.563,\\
&&\underline{{\rm For}~~r=0.005}~~\Longrightarrow~~ ~~~~~~~-1.554\le \frac{{\rm Im}\left[\gamma^{PB,NB}\right]}{\lambda^{NB}}\le -1.549,\\
&&\underline{{\rm For}~~r=0.0005}~~\Longrightarrow~~ ~~~~~~-1.552 \le \frac{{\rm Im}\left[\gamma^{PB,NB}\right]}{\lambda^{NB}}\le -1.547.~~~~~~~~~~\eea   

\item In fig.~(\ref{fig:9}\textcolor{Sepia}{(a)}) and fig.~(\ref{fig:9}\textcolor{Sepia}{(b)}),  we have plotted the behavior of the {\it Normalized Pancharatnam Berry phase} with tensor-to-scalar ratio without having Bell's inequality violation in sub and super Hubble regime for massless case respectively.  From this analysis we found for the massless field case without having the Bell's inequality violation at the value of tensor-to-scalar ratio,  $r=0.05$, $r=0.005$ and $r=0.0005$ that the constrained value of {\it Normalized Pancharatnam Berry phase} from the plots drawn in the sub and super Hubble region are perfectly match with the predictions obtained from fig.~(\ref{fig:1}\textcolor{Sepia}{(a)}) and fig.~(\ref{fig:1}\textcolor{Sepia}{(b)}) respectively.  This cross-check actually justify the correctness of our analysis and the obtained constraint on the value of {\it Normalized Pancharatnam Berry phase} in the sub and super Hubble region for the massless field case.

\item In fig.~(\ref{fig:10}\textcolor{Sepia}{(a)}) and fig.~(\ref{fig:10}\textcolor{Sepia}{(b)}),  we have plotted the behavior of the real and imaginary part of complex {\it Normalized Pancharatnam Berry phase} with tensor-to-scalar ratio without having Bell's inequality violation at the sub to super Hubble crossing point for massless case respectively.   From this analysis we found for the massless field case without having the Bell's inequality violation at the value of tensor-to-scalar ratio,  $r=0.05$, $r=0.005$ and $r=0.0005$ that the constrained value of {\it Normalized Pancharatnam Berry phase} from the plots drawn at the sub to super Hubble horizon crossing point are perfectly match with the predictions obtained from fig.~(\ref{fig:2}\textcolor{Sepia}{(a)}) and fig.~(\ref{fig:2}\textcolor{Sepia}{(b)}) respectively.  This cross-check actually justify the correctness of our analysis and the obtained constraint on the value of {\it Normalized Pancharatnam Berry phase} at the sub to super Hubble horizon crossing point for the massless field case. 

\item In fig.~(\ref{fig:11}),  we have plotted the behavior of the {\it Normalized Pancharatnam Berry phase} with tensor-to-scalar ratio with having Bell's inequality violation in the sub Hubble region for massive/heavy field case.  From this analysis we found for the massless field case without having the Bell's inequality violation at the value of tensor-to-scalar ratio,  $r=0.05$, $r=0.005$ and $r=0.0005$ that the constrained value of {\it Normalized Pancharatnam Berry phase} from the plots drawn in the sub Hubble region are perfectly match with the predictions obtained from fig.~(\ref{fig:3}).  This cross-check actually justify the correctness of our analysis and the obtained constraint on the value of {\it Normalized Pancharatnam Berry phase} in the sub Hubble region for massive/heavy field case. 




\item In fig.~(\ref{fig:12}\textcolor{Sepia}{(a)}) and fig.~(\ref{fig:12}\textcolor{Sepia}{(b)}),  we have plotted the behavior of the real and imaginary part of complex {\it Normalized Pancharatnam Berry phase} with tensor-to-scalar ratio with having Bell's inequality violation in the super Hubble region for massive field case respectively.   From this analysis we found for the massive field case without having the Bell's inequality violation at the value of tensor-to-scalar ratio,  $r=0.05$, $r=0.005$ and $r=0.0005$ that the constrained value of {\it Normalized Pancharatnam Berry phase} from the plots drawn in the super Hubble region are perfectly match with the predictions obtained from fig.~(\ref{fig:5}\textcolor{Sepia}{(a)}) and fig.~(\ref{fig:5}\textcolor{Sepia}{(b)}) respectively.  This cross-check actually justify the correctness of our analysis and the obtained constraint on the value of {\it Normalized Pancharatnam Berry phase} in the super Hubble region for massive/heavy field case.   

\item In fig.~(\ref{fig:13}\textcolor{Sepia}{(a)}) and fig.~(\ref{fig:13}\textcolor{Sepia}{(b)}),  we have plotted the behavior of the real and imaginary part of complex {\it Normalized Pancharatnam Berry phase} with tensor-to-scalar ratio with having Bell's inequality violation in the sub Hubble region for partially massless field case respectively.   From this analysis we found for the partially massless field case without having the Bell's inequality violation at the value of tensor-to-scalar ratio,  $r=0.05$, $r=0.005$ and $r=0.0005$ that the constrained value of {\it Normalized Pancharatnam Berry phase} from the plots drawn in the sub Hubble region are perfectly match with the predictions obtained from fig.~(\ref{fig:4}\textcolor{Sepia}{(a)}) and fig.~(\ref{fig:4}\textcolor{Sepia}{(b)}) respectively.  This cross-check actually justify the correctness of our analysis and the obtained constraint on the value of {\it Normalized Pancharatnam Berry phase} in the sub Hubble region for partially massless field case.   

\item In fig.~(\ref{fig:14}),  we have plotted the behavior of the purely imaginary part of the {\it Normalized Pancharatnam Berry phase} with tensor-to-scalar ratio with having Bell's inequality violation in the super Hubble region for partially massless field case respectively.   From this analysis we found for the partially massless field case without having the Bell's inequality violation at the value of tensor-to-scalar ratio,  $r=0.05$, $r=0.005$ and $r=0.0005$ that the constrained value of {\it Normalized Pancharatnam Berry phase} from the plots drawn in the super Hubble region are perfectly match with the predictions obtained from fig.~(\ref{fig:6}).  This cross-check actually justify the correctness of our analysis and the obtained constraint on the value of {\it Normalized Pancharatnam Berry phase} in the super Hubble region for partially massless field case.   

\item In fig.~(\ref{fig:15}\textcolor{Sepia}{(a)}) and fig.~(\ref{fig:15}\textcolor{Sepia}{(b)}),  we have plotted the behavior of the real and imaginary part of complex {\it Normalized Pancharatnam Berry phase} with tensor-to-scalar ratio with having Bell's inequality violation at the sub to super Hubble horizon crossing point for massive field case respectively.  From this analysis we found for the massive field case without having the Bell's inequality violation at the value of tensor-to-scalar ratio,  $r=0.05$, $r=0.005$ and $r=0.0005$ that the constrained value of {\it Normalized Pancharatnam Berry phase} from the plots drawn at the horizon crossing point are perfectly match with the predictions obtained from fig.~(\ref{fig:8}\textcolor{Sepia}{(a)}) and fig.~(\ref{fig:8}\textcolor{Sepia}{(b)}) respectively.  This cross-check actually justify the correctness of our analysis and the obtained constraint on the value of {\it Normalized Pancharatnam Berry phase} at the horizon crossing point for massive field case.    

\item Last but not the least,  in fig.~(\ref{fig:16}\textcolor{Sepia}{(a)}) and fig.~(\ref{fig:16}\textcolor{Sepia}{(b)}),  we have plotted the behavior of the real and imaginary part of complex {\it Normalized Pancharatnam Berry phase} with tensor-to-scalar ratio with having Bell's inequality violation at the sub to super Hubble horizon crossing point for partially massless field case respectively.  From this analysis we found for the partially massless field case without having the Bell's inequality violation at the value of tensor-to-scalar ratio,  $r=0.05$, $r=0.005$ and $r=0.0005$ that the constrained value of {\it Normalized Pancharatnam Berry phase} from the plots drawn at the horizon crossing point are perfectly match with the predictions obtained from fig.~(\ref{fig:7}\textcolor{Sepia}{(a)}) and fig.~(\ref{fig:7}\textcolor{Sepia}{(b)}) respectively.  This cross-check actually justify the correctness of our analysis and the obtained constraint on the value of {\it Normalized Pancharatnam Berry phase} at the horizon crossing point for partially massless field case.

\end{itemize}

\section{Conclusion} 
\label{CON}

The concluding remarks of this paper are appended below point-wise:

\begin{itemize}
\item In this paper,  we have started our discussion with a mini review on finding the geometric phase,  which is commonly known as the {\it Pancharatnam Berry phase} using the well known {\it Lewis Riesenfeld invariant quantum operator method} to find continuous eigen values within the framework of inverted Harmonic oscillator having time dependent effective frequency.  This general discussion will be helpful for the development and construction of rest of the paper which is devoted to study to find out the {\it Pancharatnam Berry phase} within the framework of primordial cosmological perturbation theory with scalar modes.

\item After giving a mini review of the framework within the context of quantum mechanics we have discussed the outcomes of the non-Bell's inequality violating and  Bell's inequality violating scenario at the classical level by solving the classical equation of motion,  which is commonly referred as the {\it Mukhanov Sasaki equation} within the framework of cosmology.

\item Before going to the further details in the computation we have considered the sub Hubble,  super Hubble and sub to super Hubble transition separately to understand the internal underlying features of the Hamiltonian of the primordial cosmological perturbation before quantization  both in the absence and presence of Bell's inequality violation.

\item Next we promote all the classical conformal time dependent variables to quantum mechanical operator and without explicit using the oscillator algebra written in terms of the annihilation and creation operators we promote the classical Hamiltonian as  quantum Hamiltonian. Then we have explicitly demonstrated that using the well known {\it Lewis Riesenfeld invariant quantum operator method} how one can able to explicitly compute the conformal time dependent amplitude part of the wave function which is treated to be the eigenfunction in this context.

\item Further using the {\it Milne Penny equation} we have written the explicit form of the {\it Lewis Riesenfeld} phase factor,  the dynamical phase factor and the expression for the {\it Pancharatnam Berry phase} which we have extracted using both of these expressions computed for having no Bell's inequality violation and an explicit Bell's inequality violation within the framework of primordial cosmological perturbation theory of scalar modes.

\item Expressing every contribution in terms of the {\it slow roll parameters} is the most useful and realistic part of the analytical computations of {\it Pancharatnam Berry phase} using {\it Lewis Riesenfeld invariant quantum operator} formalism.  The prime reason is,  this identification actually helps us to connect this computed results with the observables like the amplitude of power spectrum and spectral index/tilt from scalar modes and tensor-to-scalar ratio,  which further helps us to give stringent numerical constraint on the theoretically computed {\it Pancharatnam Berry phase} in the present context of discussion.

\end{itemize}

The further future prospects of the presented work is appended below point-wise:
\begin{itemize}

\item The present study of finding {\it Normalized Pancharatnam Berry phase} for  the massless,  partially massless and massive/heavy field case is performed in presence of pure quantum state.  It is possible to extend this analysis for mixed quantum state which allows to construct the density matrix operators $\rho$ and $\sigma$ and hence a very important quantum information theoretic measure,  which is knows as {\it Fidelity},  ${\cal F}(\rho,\sigma)$ through which it is possible to measure the closeness of two quantum states in the present framework.  In a more technical language this measure of closeness of two quantum states quantified by the following expression:  
\bea {\cal F}(\rho,\sigma):=\bigg({\rm Tr}\left[~\sqrt{\rho}~\sigma~\sqrt{\rho}~\right]\bigg)^2=\left|\langle \Psi_{\rho}|\Psi_{\sigma}\rangle\right|^2~~({\rm Using~Uhlmann's~theorem})\eea
where the corresponding density matrices $\rho$ and $\sigma$ are defined as, $\rho=|\Psi_{\rho}\rangle\langle \Psi_{\rho}|~~\&~~\sigma=|\Psi_{\sigma}\rangle\langle \Psi_{\sigma}|.$
Here the {\it Fidelity} have the following properties which are the immediate consequences of the {\it Uhlmann's theorem} for mixed quantum states:
\begin{enumerate}
\item \underline{Symmetric:}~~~${\cal F}(\rho,\sigma)={\cal F}(\sigma,\rho)$. 
\item \underline{Positivity Bound:}~~$0\leq {\cal F}(\rho,\sigma)\leq 1$  which can be shown using {\it Cauchy–Schwarz  inequality}.
\item \underline{Normalization Condition:}~~~${\cal F}(\rho,\rho)=1$.
\item \underline{Invariance under Unitary transformation:}~~~\\
${\cal F}(\rho,\sigma)~~~\xrightarrow[]{{\cal U}}~~~\widetilde{{\cal F}(\rho,\sigma)}:={\cal F}({\cal U}\rho{\cal U}^{\dagger},{\cal U}\sigma{\cal U}^{\dagger})={\cal F}(\rho,\sigma)$ where ${\cal U}$ is the Unitary operator.
\end{enumerate} 
The most important outcome is in this computation for the mixed quantum states a generalized geometric phase arises,  which is known as the {\it Uhlmann phase} \cite{1986RpMP...24..229U,Kirklin:2020zic,Kirklin:2020qtv,Kirklin:2019ror,Kirklin:2021ipb,PhysRevLett.112.130401,PhysRevA.67.032110} which can be expressed in terms of {\it Fidelity} by the following expression:
\bea \gamma_{\rm Uhlmann}:={\rm arg}\bigg[\bigg({\rm Tr}\left[~\sqrt{\rho}~\sigma~\sqrt{\rho}~\right]\bigg)\bigg]=\sqrt{{\cal F}(\rho,\sigma)}.\eea
In future it is possible to generalize the present computation for mixed states in primordial cosmological perturbation theory set up and it be really interesting explicitly find the corresponding {\it Uhlmann phase} \cite{1986RpMP...24..229U,Kirklin:2020zic,Kirklin:2020qtv,Kirklin:2019ror,Kirklin:2021ipb,PhysRevLett.112.130401,PhysRevA.67.032110} from the computation of {\it Fidelity}.  Additionally,  for this extended mixed state framework one can further study the relationship among {\it Uhlmann phase},  quantum speed limit and the trace distance which gives an additional constraint:
\bea 1-\gamma_{\rm Uhlmann}\leq {\cal D}(\rho,\sigma):=\frac{1}{2}||\rho-\sigma||_{\rm Tr}\leq \sqrt{1-\gamma^2_{\rm Uhlmann}}.\eea

\item In the present work we have only considered canonical single field model in the spatially flat De Sitter background.  This computation can be further generalized in presence of non-minimal theories like,  $P(X,\phi)$ where $X=-(\partial \phi)^2/2$.  Also this idea can be further generalized for all classes of {\it Effective Field Theory} \cite{Cheung:2007st,Weinberg:2008hq,Choudhury:2017glj,Gubitosi:2012hu,Giblin:2017qjp,Naskar:2017ekm,Delacretaz:2015edn,Delacretaz:2016nhw,Choudhury:2015pqa,Choudhury:2014sua,Choudhury:2014kma,Choudhury:2014uxa,Choudhury:2014sxa,Choudhury:2013iaa,Choudhury:2015yna, Choudhury:2012yh,Chen:2006nt,Chen:2009zp,Chen:2009we,Chen:2012ge} framework as well.  In both the cases an additional contribution of effective sound-speed parameter $c_S$ play significant role which will going to appear in the expressions derived for {\it Normalized Pancharatnam Berry phase} in absence and presence of Bell's inequality violation.

\item Till now we have not considered the effect of quantum entanglement \cite{Maldacena:2012xp,Akhtar:2019qdn,Bhattacherjee:2019eml,Choudhury:2020ivj,Choudhury:2017qyl,Choudhury:2017bou,Banerjee:2020ljo,Kanno:2014lma,Kanno:2014ifa,Kanno:2014bma,Kanno:2015lja,Kanno:2015ewa,Kanno:2016gas,Kanno:2016qcc,Albrecht:2018prr,Kanno:2019gqw,Kanno:2021gpt} without and with having squeezed states \cite{Martin:2007bw,Lewis:1968tm,Albrecht:1992kf,Grishchuk:1989ss,Grishchuk:1990bj,Caves:1985zz,Colas:2021llj,Choudhury:2020hil,Parikh:2020fhy,Parikh:2020kfh,Haque:2020pmp,Martinez-Tibaduiza:2020vwr,Bhargava:2020fhl,Bhattacharyya:2020kgu,Bhattacharyya:2020rpy,Martin:2019oqq,Martin:2019wta,Gubitosi:2017zoj,Bianchi:2016tmw,Bianchi:2015fra,Martin:2016nrr,Grain:2019vnq,Martin:2016tbd,Adhikari:2021pvv,Haque:2020pmp} in the primordial cosmological set up.  In presence of this effect it also important to study the corresponding geometric phase appearing in the present cosmological framework.

\item Further generalization for the two field or multi-field model and for the {\it Effective Field Theory} \cite{Senatore:2010wk} of multi-field scenario can be also very useful to quantify the corresponding geometric phase appearing in the present cosmological framework. 

\item Last but not the least,  in this paper we have not discussed anything related to quantum quench \cite{Mandal:2015kxi,Kulkarni:2018ahv,Mandal:2013id,Banerjee:2019ilw,Das:2019cgl,Das:2019qaj,Caputa:2017ixa,Das:2016lla,Das:2015jka,Das:2014hqa,Das:2014jna,Basu:2013soa,Banerjee:2021lqu} and its direct impact in the {\it Normalized Pancharatnam Berry phase}.  In near future we will explore this possibility.

\end{itemize}

\acknowledgments
The research fellowship of SC is supported by the J.  C.  Bose National Fellowship of Sudhakar Panda.  SC also would line to thank School of Physical Sciences, NISER,  Bhubaneswar for providing the work friendly environment. SC also thank all the members of our newly formed virtual international non-profit consortium Quantum Structures of the Space-Time \& Matter (QASTM) for elaborative discussions.  Last but not the least,  we would like to acknowledge our debt to the people belonging to the various part of the world for their generous and steady support for research in natural sciences.

\phantomsection
\addcontentsline{toc}{section}{References}
\bibliographystyle{utphys}
\bibliography{references_CPBP}

\end{document}